%% file: final_arXiv.tex
\documentclass{jfm}
\usepackage{graphicx}
\usepackage{epstopdf, epsfig}
\usepackage[utf8]{inputenc}
\usepackage{amsmath}
\usepackage{float}
\usepackage{amssymb}
\usepackage{subcaption}
\usepackage{natbib}
\usepackage{mathrsfs,xcolor}

\newcommand{\pphi}{\boldsymbol{\phi}}
\newcommand{\uu}{\boldsymbol{u}}

\newcommand{\dUdy}{\frac{dU}{dy}}
\newcommand{\dupdt}{\frac{\partial u^{p}}{\partial t}}

\newcommand{\ddx}[2]{\frac{\partial #1}{\partial x_{#2}}}
\newcommand{\deltau}[1]{\frac{\partial^{2} #1}{\partial x_{j} \partial x_{j}}}
\newcommand{\oL}[2]{#1 \slash L_#2}
\newcommand{\toT}{t\slash T}

\newcommand{\expbig} {e^{ 2 \pi \mathrm{i} (l \oL{x}{x} + k \oL{z}{z} + f \toT)}}
\newcommand{\expmbig}{e^{-2 \pi \mathrm{i} (l \oL{x}{x} + k \oL{z}{z} + f \toT)}}
\newcommand{\bxi}  {\boldsymbol{\xi}}
\newcommand{\bu}   {\boldsymbol{u}}

\newcommand{\bx}   {\boldsymbol{x}}

\newcommand{\nt}    {n_t}
\newcommand{\ntmone}   {\left(\nt - 1\right)}

\newcommand{\TDMD}   {\Delta t}

\newcommand{\icomp} {\mathrm{i}}

\newcommand{\transpose}		  {\mathsf{T}}

\DeclareMathOperator*{\argmax}	{{\mathrm{ arg\,max}}}
\newcommand{\card}[1]		{\vert{#1}\vert}
\newcommand{\be}	{\begin{equation}}
\newcommand{\ee}	{ \end{equation}}
\newcommand{\ra}		{\rightarrow}

\newcommand{\berengeredeux}[1]   {{\color{black}  {#1}}}
\newcommand{\srikanth}[1] {{{#1}}}
\newcommand{\berengere}[1]   {{{#1}}}
\newcommand{\Lionel}[1]   {{\color{black} {#1}}}

\shorttitle{Spatio-temporal POD of turbulent channel flow}
\shortauthor{S. Derebail Muralidhar, B. Podvin, L. Mathelin and Y. Fraigneau}

\title{Spatio-temporal Proper Orthogonal Decomposition of turbulent channel flow}

\author{Srikanth Derebail Muralidhar\aff{1}, 
Bérengère Podvin\aff{1}\corresp{\email{Berengere.Podvin@limsi.fr}}, \\ 
Lionel Mathelin\aff{1}$^{,}$\aff{2} \and
Yann Fraigneau\aff{1}}

\affiliation{
\aff{1}LIMSI, CNRS, Université Paris-Saclay, 91403 Orsay Cedex, France
\aff{2} Dpt. Applied Mathematics, Univ. Washington, Seattle, WA, USA
}

\begin{document}

\maketitle

\begin{abstract}
\berengere{An extension of Proper Orthogonal Decomposition is applied
to the wall layer of a turbulent channel flow
($\mathrm{Re}_{\tau}=590$), so that empirical eigenfunctions are defined in
 both space and time.}
Due to the statistical symmetries of the flow, the eigenfunctions are associated
 with individual wavenumbers and frequencies.
\berengere{Self-similarity of the dominant eigenfunctions,
consistent with wall-attached
structures transferring energy into the core region,
 is established.
 The most energetic modes
 are characterized by a fundamental time scale in
the range 200-300  viscous wall units.
 The full spatio-temporal decomposition} provides a natural measure of the convection velocity of structures, with a characteristic value of 12$u_{\tau}$ in the wall layer.
 Finally, we show that the energy budget can be split into 
specific contributions for each mode, \berengeredeux{which provides a closed-form expression for nonlinear effects.}
% and provides a new closure formulation for the quadratic terms.

\end{abstract}

\begin{keywords}
%Authors should not enter keywords on the manuscript, as these must be chosen by the author during the online submission process and will then be added during the typesetting process

%in an interval $[b,c]$
\end{keywords}

\section{Introduction}
Proper Orthogonal Decomposition (POD) was first introduced in turbulence by \citet{kn:lumleyPOD}.
Its derivation stemmed 
from the Karhunen-Lo\`eve (KL) decomposition \citep{kn:loeve} 
which represents a square-integrable 
centered stochastic process in the time domain 
$U(t)$ as an infinite linear combination of orthogonal functions. 
If $t$ is defined over a finite range, then
\begin{equation}
U(t) = \sum_n a^{n} \chi^{n}(t),
\end{equation}
where $a^{n}$ is stochastic and $\chi^{n}$ are \berengeredeux{orthogonal, square integrable,} functions.
\berengeredeux{In all that follows the superscript refers to the mode index.}
It is important to note that $U(t)$ represents a stochastic variable and not
a sample. Realizations of $U(t)$ will be noted $u(t)$. 
The functions $\chi^{n}(t)$ are the eigenfunctions of the covariance function $K_{U}(t,t') = \mathbb{E}[U_t U_{t'}]$, where the operator $\mathbb{E}$ refers to expectation with respect to the measure of $U$. 
In the Karhunen-Lo\`eve derivation, the variable $t$ corresponds to time, but it could indicate any other variable - such as space.

\citet{kn:lumleyPOD} 
\berengeredeux{(see also \citet{berkooz1993})}  
adapted the decomposition to Fluid Mechanics: 
the samples were constituted by flow realizations, and the ergodicity assumption was used to replace the covariance function 
corresponding to an ensemble average with the one obtained 
by the time average so that the KL transform was generally applied to space.
He considered the spatial autocorrelation tensor 
$K_{U}(x,x')= \left<U(x,t) U(x',t) \right>$, where now $x$ and $x'$ 
represent the deterministic variable (space) and $\left< \cdot \right>$ represents the ensemble average (which is simply the time average here). $K_{U}(x,x')$ therefore represents the spatial autocorrelation tensor at zero time lag. 
 As pointed out by \citet{george2017}, it is important to 
realize that the spatial auto-correlation tensor $K_U(x,x')$ is different from
its sampled estimation $\frac{1}{N_s}\sum_{i=1}^{N_s} u(x,t^{i}) u(x',t^i)$, where $N_s$ is the number of samples.
In general, exact eigenfunctions of $K_U(x,x')$ cannot be computed since
the $K_U(x,x')$ can only be approximated. However, for spatially homogeneous flows such as the channel flow in horizontal directions 
\citep{kn:aubry88}, exact solutions are known {\it a priori} since 
POD modes are Fourier modes in the homogeneous directions. 

A key insight stated by \citet{george2017} is that
if the process is stationary in time, the KL or POD modes are Fourier modes in the deterministic variable $t$. 
\berengere{Combining Fourier transform in time with Proper Orthogonal Decomposition in space was first applied in experimental studies 
of free turbulent shear flows, such as mixing layers and jets. 
Spatial and frequency decomposition was performed in
 the pioneering work of \citet{kn:glauserleibgeorge}, \citet{kn:glausergeorge}
as well as in \citet{kn:arndt97}, \citet{kn:delville99}, \citet{kn:citriniti00}, \citet{kn:ukeiley01} to cite only a few. 
However, as far as we know, a full 
four-dimensional decomposition was never attempted for any flow until recently
(see the spectral POD of \citet{kn:colonius18}),
and has never been implemented for wall-bounded flows.
 The present work builds on these previous developments, and  implements
George's suggestion to apply the decomposition in the four-dimensional space,
i.e.} in both time and space, reverting to ensemble average (as in the original definition) to evaluate the covariance function.
This will allow us to identify modes that can be associated with individual spatial wavenumbers and temporal frequencies. 
Such decompositions could be useful to identify key 
features and instability mechanisms
underlying the flow dynamics and eventually attempt to control them.
One particularly promising approach for this is the resolvent analysis \citep{kn:mckeon2017}, which is 
based on the Navier-Stokes equations.
In contrast, the present approach is data-based, and can therefore
provide a complementary viewpoint.

\berengere{As mentioned above, we note that  another four-dimensional decomposition is  provided
by
the Spectral POD introduced by  \citet{kn:colonius18}.  Their method only requires stationarity in time, so that
temporal Fourier transform can be applied, while standard snapshot POD is performed in the spatial directions.
In contrast, in our implementation, we apply Fourier transform in the spatial (horizontal) directions where turbulence
is  homogeneous. The connection of the present method with the snapshot POD and the dynamic mode decomposition
(DMD) is discussed in Section 3.}

%\srikanth{The idea of extending POD in time by computing Fourier transform for stationary flows is definitely
%not new. REFER some early works here. A recent work which develops such a method for snapshots of data 
%is Spectral POD by Towne et al. This method only requires stationarity in time, and Fourier transform is first applied to the snapshot data and then standard snapshot POD is computed. However, for spatially homogeneous flows, it is known that the optimal POD modes are Fourier modes. In the current work, we propose a fully Spectral POD for spatially homogeneous and stationary flows.
%}

In the present paper, we apply fully spatio-temporal Proper Orthogonal 
Decomposition to the turbulent channel flow. 
Wall turbulence is characterized by a variety of spatio-temporal 
scales interacting in a highly complex fashion \citep{kn:robinson, DENNIS2015}.
It is well known that the flow is characterized by an alternation of high and low-speed streaks aligned with the flow, with a typical spacing 
$\lambda_{z+} \sim 100-150 $ in the spanwise direction and a length of
$600-1000$ wall units (\citet{kn:kim71}, \citet{kn:stanislas08}, \citet{jimenez2013}), which are units based on the fluid viscosity and wall-friction velocity
$u_{\tau}$, and will be denoted with a + throughout the paper.
These streaks are associated with longitudinal 
vortical structures pushing low-speed 
fluid upwards and bringing high-speed fluid downwards, which results in
a strong contribution to the Reynolds stress \citep{kn:kim71}.
This contribution is highly intermittent in space and time, with `bursts' of turbulence production.
\berengere{The time scale of the bursts is difficult to assess.
Eulerian measures of the bursts yield time scales on the order of
300 wall units \citep{kn:blackwelder83}. This value is based on
the VITA (Variable Interval Time Average) criterion applied to the
streamwise velocity and depends on the choice of a particular threshold,
which makes it difficult to provide an absolute value for the bursting
period. However the value appears to be independent of the Reynolds
number.
 Minimal flow units, i.e. periodic domains of small extent \citep{jimenez2013}
were characterized by global characteristic time scales of 300-400 wall units,
which appear to correspond to regeneration cycles of coherent structures \citep{kn:kimhamwaleffe},
but relating this time scale with that of a full-scale turbulent flow is difficult.}

Classic spatial POD has been previously used to investigate the boundary 
layer (e.g., see \citet{kn:aubry88}, \citet{moin1989}, \citet{kn:jfm98}).
Spatial eigenfunctions are determined from second-order statistics.
 The amplitudes of eigenfunctions are characterized by a mixture of
time scales and can only be computed by projecting
 the full instantaneous field onto the spatial eigenfunctions.
In contrast, the new decomposition directly provides 
spatio-temporal patterns.

In this paper we focus on a relatively moderate Reynolds number $\mathrm{Re}_{\tau}=590$, based on the fluid 
viscosity, channel half-height and friction velocity. 
At these moderate Reynolds numbers our focus will be on the wall layer
as large scales such as those observed by
\cite{kn:smits11} are not present in the flow. 
%We focus on the wall region $y_+ < 100$, where $+$ units are based on the friction velocity and fluid viscosity.
The rest of the paper is organized in the following manner. Section 2 gives details of the methodology for the \srikanth{spatio-temporal} POD. Section 3 discusses the connections of our approach with snapshot POD and DMD. Section 4 presents the POD results while the contributions to the turbulent kinetic energy equation 
associated with each mode are examined in Section 5, followed by a conclusion in Section 6 which summarizes the key observations and results.

\section{Description of the procedure }

%\Lionel{I would shift this Italic portion later in the paper, say, in a section 4.0} 

\subsection{Full spatio-temporal POD}

%In traditional studies of channel flow, POD is only applied to space.
%One of the theoretical properties of POD is that it is equivalent to Fourier transform in the homogeneous directions of the flow, and in time if the flow is statistically stationary. The turbulent channel flow is statistically homogeneous along $x$ and $z$. In most of the POD-based studies on turbulent channel flow, POD is implemented in the following manner (see e.g. \citep{aubry1991}, \citep{podvin2001}). 
%The POD decomposition of the velocity fields along $x,z$ are obtained using Fourier transform along these direction. 
%Let $u_{lk}^i(y,t)$ denote the $i$th velocity component in Fourier space corresponding to the streamwise mode $l$ and spanwise mode $k$. POD is applied to each mode pair, i.e. $(l,k)$, thereby obtaining a suitable basis for representing the velocity field. 
%\begin{equation}
%u_{lk}^i(y,t) = \sum_0^{\infty} a_{lk}^{n}(t) \phi_{lk} ^{n,i}(y)
%\end{equation}
%Thus, the velocity field is decomposed as a spatial structure $\phi_{lk} ^{i,(n)}(y)$ 
%and a time-dependent coefficient $a_{lk}(t)$. 

%\Lionel{This section requires some work to extend and polish it. This is where we present the main contribution of the paper and it seems to me that it falls a bit short for now.}

Owing to the homogeneity of the statistics in the spatial directions $x$ and $z$ and their stationarity, POD modes are Fourier modes in the horizontal directions as well as in time, \berengeredeux{\citep{berkooz1993}}. 
Here, for each configuration, 
the Fourier transform of the velocity is computed in the temporal and
the homogeneous spatial directions for each sample corresponding to a set of fields. 
The $p$th component of the velocity field ($u^{p}(x,y,z,t)$) in the physical space is then written as 
\begin{equation}
u^{p}(x,y,z,t)= \sum_{l} \sum_{k} \sum_{f} u_{lkf}^p(y) \expbig,
\end{equation}
where $u_{lkf}^p(y)$ denotes the $p$th velocity component in the Fourier space corresponding to the streamwise wavenumber $l$, spanwise wavenumber $k$ and frequency $f$. 
Proper Orthogonal Decomposition is then applied in the wall-normal direction for each triad $(l,k,f)$:
\begin{equation}
u_{lkf}^p(y) = \sum_{n} a_{lkf}^{n} \phi_{lkf}^{n,p}(y),
\end{equation}
with, for any two $4$-tuples $(l,k,f,n)$ and $(l,k,f,m)$,  
\begin{equation}
\left<a_{lkf}^n a_{lkf}^{m*}\right>= \delta_{n m} \lambda_{lkf}^{n}.
\end{equation}
Here, $\left< \cdot \right>$ represents the usual ensemble average over the space of all possible flow realizations, $\delta_{mn}$ is the Kronecker-delta function, $n$ denotes a specific POD mode for each triad, and * denotes complex conjugation. Note that $a_{lkf}^{n}$ is independent of $p$ as the three velocity components are stacked into a single vector.
The complex, stochastic coefficients $a_{lkf}^{n}$
are uncorrelated and their variance is equal to $\lambda_{lkf}^{n}$.

The eigenfunctions $\phi_{lkf}^{n,p}$ and eigenvalues $\lambda_{lkf}^{n}$
are obtained by solving the following eigenproblem, where
the autocorrelation is estimated from taking an ensemble average: 
\begin{equation} 
\int_{0}^{Y} \left< \uu_{lkf}(y) \uu_{lkf}^{*}(y') \right> \pphi_{lkf}^{n}(y') \mathrm{d}y' = 
\lambda_{lkf}^{n} \pphi_{lkf}^{n}(y). \label{pod_eigprob}
\end{equation} 
Here, $\uu_{lkf}(y)$ represents the single velocity vector with the three components stacked in it. Upon discretization in $y$, $\pphi_{lkf}^{n}(y)$ represents the single eigenvector containing the three components. By construction, the eigenvectors are orthonormal with respect to the Euclidean inner product, and $\lambda_{lkf}^{n}$ can be interpreted as the energy content in each mode. Similarly, for each triad $(l,k,f)$, the modes (i.e. different $n$'s) are sorted by energy. Since the left-hand-side of Equation~\eqref{pod_eigprob} is a Hilbert-Schmidt integral operator, its eigenvalues are real and non-negative. 
%The entire procedure is schematically represented in Figure~\ref{schema}. 
The eigenvalues $\lambda_{lkf}^{n}$ associated with the triplets $(l,k,f)$ can then be gathered and sorted globally 
according to their magnitude.  
We will call {\it mode number} the global index $N(l,k,f,n)$ associated with the sorted modes 
over spatial wavenumbers $(l,k)$, frequency $(f)$ and quantum number $n$, with lower {\it mode number} indicating a larger eigenvalue. In all that follows, we focus only on the most energetic modes.

\subsection{Numerical implementation}

\begin{figure}
\begin{center}
% \begin{subfigure}[b]{0.8\textwidth}
% \includegraphics[width=\linewidth]{schematic-eps-converted-to.pdf}
%\includegraphics[width=0.8\linewidth]{schematic_LM.png}
\includegraphics[width=0.8\linewidth]{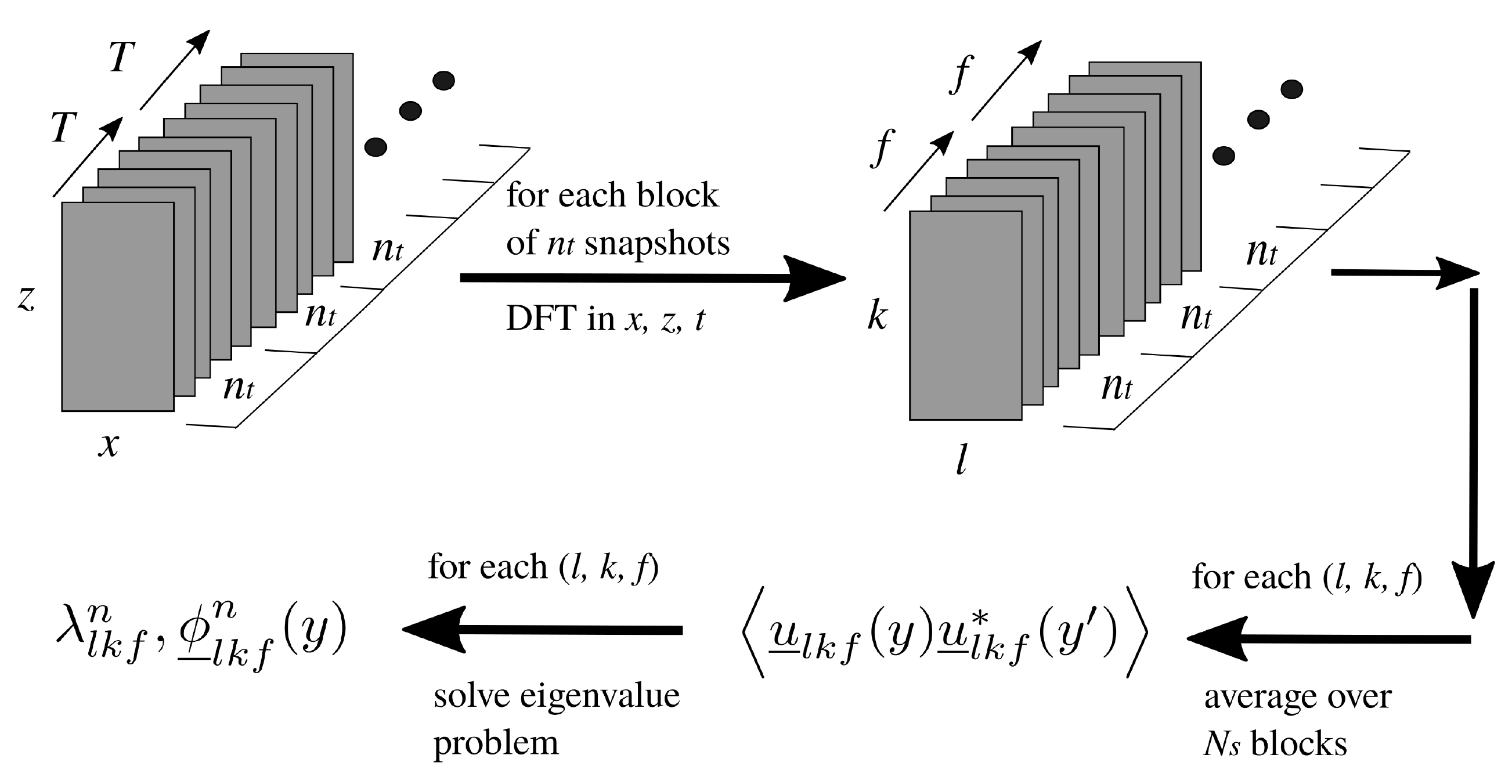}
\end{center}
% \end{subfigure}
\caption{Schematic of the methodology. Each grey slab on the left represents an instantaneous realization of the flow
in the physical domain (the wall-normal direction corresponding to the thickness of the slab), while each slab on the 
right corresponds to the Fourier transform of a block of $n_t$ realizations in time as well as in the horizontal 
directions $x$ and $z$. For each Fourier triad $(l,k,f)$ the autocorrelation tensor is computed 
by averaging over the $N_s$ blocks and an independent eigenproblem is solved.} 
\label{schema}
\end{figure}

The methodology is summarized in Figure~\ref{schema}.
Since POD is also applied in time, the autocorrelation tensor needs to be computed from several independent realizations of the same experiment. 
%Each realization is represented as a rectangular plane along $x$ and $z$, with the thickness of the plane indicating the coordinate $y$. 
An ideal sample would consist of the Fourier transform in space and time of the velocity field corresponding to different databases
obtained at the same Reynolds number.
In practice, owing to the cost of the simulation, we split a single database into several contiguous chunks of length $T$, each of which constitutes a sample. 
The samples are therefore not independent realizations, but we assume that the period $T$ is sufficiently larger than 
the characteristic time scales of the flow, or at least sufficiently large to allow separation of the dominant time scales.
Yet $T$ cannot be too large in order to allow for a reasonable 
number of samples $N_s$ to be constituted. 
We note that as in \citet{moin1989} and \citet{podvin2001},
the number of samples is doubled by considering spanwise reflections and quadrupled 
by considering both lower and upper parts of the channel (including spanwise reflections).

Each sample then consists of a block of $n_t$ velocity fields uniformly sampled 
at a rate $ \delta t $. \srikanth{The value of $\delta t$ is chosen such that the Nyquist frequency
is well above the range of characteristic frequencies of the flow.} 
%\Lionel{[I do not understand this sentence. Is not it the opposite?} 
\srikanth{It is however limited by computational
tractability, as large values of $n_t$ will increase memory requirements
for the samples.}
For most of the test cases, we have fixed the value of $T$ to be
3000 wall units. %(500 fields spaced by 5.9 viscous time units). 
%About 120 samples using symmetries could be extracted from the simulation. 

We use the finite-volume code SUNFLUIDH to simulate the incompressible flow in a channel.
Further details on the configuration and numerical scheme can be found in \citet{podvin2014}. 
Periodic boundary conditions are enforced in the streamwise ($x$) and the spanwise direction ($z$), and the channel lengths along these directions are $L_x = 2\pi h$ and $L_z = \pi h$, where $h$ is the channel half-height. 
The total domain is discretized using a grid of size $256 \times 256 \times 256$, with a uniform grid for the horizontal directions $x$ and $z$ and a hyperbolic tangent-distributed mesh for the wall-normal direction $y$. 
The velocity components in the streamwise, wall-normal and spanwise
direction will be denoted respectively by $u,v,w$ or 
equivalently by $u_{1}, u_{2}, u_{3}$.
The simulation is conducted at $\mathrm{Re}_{\tau} = 590$, based on the friction velocity $u_{\tau}$ and channel half-height $h$, which corresponds to the simulation of \citet{moser1999}. 
%(details of scaling are in section~\ref{scaling}). 
The database consists of 16500 flow realizations, separated by $\delta t_+=5.9$. The time spanned by the database represents about $10^5$ viscous time units. 
% where $+$ refers to viscous or wall units based on the fluid viscosity and friction velocity

 In most of the paper, we focus on the
domain $y_+ < Y_+=80$, but we also considered the full 
boundary layer where $Y_+=590$. 
%, and we denote this maximum height in $y_+$ in the wall layer by $Y$.
%Due to the moderate Reynolds numbers, we can assume that no large scales
%or very large scales are present in the flow \cite{kn:stanislas08}. 
Due to the strong decrease of the energy spectra with the streamwise wavenumber,
and given the typical extent of longitudinal streaks in the wall layer 
(about 600-1000 wall units \citep{jimenez2013}), the 
streamwise extent of the domain was limited to $L_x=\pi/2h$ 
representing 900 wall units.
The full spanwise extent of the domain (about 1800 wall units) was considered.

Results are reported for four datasets extracted from the single database, 
 which represent different domain sizes and sampling rates.
The characteristics of the different configurations 
are given in Table~\ref{table_config}.
%The first data set (D1) corresponds to a domain limited in the horizontal directions, and
%a relatively low sampling rate.
The sampling rate
and number of samples were varied in the configurations D1-D3. 
These datasets correspond to the wall layer, $Y_+ = 80$.
The fourth one (D4) corresponds to the full boundary-layer height, $Y_+ = 590$.

\begin{table}
\begin{center}
\begin{tabular}{c c c c c c c c}
 $\mathrm{Name} $ & $L_x \slash h$ & $L_z \slash h$ & $Y\slash h$ & $n_t$ & $N_s$ & $\delta t_+$ & $T_+$\\ 

 %D1 & $\pi/4$ & $\pi/4$ & 0.14 & 50 & 120 & 29.5 & 1475 \\
 D1 & $\pi/2$ & $\pi$ & 0.14 & 100 & 60 & 29.5 & 2950 \\
 D2 & $\pi/2$ & $\pi$ & 0.14 & 100 & 120 & 29.5 & 2950\\
 D3 & $\pi/2$ & $\pi$ & 0.14 & 500 & 132 & 5.9 & 2950\\
 D4 & $\pi/2$ & $\pi$ & 1 & 500 & 132 & 5.9 & 2950

 \end{tabular}
\end{center}
\caption{Description of different test cases; $n_t$ denotes the number of snapshots in a given block, $N_s$ denotes the number of samples, $\delta t_+$ is the time gap between successive snapshots, and $T_+$ is the total time period for a given block (expressed in viscous units).}
\label{table_config}
\end{table}

\section{Connections with alternative decomposition methods}

%\srikanth{Dynamic mode decomposition (DMD) is another recent technique that aims at obtaining
%coherent structures from data that capture the temporal information thereby overcoming the
%drawbacks of classic spatial POD. Towne et al.  have shown that Spectral POD is closely 
%related to DMD, such that they can be considered as ensemble averaged DMD modes in the case of
%stationary flows. Classic DMD is applied to one flow realization, and thus a mode computed at
%given frequency will have statistical variability over a set of realizations. Spectral POD
%can be viewed as an optimal basis that account for the variability of modes over an ensemble of
%realizations. Towne et al. have also shown connections between Spectral POD and resolvent analysis.
%We would like to note that spatio-temporal POD is just a fully spectral POD as the flow is homogeneous along streamwise and spanwise directions and thus the Fourier transform is applied in both space and time, and correlation function is
%computed in the only inhomogeneous wall-normal direction.
%}

\berengere{
\input{Lionel1rev}}

\berengere{
\input{Lionel2rev}}

\section{Results from spatio-temporal POD}

\subsection{POD eigenvalues }

\begin{figure}
\begin{subfigure}[b]{0.5\textwidth}
\includegraphics[width=\linewidth]{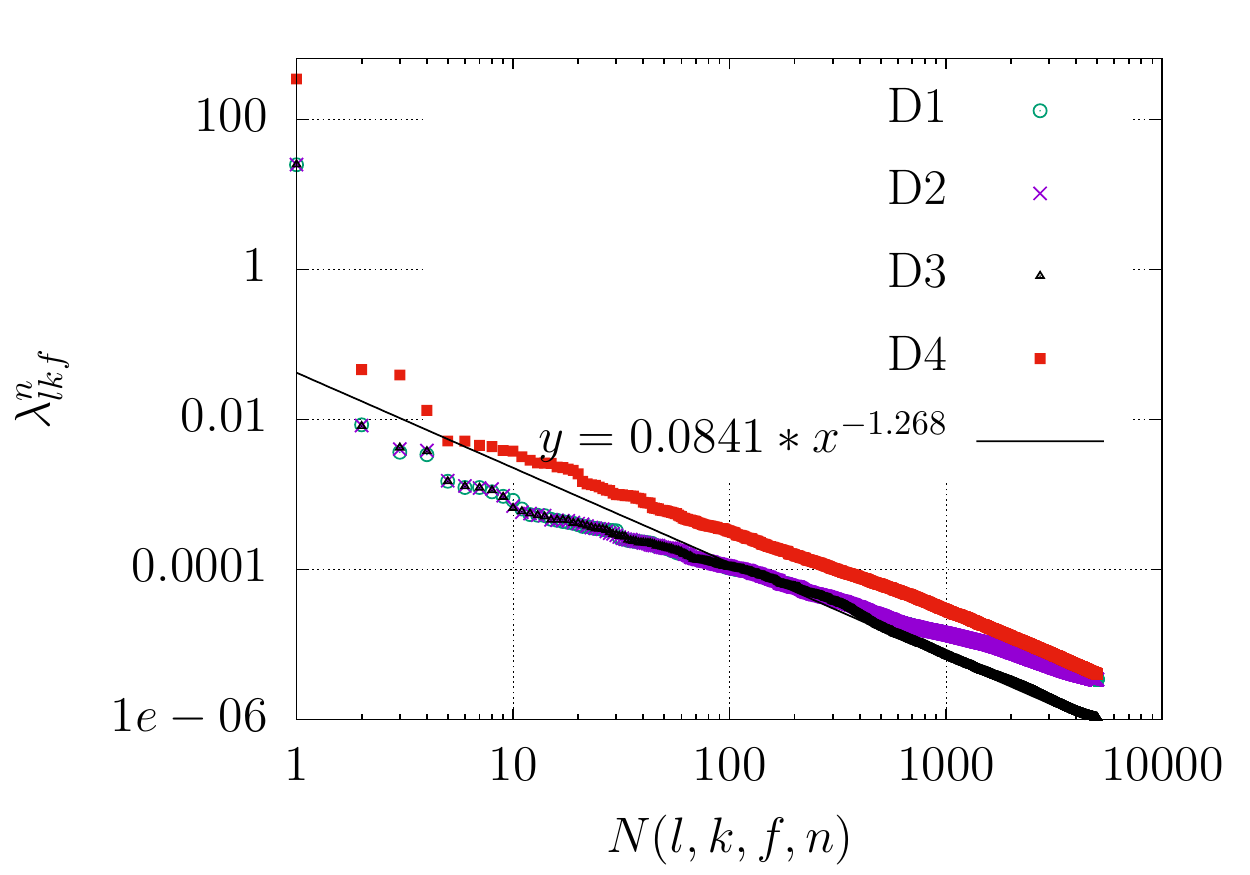}
\caption{}
\label{spectracomp}
\end{subfigure}
\begin{subfigure}[b]{0.5\textwidth}
\includegraphics[width=\linewidth]{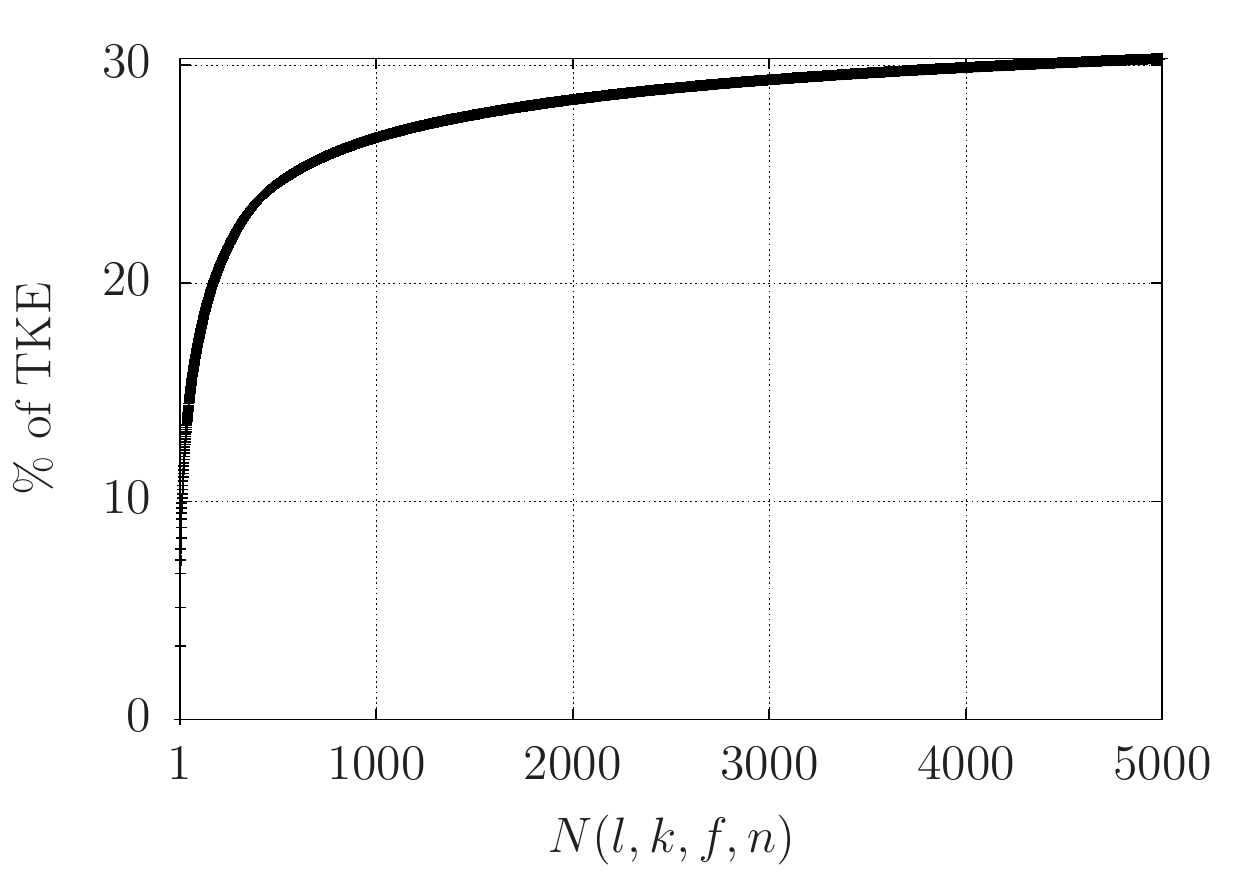}
\caption{}
\label{spectracumul}
\end{subfigure}
\caption{a) Log-log plot of the top 5000 eigenvalues for D1-D4 configurations. An asymptotic fit for the decay rate of the spectra for D3 is also shown. b) Percentage of turbulent kinetic energy (TKE) captured by the $N$ most energetic modes for D3.}
\label{spectra_full}
\end{figure}

Figure~\ref{spectracomp} shows the top 5000 eigenvalues for each configuration.
We note that this represents a tiny fraction of the total number of 
eigenvalues defined in the spatio-temporal space, which is
$3 n_t \times N_l \times N_k \times N_y$, 
where $n_t$ is the number of instantaneous fields contained in a sample,  
$N_l$ and $N_k$ are respectively the 
numbers of streamwise and spanwise wavenumbers, and 
$N_y$ is the number of grid points in the wall-normal direction. 
For D3 this corresponds to about $4 \times 10^8$ degrees of freedom. 
The largest eigenvalue $N(l,k,f,n)=1$ where $l=k=f=0$ and $n=1$ 
corresponds to that of the mean mode (discussed in the next section).
\srikanth{For each configuration, we have 
\berengere{applied spatio-temporal POD to
the data without removing the sample mean and compared the results
to the case where the sample mean was removed. In both cases} 
 the eigenvalue distribution was very similar, 
and the deviations in eigenfunctions were negligibly small \berengere{at least for the most energetic modes}.}
In the rest of the paper we will focus on the $n=1$ eigenfunctions. 
We note that the eigenfunctions corresponding to 
$n=1$ capture about $80\%$ of the turbulent kinetic energy,
which is in agreement with \cite{moin1989}'s results.
%If only the streamwise-averaged $(l=0)$ fluctuations are considered, 85\% of
%the fluctuating streamwise-averaged energy is captured by the modes with $n=1$
%and $l=0$.
%As the eigenvalues are related to the kinetic energy of the field, this indicates that mean contains most of the turbulent kinetic energy, which is expected in a highly turbulent flow. The fluctuations have lower kinetic energy, and a fast initial decay is observed followed by a slower decay as the number of modes increase.

Figure~\ref{spectra_full}a shows that the \berengeredeux{spectrum eigenvalues} in D1 and D2, which correspond
to a different number of samples, are essentially indistinguishable from each other. 
This indicates that the number of samples of 
ensemble averaging appears to be sufficient for the convergence of the dominant eigenvalues.
%increasing $n_e$ to 66 doesn't produce much differences in the spectra. 
The \berengeredeux{eigenvalues} in D1 and D2 are similar to that in D3 at low values of $N$, but have 
higher energy levels at large values of $N$, which is likely to be due to 
aliasing effects as the fields are sampled at higher rates there than in D3.
The spectrum has an asymptotic decay of $\lambda(N) \sim N^{-1.27}$.
Unlike the other configurations, D4 corresponds to the full boundary layer $0 < y_+ < 590$. 
As expected, energy levels are higher but the shape of the spectrum and its 
asymptotic decay rate are the same as in the wall region $0 < y_+ < 80$, which suggests 
self-similarity (see next section). 
Figure~\ref{spectra_full}b shows the fraction of turbulent kinetic
energy (TKE) captured by the first $N$ modes for each $N$ (as we are considering fluctuations, 
the contribution from the mean mode was set to zero). 
The fraction of TKE captured by the most energetic $100$ 
mode numbers is relatively high, but increases only slightly for mode numbers 
higher than $1000$, which highlights the complexity of the flow.
\berengere{The slow convergence in Fourier space also indicates that the structures are highly localized
in space and time, owing to the fundamentally intermittent nature 
of turbulence \citep{frisch}.}
This is confirmed by Table~\ref{table_cumulenergy}, which shows that the first $200$ eigenvalues capture about $21\%$ of the TKE which is increased only to $32\%$ when $5000$ modes are included, for the D3 configuration. The trend remains the same for the D4 configuration, where $200$ eigenvalues capture around $24\%$ of TKE which increases only to $34\%$ when $5000$ eigenvalues are included.

\begin{table}
\begin{center}
\begin{tabular}{ccc} %\hline
 Number of eigenvalues ($N$) & \% of TKE (D3) & \% of TKE (D4)\\ %\hline 
 200 & $21$ & $24$ \\
 3000 & $29$ & $33$ \\
 5000 & $30$ & $34$  %\hline
 \end{tabular}
\end{center}
\caption{Percentage of TKE in the first $N$ eigenvalues for D3 and D4 configurations.}
\label{table_cumulenergy}
\end{table}

%Again, the convergence is observed as numerical resolutions are improved. A dominant eigenmode is observed at $\bar{k}_+ = 0.001$, which persists for the three different cases and corresponds to
%a spanwise wavelength of about 618. 
%This is in agreement with other studies (see for instance \citep{stanislas08}, \citep{kn:jimenez2013}).
%We note a slight increase in the low-wavenumber range for the case D5, which is consistent with an increase in the spanwise size of the structures as the distance from the wall increases (see for e.g. \citet{delalamo}, \citet{podvin2017}). 
%
%Note that the dominant eigenmode observed at $k = 0.006$ does not correspond to the dominant mode observed at $f = 0.0044$ in figure~\ref{freq_comp}.
\begin{figure}
\begin{subfigure}[b]{0.5\textwidth}
\includegraphics[width=\linewidth]{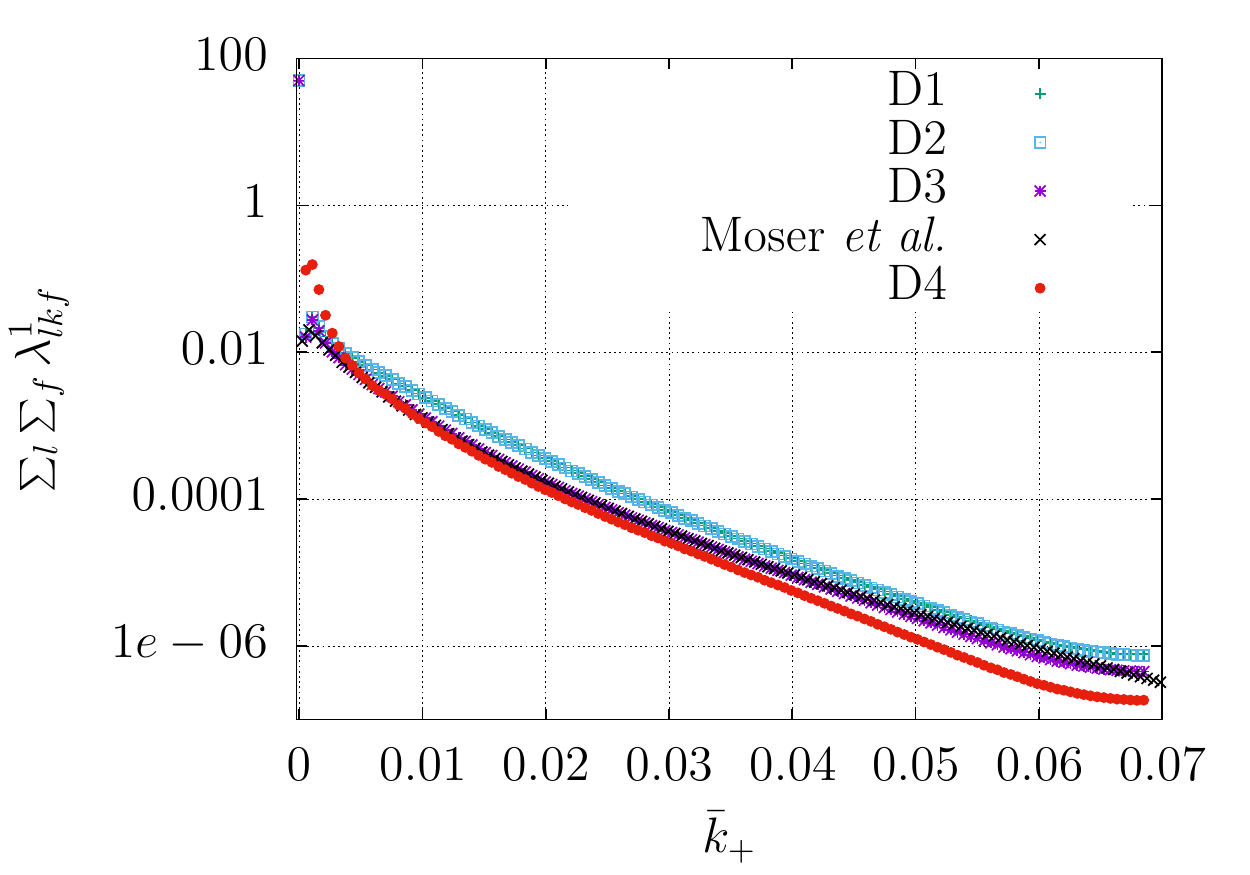}
\caption{}
\label{fig:k_dist1b}
\end{subfigure}
\begin{subfigure}[b]{0.5\textwidth}
%includegraphics[width=\linewidth]{fdist_comp-eps-converted-to.pdf}
\includegraphics[width=\linewidth]{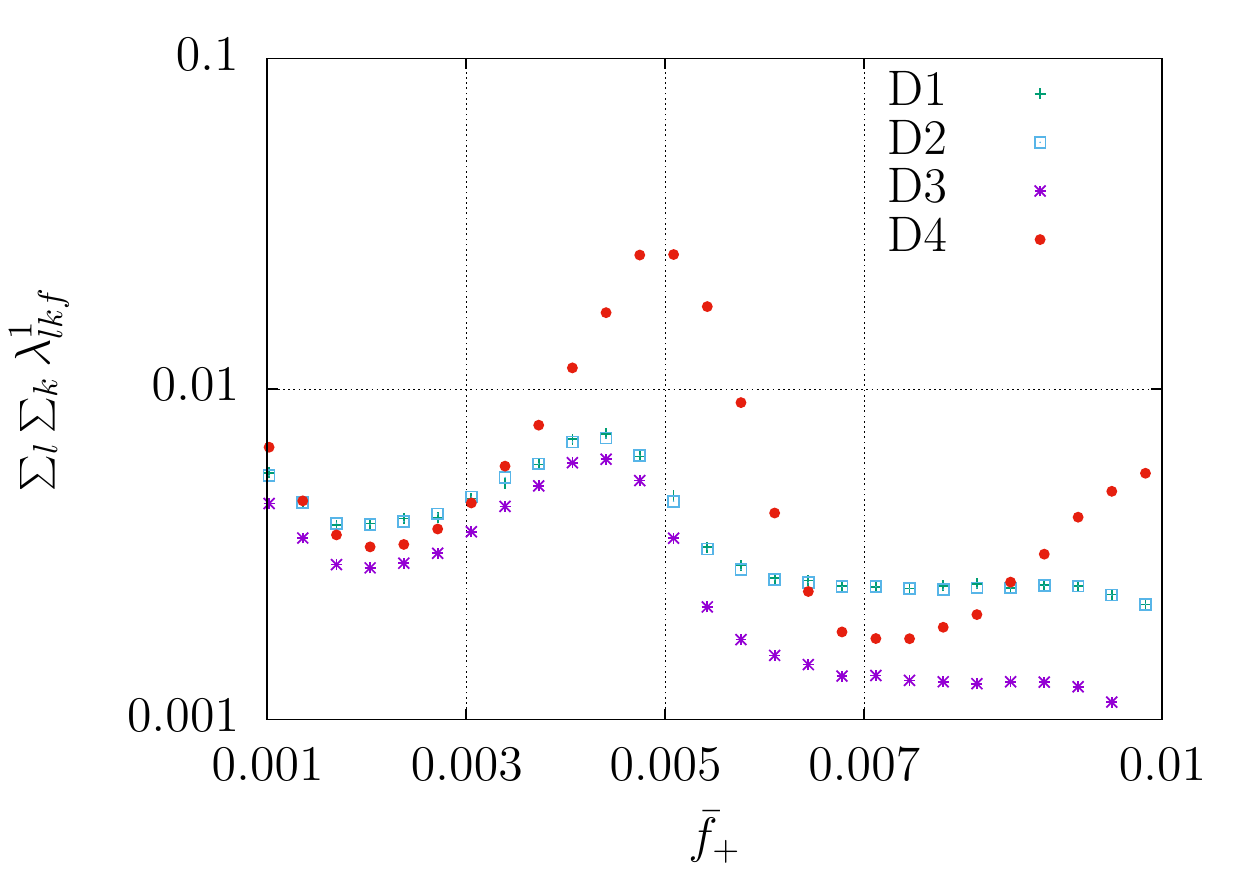}
\caption{}
\label{fig:k_dist4c}
\end{subfigure}
\caption{
a) \berengeredeux{Integrated} energy spectrum ($\sum_l \sum_f \lambda_{lkf}^{1}$) as a function of $\bar{k}_+$; the data is compared with 
\citet{moser1999}'s energy spectrum at $y_+=80$ (with appropriate rescaling).
b) \berengeredeux{Integrated} energy spectrum ($\sum_l \sum_k \lambda_{lkf}^{1}$) as a function of $\bar{f}_+$ for
$0 < y_+ < 80$ (D1-D4).}    
\label{k_comp}
\end{figure}
Figure~\ref{k_comp} shows  
the dependence of the integrated spectra with respect to the frequency and spanwise wavenumber
for the different configurations. 
%The frequency $\bar{f}_+$ and spanwise wavenumber $\bar{k}_+$ are in wall units, and are proportional to $f$ and $k$. 
For the D3 configuration, the spatial wavenumbers 
and the frequency $(\bar{l},\bar{k},\bar{f})$, expressed in wall units, 
are related to the mode indices $(l,k,f)$ as 
%\begin{equation}
%\bar{l}_+ = \frac{l}{927.142}, \quad
%\bar{k}_+ =\frac{k}{1854.286}, \quad
%\bar{f}_+ = \frac{f}{2950}.
%\end{equation}
\begin{equation}
\bar{l}_+ = \frac{l}{927}, \quad
\bar{k}_+ =\frac{k}{1854}, \quad
\bar{f}_+ = \frac{f}{2950}.
\end{equation}

Figure~\ref{k_comp}a shows how the sum of the eigenvalues over streamwise wavenumber and frequencies ($\sum_l \sum_f \lambda_{lkf}^{1}$) varies as a function of $\bar{k}_+$. 
Results show a good agreement with the rescaled standard energy spectrum obtained by \citet{moser1999} at $y_+=80$. 
As expected, the spanwise spectrum over the full layer 
has more energy in the lower wavenumbers 
and less energy at higher wavenumbers compared to that in the wall layer.

Figure~\ref{k_comp}b shows the sum of the dominant $n=1$ eigenvalues over all 
$(l,k)$ modes (i.e. $\sum_l \sum_k \lambda_{lkf}^{1}$) as a function of frequency $\bar{f}_+$ for 
the different configurations D1-D4. 
%The range in $\bar{f}_+$ has been limited to $0.01$ to match the maximum frequency contained in D1. 
As observed previously, aliasing effects can be observed for the fields sampled in time 
at a lower rate (D1-D2), but the trends are very similar.
 A marked increase in the energy is observed in the frequency range of $0.003-0.005$, which corresponds to time scales of 200-300 viscous units. 
This value is in good agreement with the duration of the regeneration cycle identified by \cite{kn:kimhamwaleffe}
and \cite{kn:jimemoin} in minimal domains, as well as the investigations of larger-scale domains reported in \cite{jimenez2013}.
A similar frequency peak is observed in the full boundary layer (D4), but with a shift towards 
slightly higher
frequencies: the maximum is located at $\bar{f}_{+}=0.005$, while it is located at $\bar{f}_+ = 0.0044$,
in the region $y_+ < 80$.

\berengere{As far as we know this is the first evidence of an objective time scale identified
in the channel configuration.
The time scale did  not appear to be affected by spatial variability: it did not
depend on the spanwise extent (different  domain widths
were considered) or the streamwise extent of the domain (two domains of different streamwise sizes
were compared).
We note that the dominant frequency of the nonlinear terms (e.g. the Reynolds stresses) should be on the order of
100-150 wall units.   An interesting connection, which will need further exploration,
can be made with wall reduction control schemes
based on wall spanwise oscillations such as those of \citet{kn:quadrizioricco04} and \citet{kn:kschoi98},
where  it was shown that the optimal oscillation period is about 100-150 wall units.}

%We also note that second and third harmonics of $\bar{f}_{c+}$ can be observed in the spectrum. 

The predominance of the characteristic frequency $\bar{f}_{c+}$ at $\bar{f}_+=0.0044$ is confirmed by
Table~\ref{top_eig}, which shows the top 30 eigenvalues (denoted by $\lambda_{lkf}^{n}$) with 
the corresponding Fourier mode indices $(l,k,f)$ in space and time and quantum mode $n$ for the configuration D3. 
%$(\bar{l},\bar{k},\bar{f})$ and $n$ corresponds to the $n$th eigenvalue for a particular mode pair. 
Although a large fraction (18 modes) of the first 30 modes are associated to $f \le 2$, which corresponds 
to long time scales that are outside the scope of our analysis, 
%, and spatial wavenumbers in the range 264-1854 wall units ($k=1-6$).
all the other modes are characterized by frequencies between $11$ and $15$, which correspond
to the previously identified time scale of about 200-300 viscous units. 
The most energetic of these modes corresponds to $f=13$, i.e. $\bar{f}_{c+}$.  
We note that all the modes in the table are characterized 
by $n=1$, which corresponds to the largest eigenvalue for a triad $(l,k,f)$,
and $l=0$, which corresponds to the streamwise-averaged flow.

\begin{table}
\begin{center}
\begin{tabular}{cccccc}
$N$ & $l $ & $k$ & $f$ & $n$ & $\lambda_{lkf}^n$ \\ 
%1& 0 & $ 0 $ & $ 0 $ & 1 & 24.97645 \\
1& 0 & $ 0 $ & $ 0 $ & 1 & 24.98 \\
2& 0 & $ 2 $ & $ 0 $ & 1 & 0.008133 \\
3& 0 & $ 1 $ & $ 0 $ & 1 & 0.004205 \\
4& 0 & $ 3 $ & $ 0 $ & 1 & 0.003754 \\
5& 0 & $ 4 $ & $ 0 $ & 1 & 0.001489 \\
6& 0 & $ 3 $ & $ 1 $ & 1 & 0.001277 \\
7& 0 & $ 2 $ & $ 1 $ & 1 & 0.001222 \\
8& 0 & $ 1 $ & $ 1 $ & 1 & 0.001128 \\
9& 0 & $ 4 $ & $ 1 $ & 1 & 0.000923 \\
10& 0 & $ 5 $ & $ 0 $ & 1 & 0.000655 \\
11& 0 & $ 3 $ & $ 13 $ & 1 & 0.000591 \\
12& 0 & $ 2 $ & $ 14 $ & 1 & 0.000552 \\
13& 0 & $ 5 $ & $ 0 $ & 1 & 0.000523 \\
14&  0 & $ 2 $ & $ 13 $ & 1 & 0.000505 \\
15& 0 & $ 3 $ & $ 12 $ & 1 & 0.000457 
 \end{tabular}
 \quad \quad
 \begin{tabular}{cccccc}
$N$ & $l $ & $k$ & $f$ & $n$ & $\lambda_{lkf}^n$ \\ 
16& 0 & $ 3 $ & $ 2 $ & 1 & 0.0004568 \\
17& 0 & $ 2 $ & $ 2 $ & 1 & 0.0004559 \\
18& 0 & $ 6 $ & $ 0 $ & 1 & 0.000451 \\
19& 0 & $ 4 $ & $ 12 $ & 1 & 0.000415 \\
20& 0 & $ 3 $ & $ 14 $ & 1 & 0.000412 \\
21& 0 & $ 4 $ & $ 2 $ & 1 & 0.000396 \\
22& 0 & $ 6 $ & $ 1 $ & 1 & 0.0003806 \\
23& 0 & $ 0 $ & $ 1 $ & 1 & 0.0003643 \\
24& 0 & $ 5 $ & $ 2 $ & 1 & 0.000349 \\
25& 0 & $ 4 $ & $ 11 $ & 1 & 0.0003475 \\
26& 0 & $ 2 $ & $ 12 $ & 1 & 0.0003445 \\
27& 0 & $ 4 $ & $ 13 $ & 1 & 0.000335 \\
28& 0 & $ 5 $ & $ 12 $ & 1 & 0.000314 \\
29& 0 & $ 2 $ & $ 15 $ & 1 & 0.000295 \\
30& 0 & $ 1 $ & $ 14 $ & 1 & 0.000289 
 \end{tabular}
\end{center}
\caption{Top 30 most energetic eigenvalues $\lambda_{lkf}^{n}$ along with their corresponding $(l,k,f,n)$.}

\label{top_eig}
\end{table}

\begin{figure}
\begin{subfigure}[b]{0.5\textwidth}
\includegraphics[width=\linewidth]{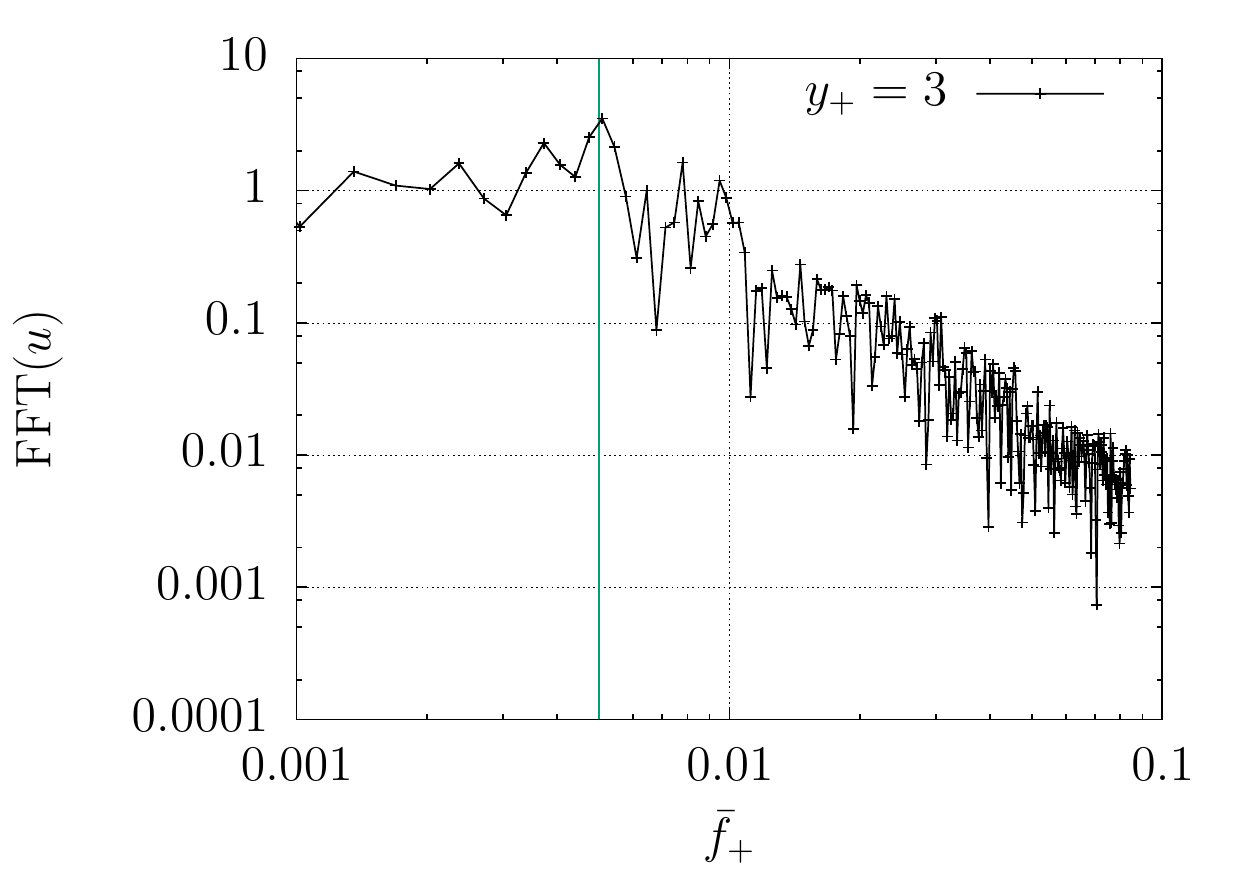}
\caption{}
\label{fft_pointa}
\end{subfigure}
\begin{subfigure}[b]{0.5\textwidth}
\includegraphics[width=\linewidth]{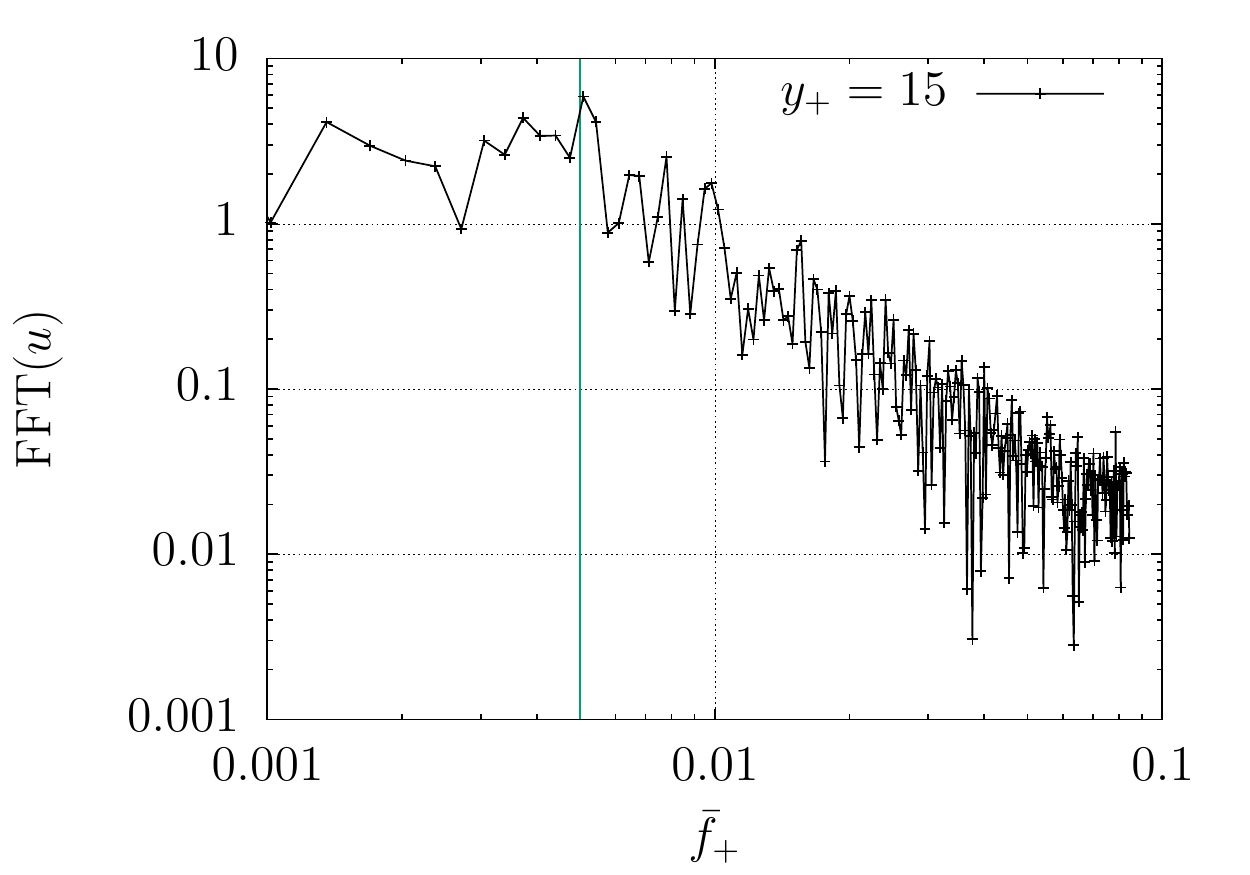}
\caption{}
\label{fft_pointb}
\end{subfigure}
\caption{Horizontally-averaged ($x$ and $z$) temporal Fourier transform of the streamwise velocity component $u$ as a function of $\bar{f}_+$ in a log-log scale at a) $y_+ = 3$ and b) $y_+ = 15$. A vertical line is shown at $\bar{f}_+ = 0.005$ to indicate the peak in both figures.}
\label{fft_point}
\end{figure}

To provide a comparison with standard analysis tools,
Figures~\ref{fft_point}a and b show the temporal Fourier transform of the streamwise 
velocity component averaged along horizontal directions for two different heights: 
$y_+ = 3$ and $15$.
\berengeredeux{Although there seems to be some energy increase around the
characteristic frequency $\bar{f}_{c+}$, evidence of a local peak is 
not clear in these figures.} 
This illustrates the usefulness of the new POD implementation.

\begin{figure}
\begin{subfigure}[b]{0.5\textwidth}
\includegraphics[width=\linewidth]{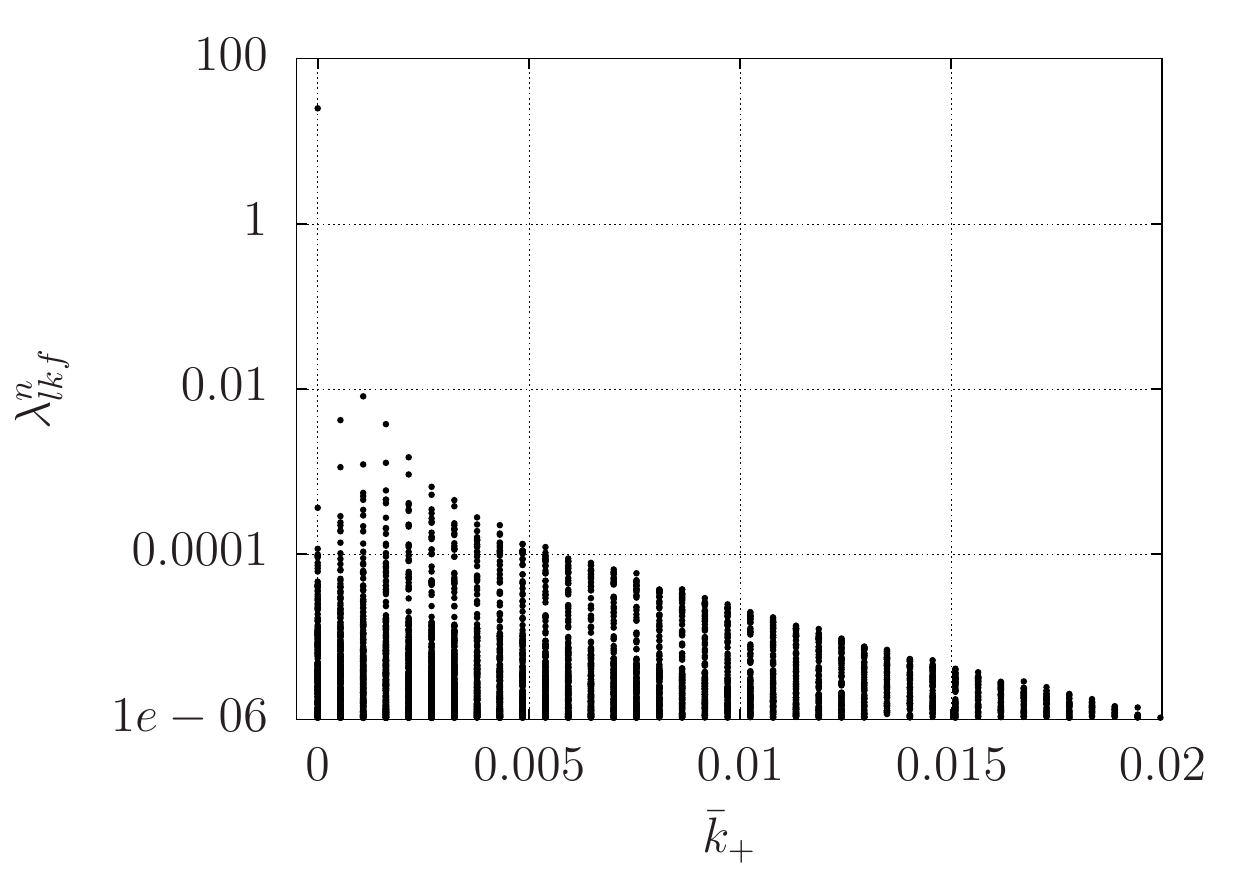}
\caption{}
\label{fig:k_dist_full}
\end{subfigure}
\begin{subfigure}[b]{0.5\textwidth}
\includegraphics[width=\linewidth]{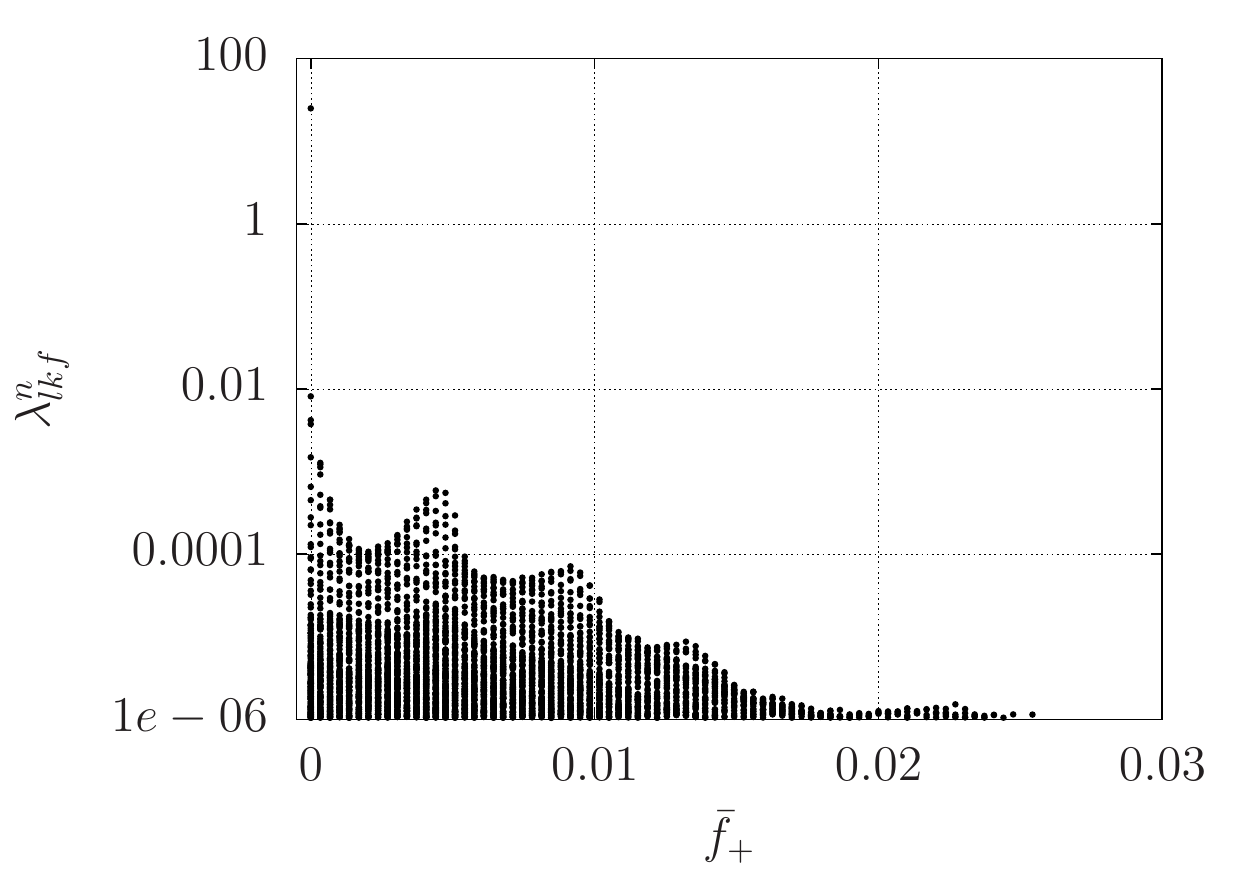}
\caption{}
\label{fig:freq_dist_full}
\end{subfigure}
\caption{Eigenvalue distribution of the top 5000 eigenvalues for D3 configuration as a function of a) $\bar{f}_+$ and b) $\bar{k}_+$.} 

\label{dist_full}
\end{figure}

\begin{figure}
\begin{subfigure}[b]{0.5\textwidth}
\includegraphics[width=\linewidth]{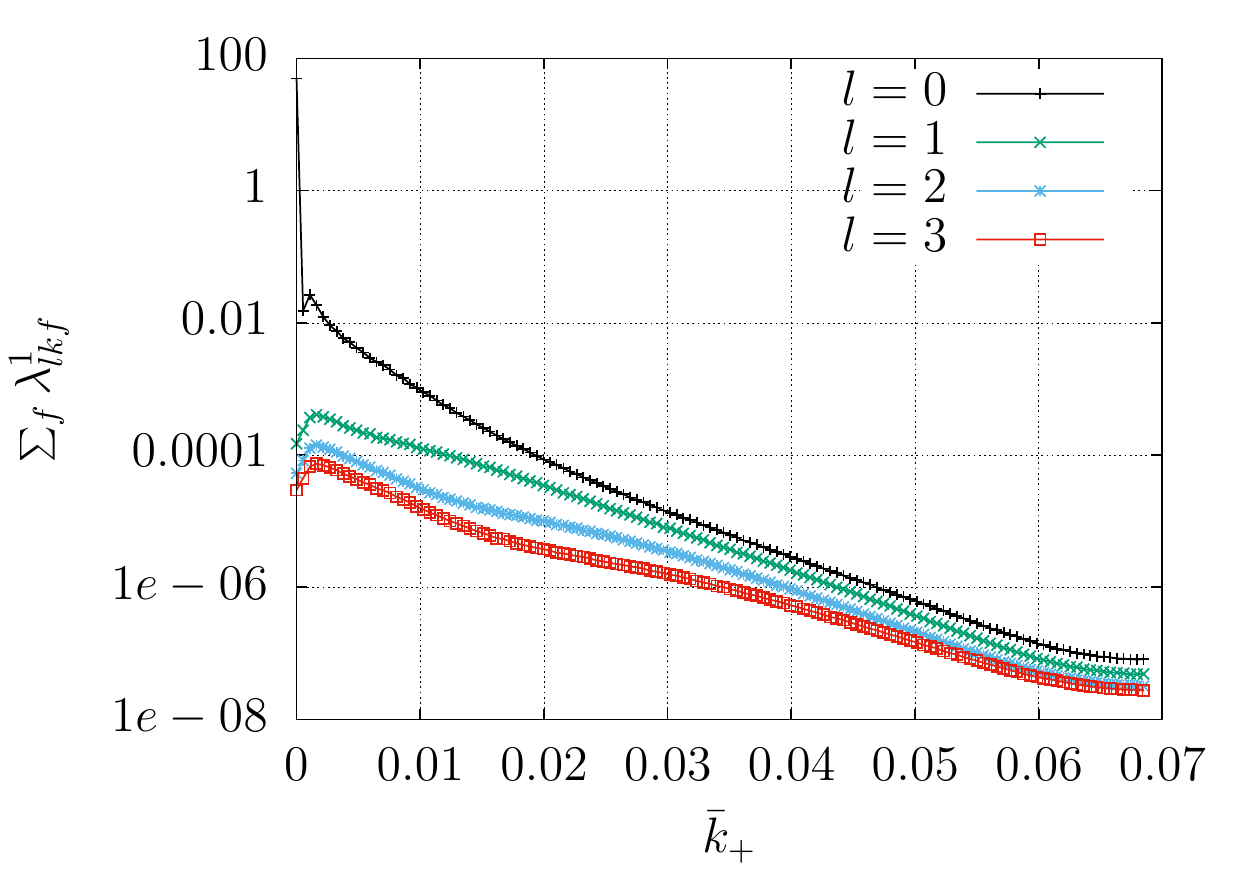}
\caption{}
\label{fig:cumul_k}
\end{subfigure}
\begin{subfigure}[b]{0.5\textwidth}
\includegraphics[width=\linewidth]{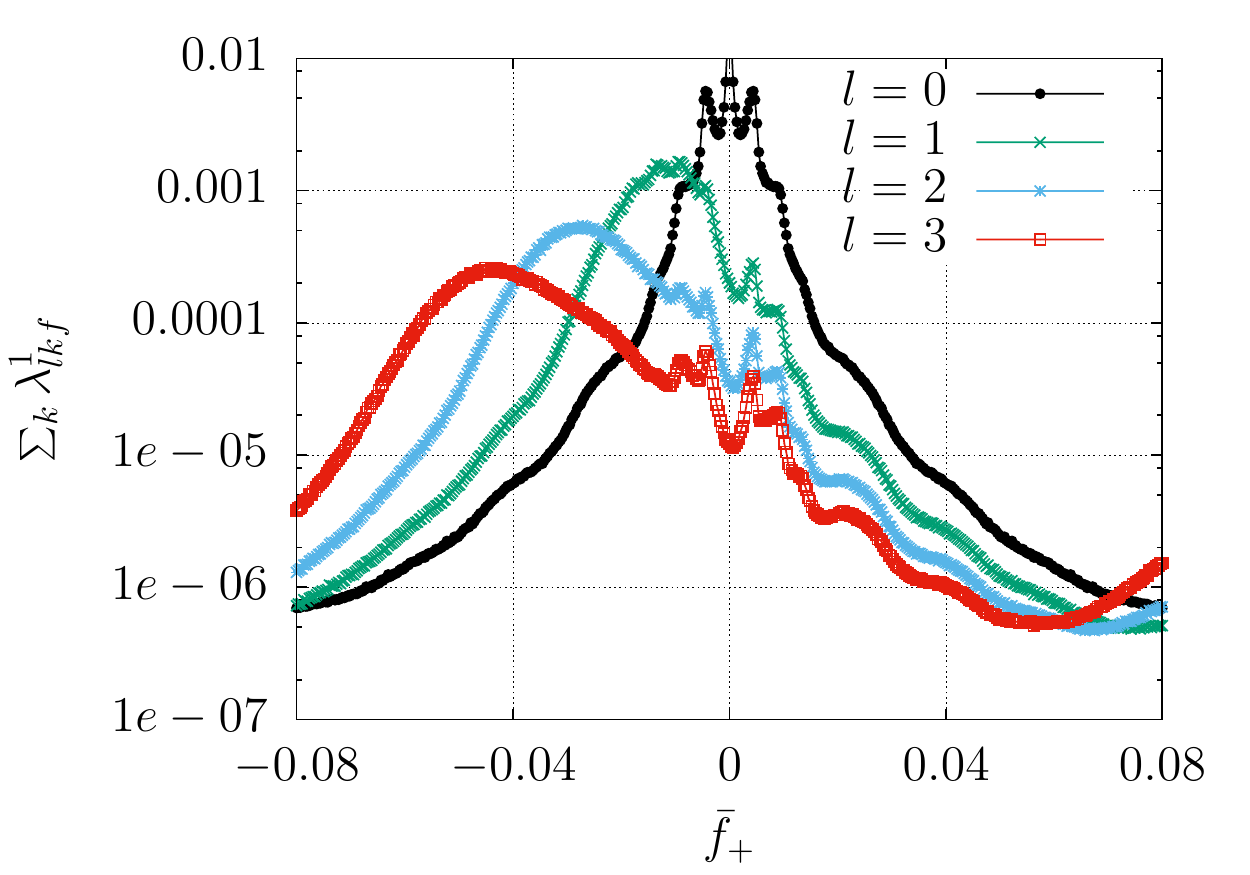}
\caption{}
\label{fig:cumul_f}
\end{subfigure}
\caption{ a) Energy spectrum { (represented in log scale)} summed over frequency $f$, $\sum_f \lambda_{lkf}^{1}$, 
for different streamwise wavenumbers $l$ as a function of $\bar{k}_+$.
 b) Energy spectrum { (represented in log scale)} 
summed over spanwise wavenumber $k$, $\sum_k \lambda_{lkf}^{1}$, for different streamwise wavenumbers $l$
as a function of $\bar{f}_+$.} 
\label{cumul}
\end{figure}

%Figure~\ref{fig:eig_dist} shows the eigenvalues as a function of both $\bar{f_+}$ and $\bar{k_+}$, to get a complete view. For better visibility of the eigenvalues, only top 200 eigenvalues are shown, without the mean mode. The size of the dot is proportional to the value of eigenvalue.

Figures~\ref{dist_full}a and \ref{dist_full}b respectively show the distribution of the top 5000 eigenvalues along $\bar{k}_+$ and $\bar{f}_+$ for the D3 configuration.
Each eigenvalue is represented by a dot.
Both figures show that the peaks observed in frequency and wavenumber space 
are not created by an accumulation of small eigenvalues, but correspond to coherent, more energetic structures.  
We note that Figure~\ref{dist_full}b also shows smaller, but noticeable 
peaks of $\bar{f}_+$ at around 0.009 and 0.0135, which correspond to the harmonics of the frequency $0.0044$.
 
Figure~\ref{fig:cumul_k} shows that the energy spectrum
integrated in frequency space $\sum_{f} \lambda_{lkf}^{1}$
slowly decreases in an apparently self-similar manner 
with respect to both streamwise and spanwise wavenumbers. 
Figure~\ref{fig:cumul_f} shows the corresponding variations 
 in the frequency space of the energy spectrum integrated in 
the spanwise wavenumber space for different streamwise wavenumbers $l$.
For the $l=0$ mode, owing to spanwise reflection invariance, the plot is symmetric 
in the frequency space, and the characteristic frequency $\bar{f}_{c+}$ 
and its harmonics can be clearly identified.
 These peaks observed at $l=0$ are still present in the $l \neq 0$ spectra, which confirms that
 the characteristic frequency $f_{c+}$ is not an artifact of the streamwise average.
 However non-zero streamwise wavenumbers are also characterized by a broad peak, which represents
 convection effects.
In general, defining a convection velocity for the turbulent wall layer is not 
straightforward (\cite{kn:krogstad98, kn:alamojimenez09}), but the question can 
be more easily addressed 
in the present framework where each POD eigenfunction is naturally 
 associated with a phase velocity $- \bar{f}/\bar{l}$. The spectra $\sum_{k} \lambda_{lkf}^{1}$ 
can be used to define 
 a global convection velocity $c$ at each streamwise wavenumber using 
 \begin{equation}
 c(\bar{l})= - \frac{\bar{f}_{\mathrm{max}}}{\bar{l}} ,
 \end{equation}
where
 \begin{equation}
\bar{f}_{\mathrm{max}}=\argmax_{\bar{f}} \left[\sum_{k} \lambda_{lkf}^{1}\right].
 \end{equation}
Peaks in the spectra are located at $\bar{f}_+=(0.013, 0.027, 0.044)$ for respective streamwise wavenumbers of
$\bar{l}_+=(0.0011, 0.0022, 0.0033)$.
This corresponds to a global convection velocity of
12$u_{\tau}$, with a slight upward shift observed with 
increasing streamwise wavenumbers. This 
agrees well with the classical results
(\cite{kn:kreplin79}, 
\cite{kn:krogstad98}, 
\cite{kn:wallace14}, 
\cite{kn:alamojimenez09}). 

%c(l)= - \frac{1}{\bar{l}} arg max_{\bar{f}_+} [\sum_{k} \lambda_{lkf}^{1}] 
%One can see peaks for non-zero values of the streamwise
%This shows that one can recover the eigenvalues obtained by the classical POD by summing over all frequencies. Unlike the spectra in the $f$ plane, the energy spectra in $k$ plane does not present a clear peak, but rather has an initial flat region followed by a sharp decay. Sharp peaks at $k$ around $10$ are reported only for a lower $Re^+$ of $180$, which corresponds to a spanwise wavelength of 100 \citep{moser1999}. This indicates that a clear distinction of the characteristic spanwise wavelengths are a low-Reynolds number effect.
%The largest eigenvalues correspond to very low frequencies, corresponding to scales larger than the time spanned by each sample. 

%In the next section, we focus on the 
%eigenfunctions associated with the region where $\bar{f_+} \approx 0.0044$ and $\bar{k_+} \approx 0.0016$. 

\begin{figure}
\begin{minipage}{0.5\textwidth}
\includegraphics[width=0.9\textwidth]{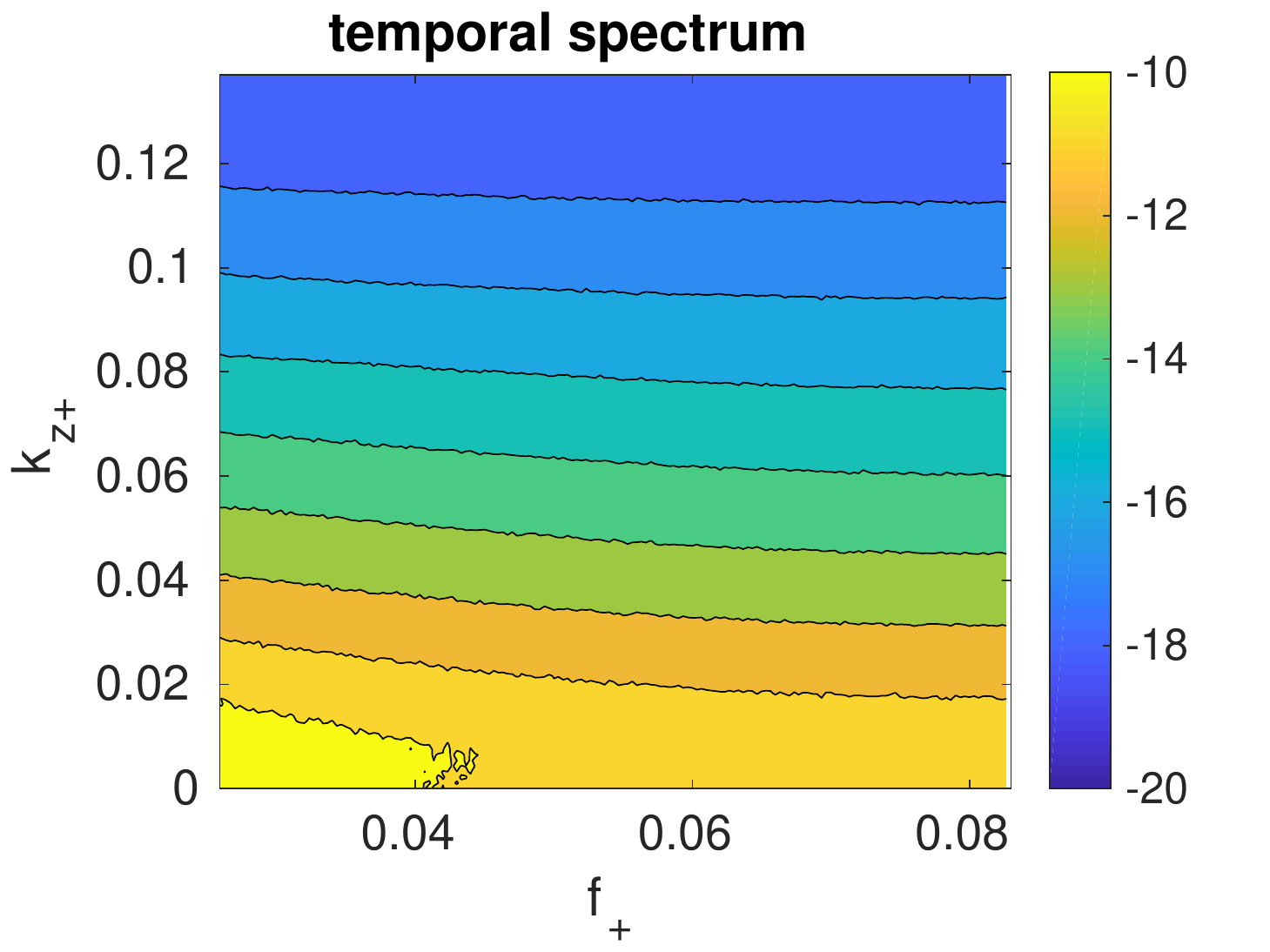}
\end{minipage}
\begin{minipage}{0.5\textwidth}
\includegraphics[width=0.9\textwidth]{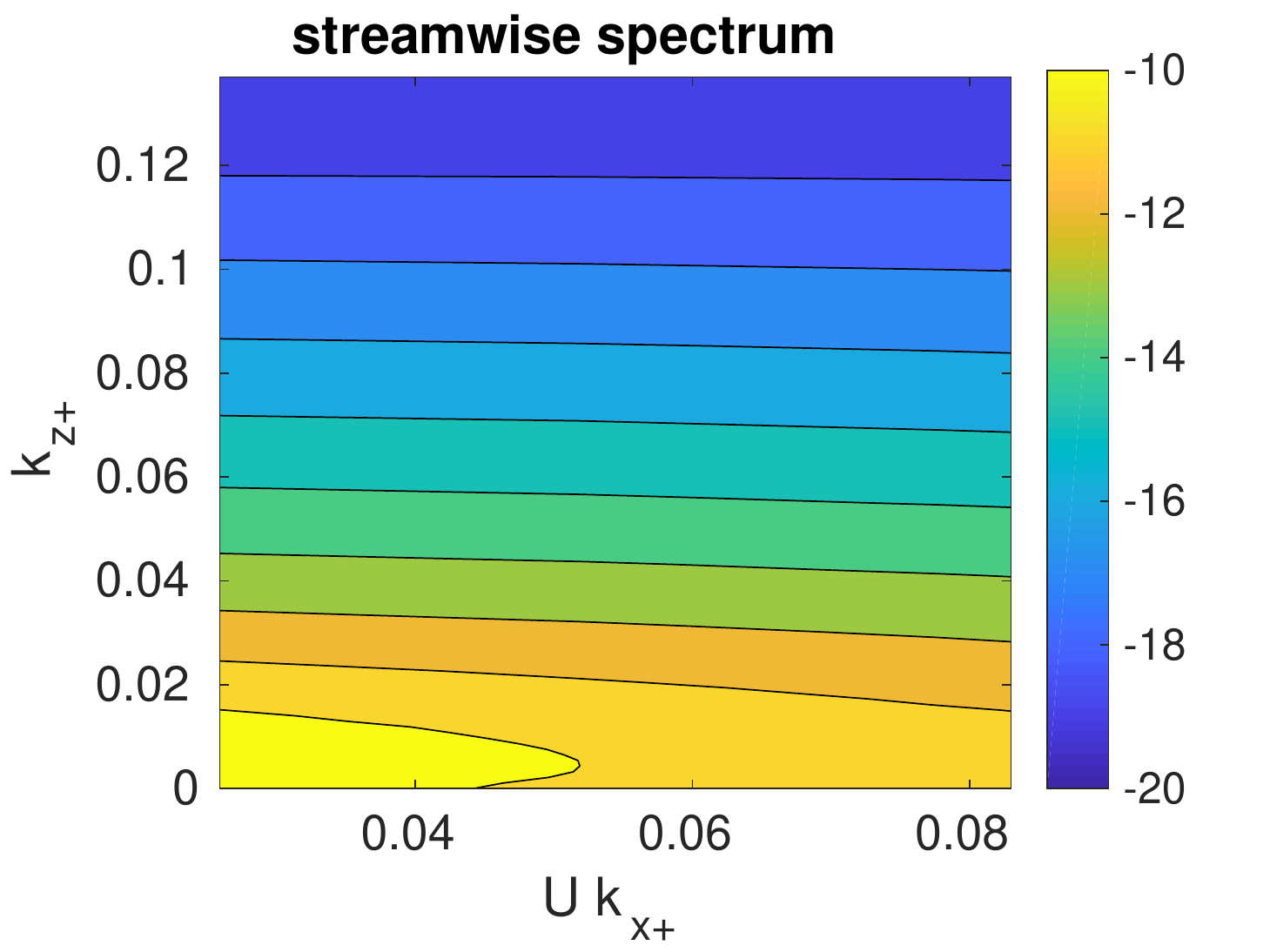}
\end{minipage}
\caption{ left: Eigenvalue spectrum $\sum_{l} \lambda_{lkf}^{1} \delta k_{x+}$ 
as a function of $f_+$ ;
right: Eigenvalue spectrum $\sum_{f} \lambda_{lkf}^{1} \delta k_{x+}$ 
as a function of $k_{x+} U_{c+}$, where $U_{c+}=U=12$.
\berengeredeux{The colorbar is based on the decadic log scale.} } 
\label{compstreamspec}
\end{figure}

\berengere{Having access to the four-dimensional space provides a way to 
test the Taylor's frozen turbulence hypothesis \citep{kn:lumley65},
which states that the spatial spectrum in the streamwise direction
$E_{k_x}$  of  $\overline{ u(x) u(x') } $ (where 
the overbar represents a spatial average)  
can be recovered from the temporal spectrum $E_{\overline{f}}$ of $ \overline{\overline{ u(t) u(t') }}$ 
(the double overbar denotes here a temporal average) where 
$\overline{f}= k U$ and  $U$ is a suitable convection velocity.
In the context of the present decomposition, this leads us to compare, for each spanwise wavenumber $k_z$ and  quantum mode $n$, the temporal spectrum
\[ \tilde{E}(\overline{f},k_z,n) = \sum_{l} \lambda_{lkf}^{n}  \delta k_x,  \]
 where $\overline{f}=f/T$ and $k_{z}=k/L_z$,}
%\Lionel{Should not this equation instead be ?:
%\[ \tilde{E}(\overline{f},k_z,n) = \sum_{l} \lambda_{l k_z \overline{f}}^{n}  \delta k_x  \] }
\berengere{with the streamwise spectrum  
for the streamwise wavenumber $k_x$
\[ E(k_x,k_z,n) = \sum_{f} \lambda_{lkf}^{n}  \delta \overline{f}, \]
 where $k_x=l/L_x$, using a suitable rescaling factor $\delta k_x / \delta \overline{f}$  which can be obtained 
from 
\[ {\cal E} =  \sum_{l} \sum_{k} \sum_{f} \lambda_{lkf}^{1} \delta k_x \delta k_z \delta \overline{f}
=   \sum_{l} \sum_{k}  E(k_x,k_z,n) \delta k_x \delta k_z 
=   \sum_{f} \sum_{k} \tilde{E}(f,k_z,n) \delta \overline{f} \delta k_z. 
 \]

For the first quantum mode $n=1$, Figure~\ref{compstreamspec} compares 
the spectra  in the temporal direction $E(k_x,k_z,1)$
with the equivalent spectrum in the streamwise direction $\tilde{E}(f,k_z,1)$. 
The frequency expressed in wall units $f_{+}$ is related to the 
the streamwise  wavenumber $k_{x+}$ such that $k_{x+} U_{c+} = f_{+}$ where we 
have taken $U_{c+}=12$.
For the range of corresponding frequencies,  
a good agreement can be observed between the spectra at each spanwise wavenumber $k_z$.}

%{\berengere new figure}

\subsection{POD eigenfunctions}

% Add the mean velocity comparison
% The mean profile is reconstructed using one POD eigenfunction for configuration D4. 

\begin{figure}
\centering
\includegraphics[width=\linewidth]{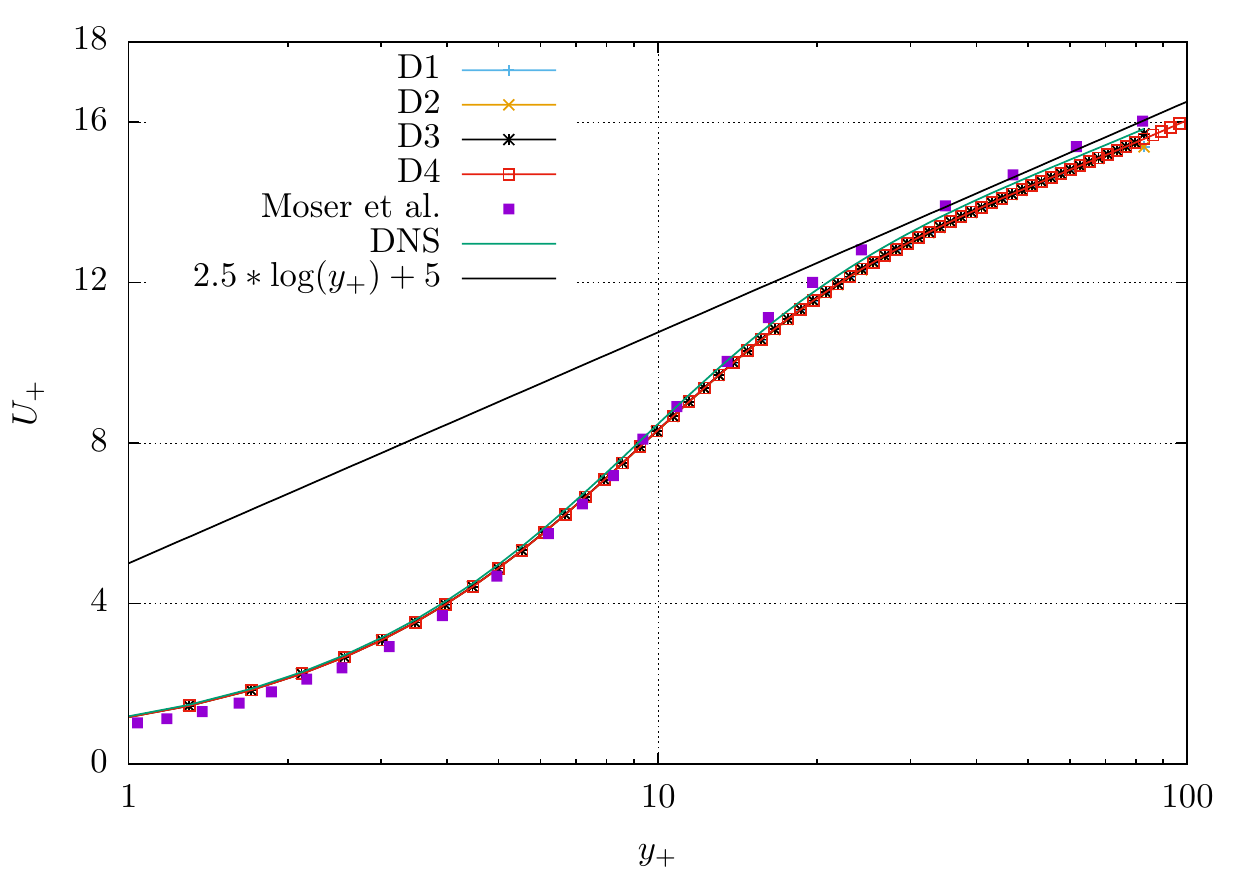}
\caption{Comparison of the mean streamwise profile reconstructed using the first 
eigenfunction for the mode pair $(0,0,0)$ with that of \citet{moser1999}, and with the mean profile obtained from the DNS.} 
\label{mean_comp}
\end{figure}
%. Furthermore, the mean profile is also compared with that obtained directly from the DNS data, and the two are in reasonable agreement. 
%The mean flow also respects the log-law as shown using the straight line.

% Eigenfunctions
\begin{figure}
\begin{subfigure}[b]{0.5\textwidth}
\includegraphics[width=\linewidth]{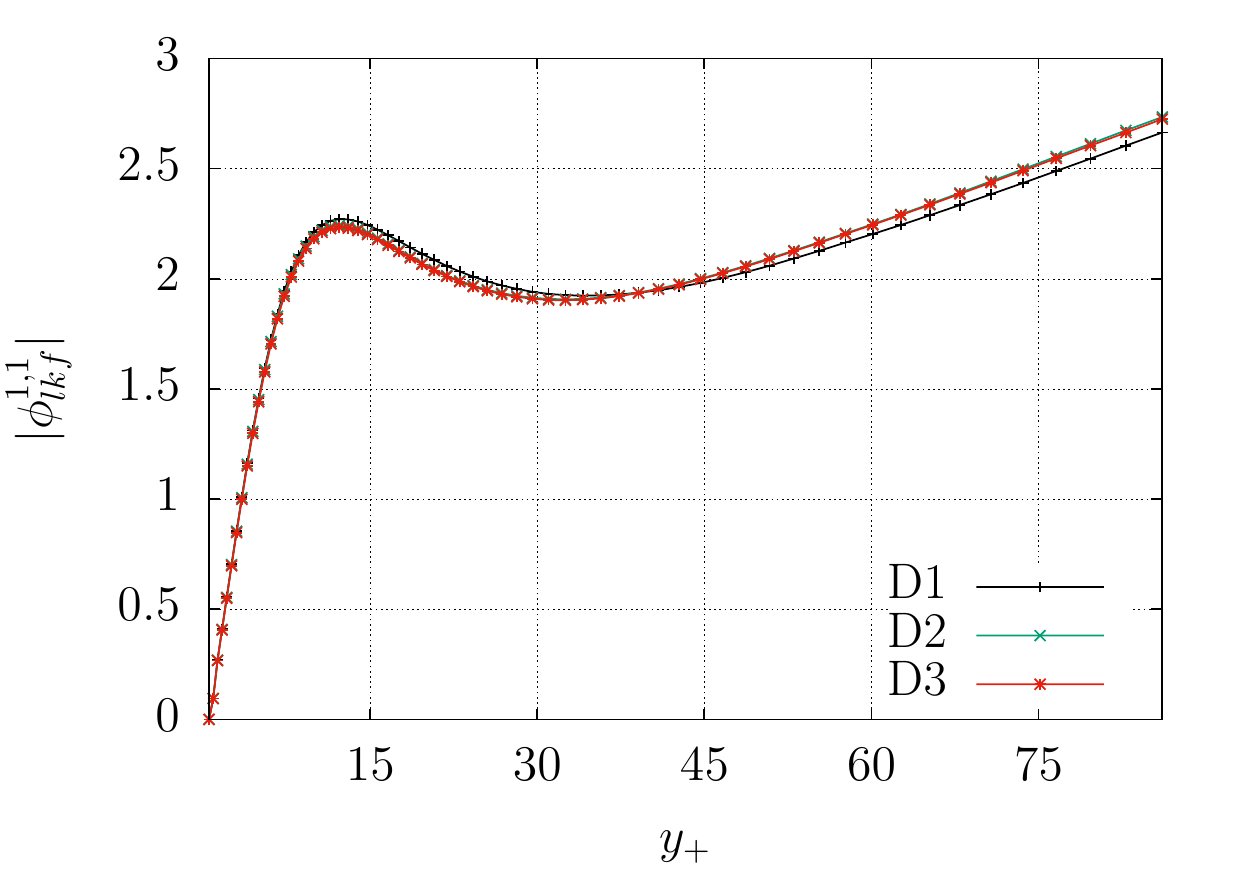}
\caption{streamwise component}
\end{subfigure}
\begin{subfigure}[b]{0.5\textwidth}
\includegraphics[width=\linewidth]{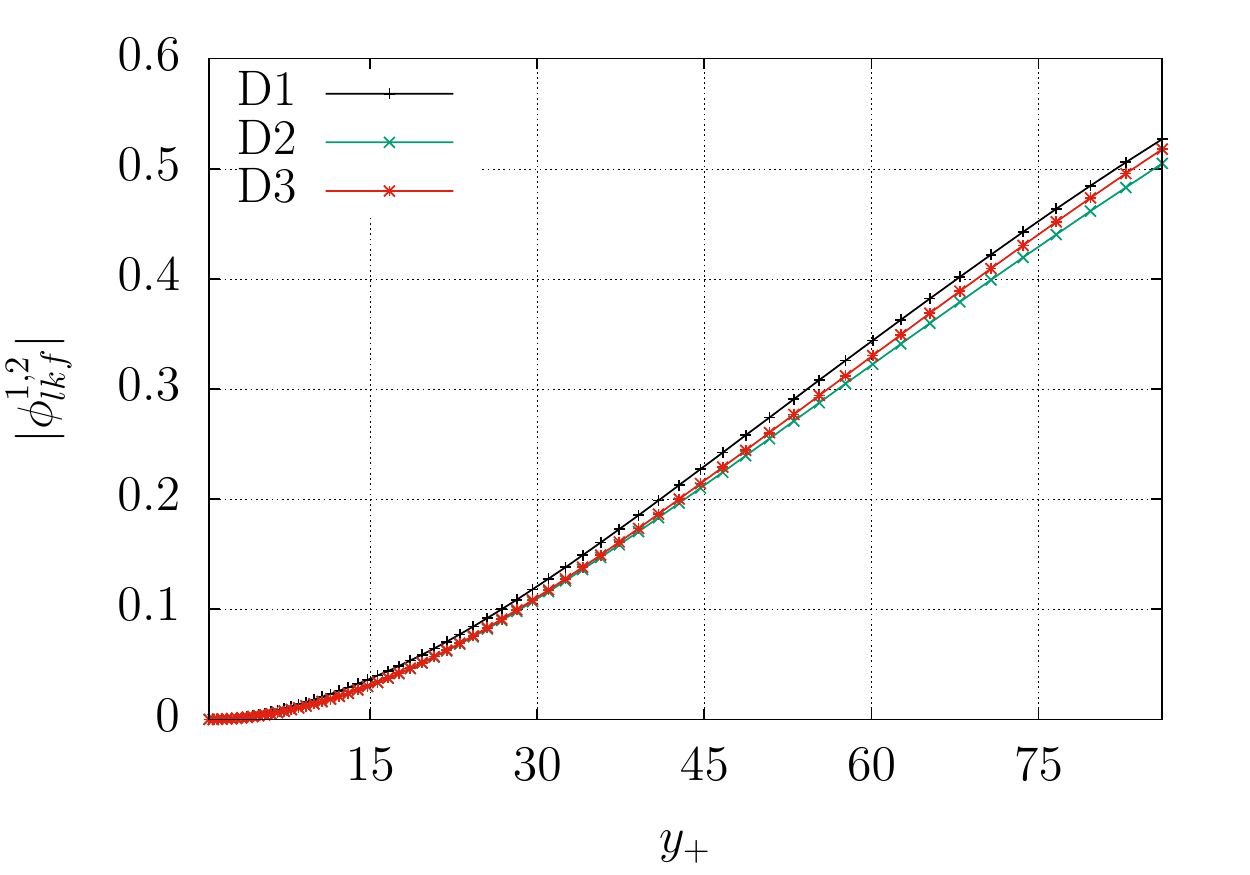}
\caption{wall-normal component}
\end{subfigure}
\begin{subfigure}[b]{0.5\textwidth}
\includegraphics[width=\linewidth]{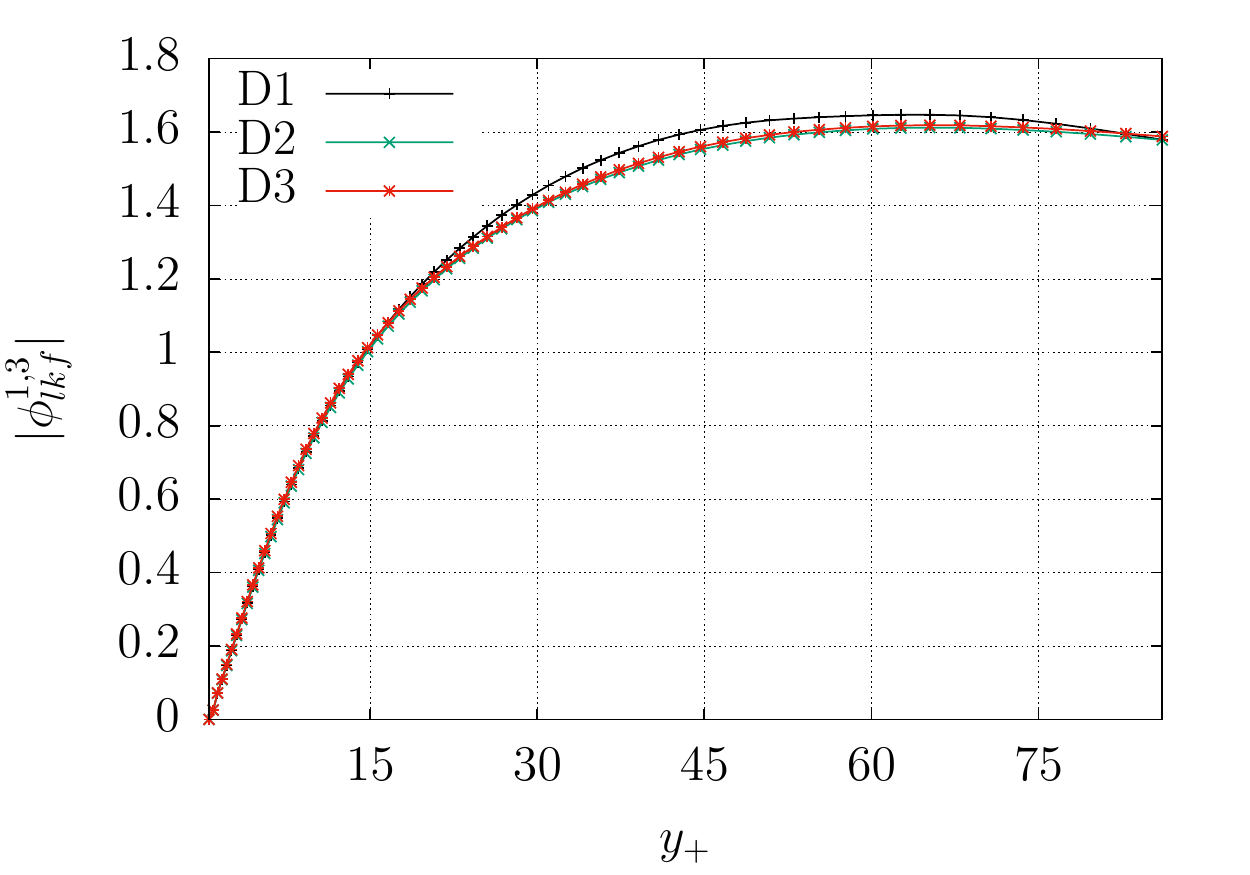}
\caption{spanwise component}
\end{subfigure}
\caption{ Comparison of the eigenfunctions $\pphi_{lkf}^{1}$ corresponding to $\bar{l}_+=0$, $\bar{k}_+=0.0016$, $\bar{f}_+=0.0044$ for 
different configurations.}
\label{convergence_eigenfun}
\end{figure}
Figure~\ref{mean_comp} shows that the reconstructed mean streamwise profile 
coincides with previous results (\citet{moser1999}) and 
with the dominant mode $(l=0,k=0,f=0,n=1)$ of the decomposition. 
Figure~\ref{convergence_eigenfun} shows the absolute value of each component of the most energetic 
mode $\phi_{lkf}^{p,n}$, for \berengeredeux{$(\bar{l}_+,\bar{k}_+,\bar{f}_+)=(0,0.0016,0.0044)$}, obtained 
for the datasets D1-D3.
As expected, there is a good agreement between the different configurations, which indicates convergence of the procedure at least for the most energetic modes.
In all that follows only results for D3 and D4 will be presented.
%As already observed in the spectra comparison in Figure~\ref{spectra_full}, the eigenfunctions for D1 do reproduce the trends, but less closely than the other three configurations, which agree well with each other. 

% This was also evident in comparison of eigenspectrum in Figure~\ref{spectra_full}.
%The eigenfunctions for the last two cases visually coincide, indicating that the Fourier transform in time has converged for $n_t=500$. However, it does not indicate that the eigenfunction shapes would remain the same if the domain size is increased along $x$ or $y$. The tests on the influence of larger domain be performed in the future. 

\begin{figure}
\begin{subfigure}[b]{0.5\textwidth}
\includegraphics[width=\linewidth]{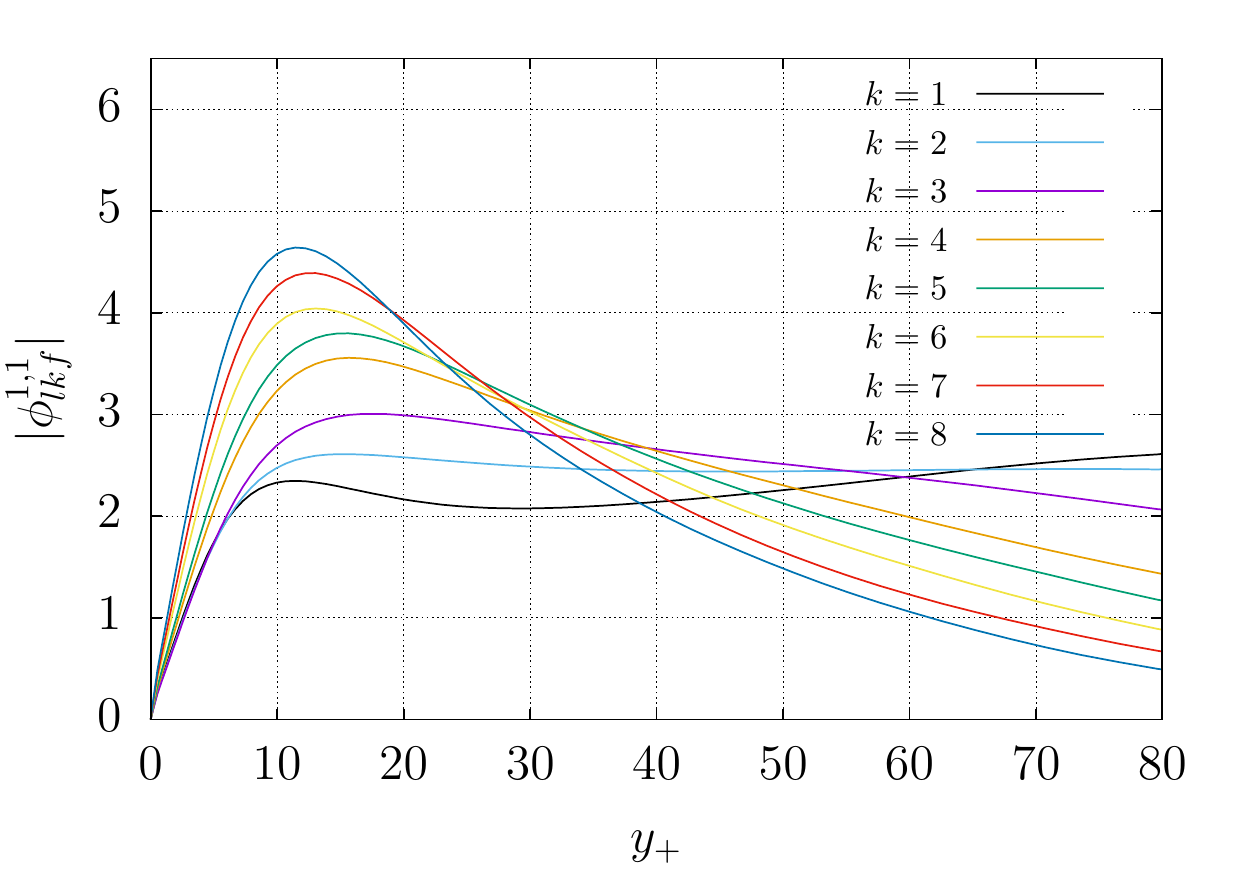}
 \caption{streamwise component}
\end{subfigure}
\begin{subfigure}[b]{0.5\textwidth}
\includegraphics[width=\linewidth]{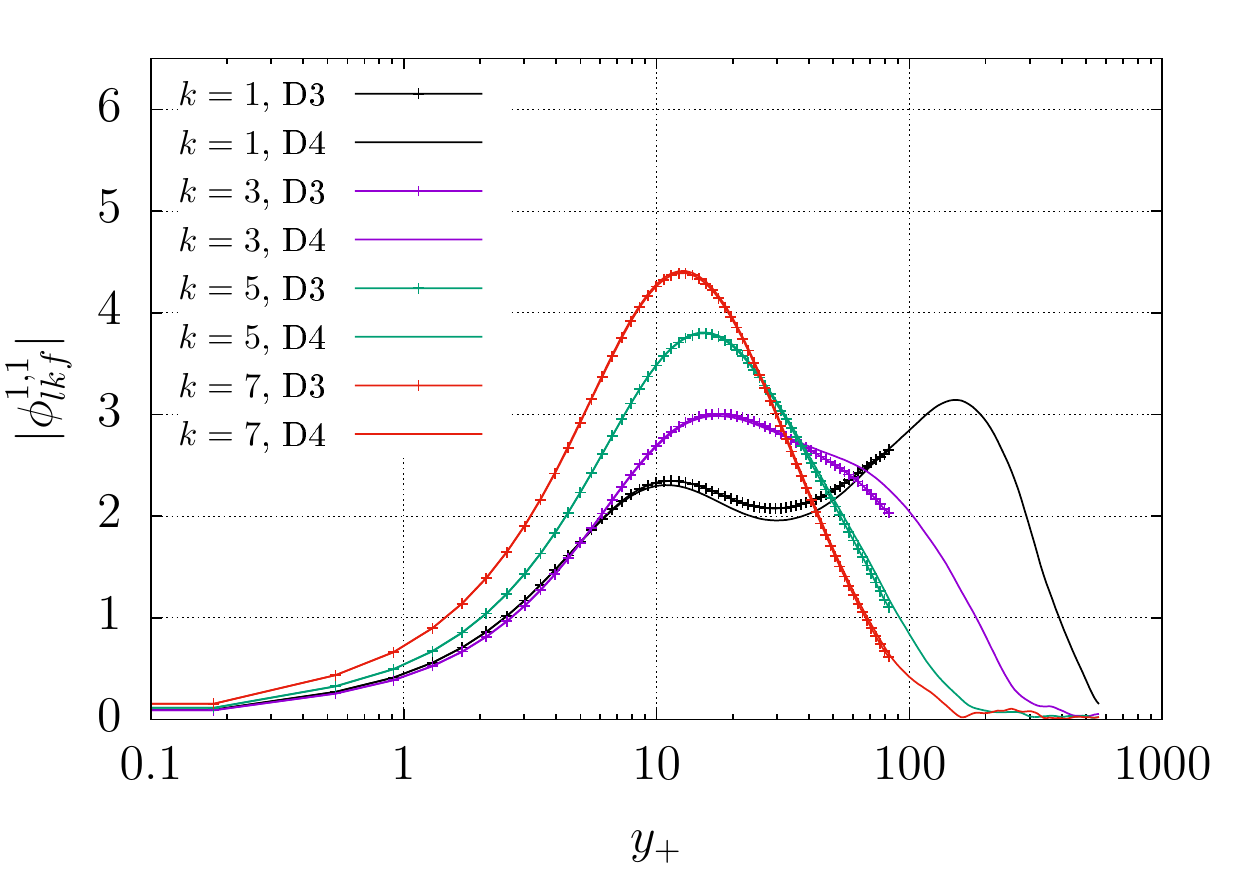}
\caption{streamwise component}
\end{subfigure}
\begin{subfigure}[b]{0.5\textwidth}
\includegraphics[width=\linewidth]{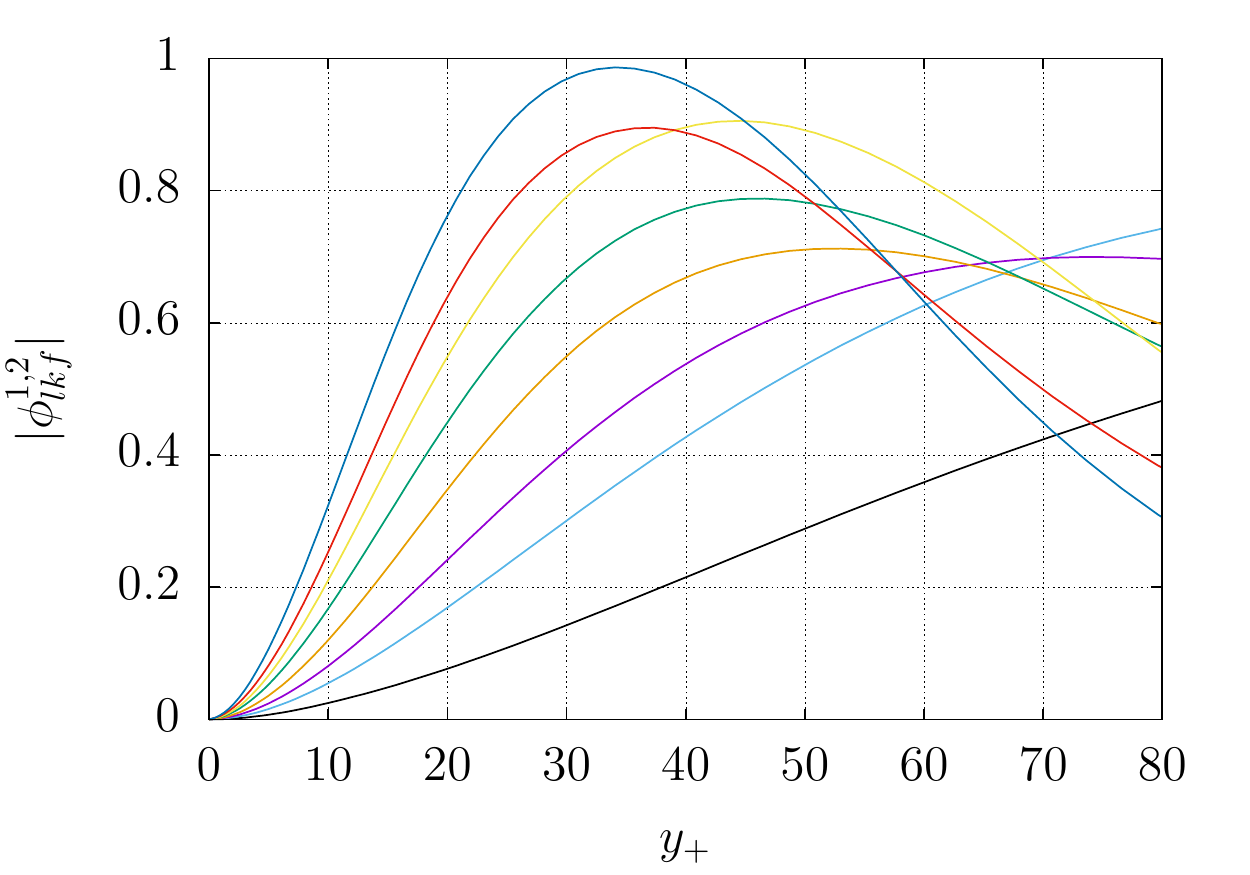}
\caption{wall-normal component}
\end{subfigure}
\begin{subfigure}[b]{0.5\textwidth}
\includegraphics[width=\linewidth]{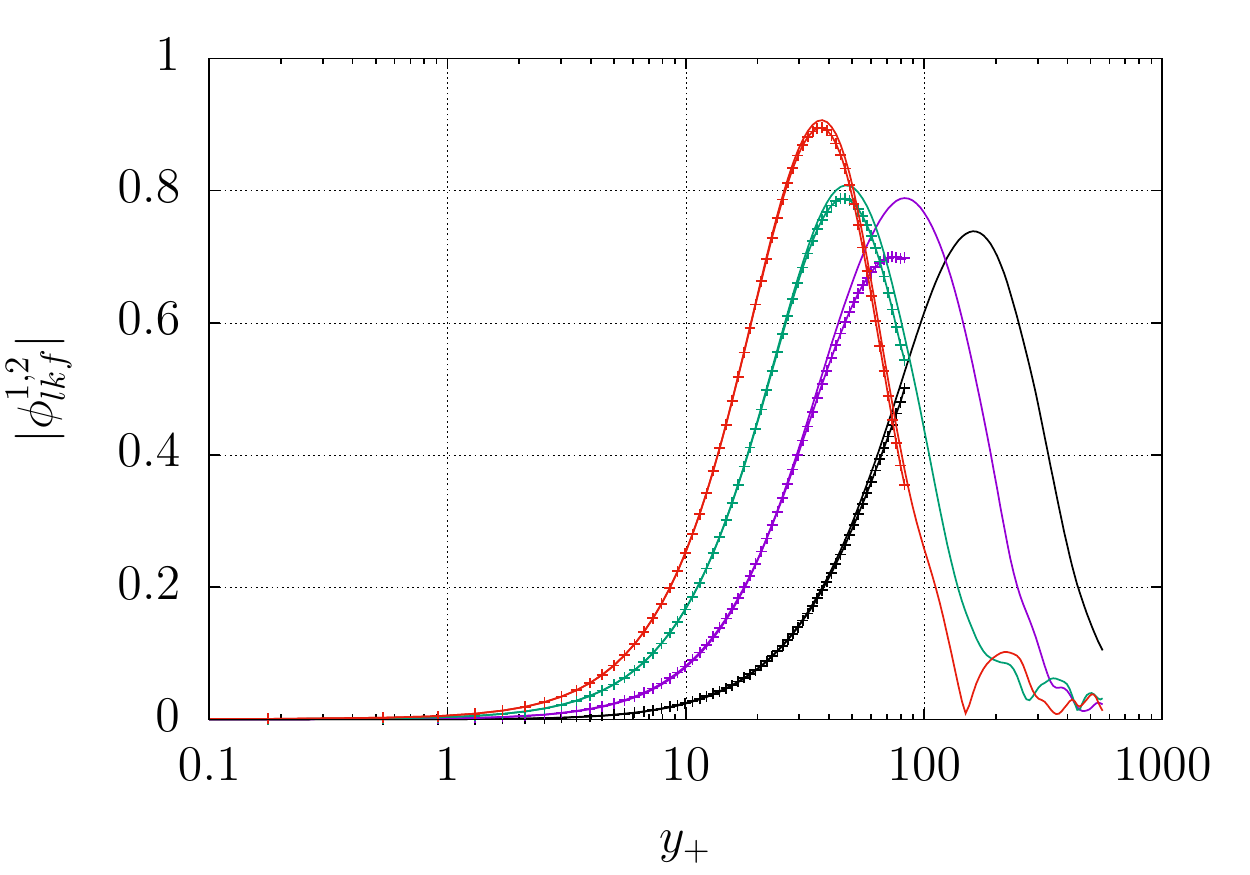}
\caption{wall-normal component}
\end{subfigure}
\begin{subfigure}[b]{0.5\textwidth}
\includegraphics[width=\linewidth]{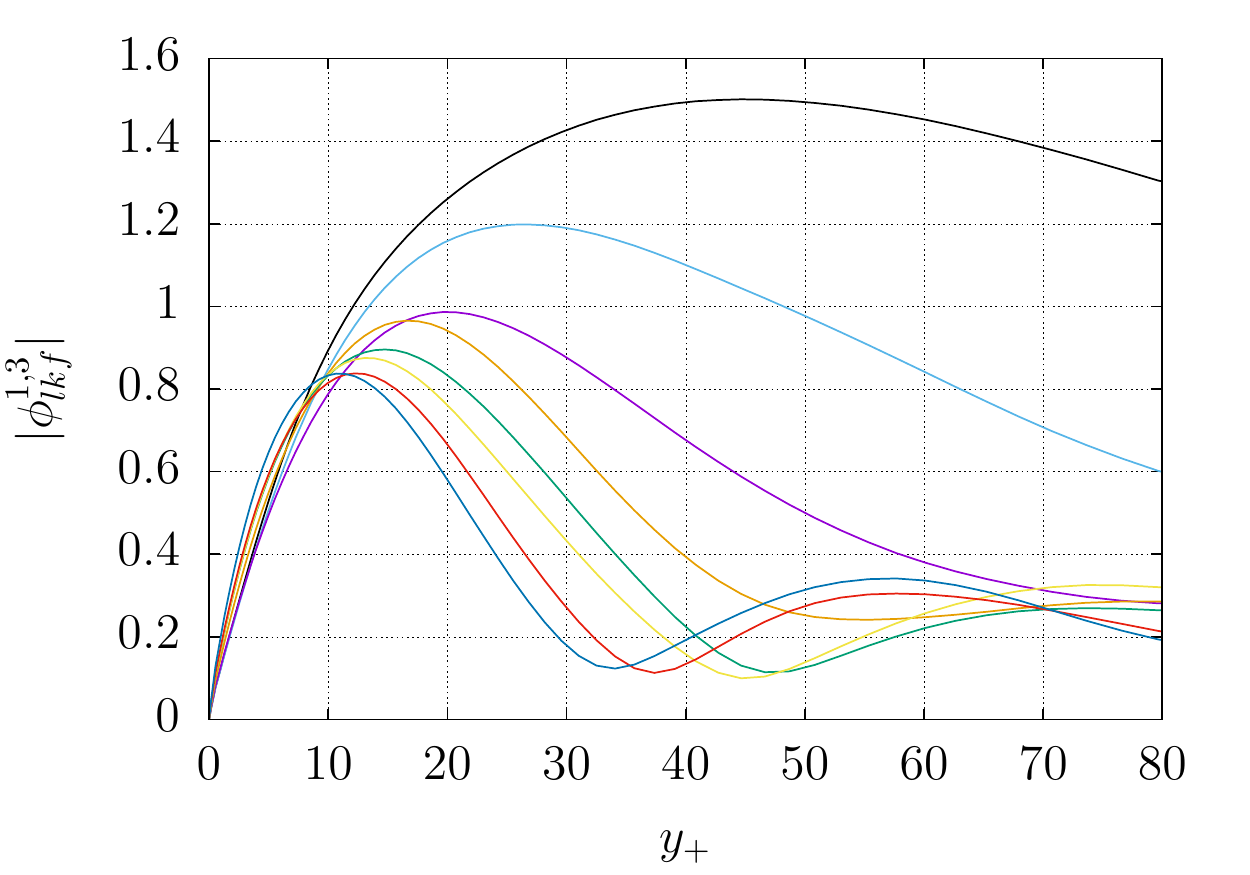}
\caption{spanwise component}
\end{subfigure}
\begin{subfigure}[b]{0.5\textwidth}
\includegraphics[width=\linewidth]{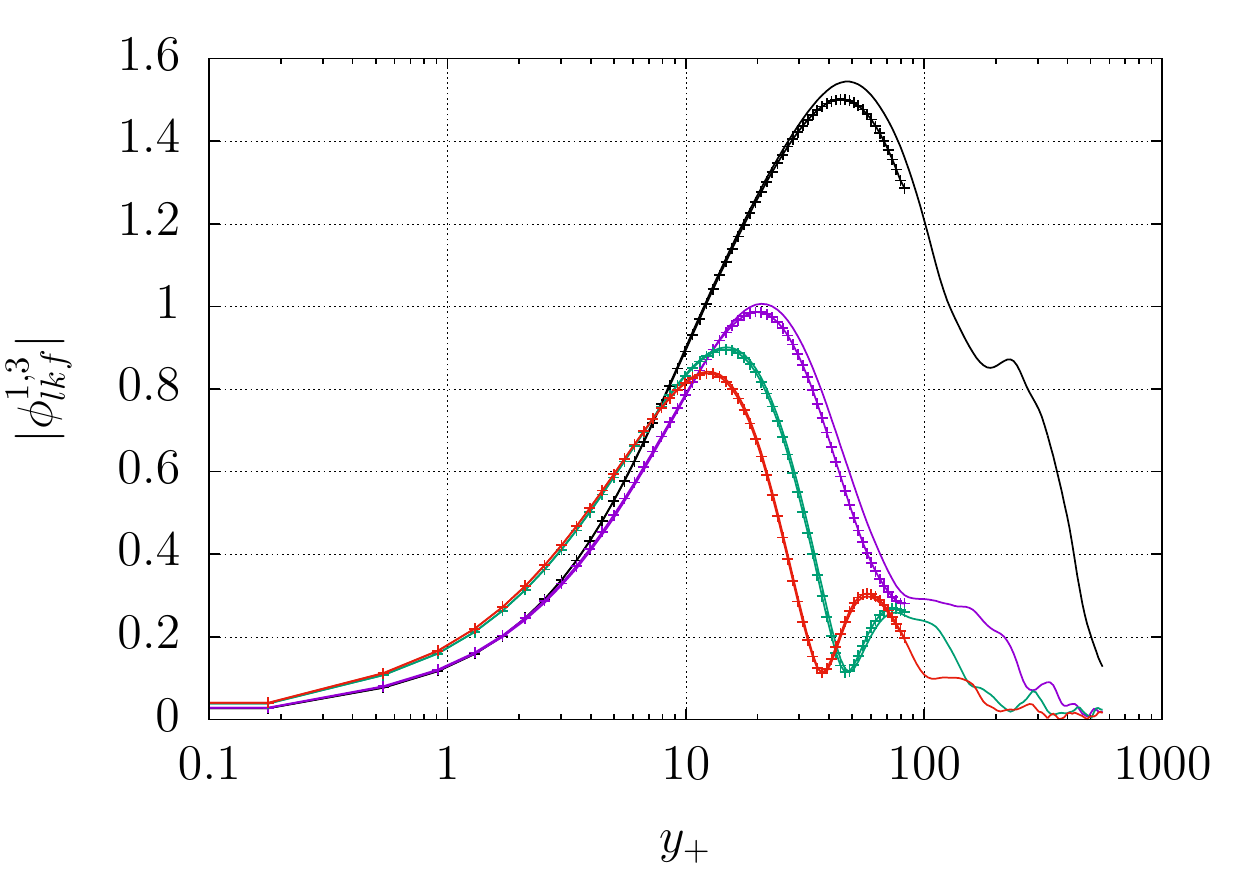}
\caption{spanwise component}
\end{subfigure}
\caption{ Dominant eigenfunctions $\pphi_{0 k f}^{1}$ corresponding to $\bar{f}_{c+}$ and for increasing values of $\bar{k}_+$.
Top row: Streamwise component; Middle row: Wall-normal component; Bottom Row: Spanwise component;
Left column: Comparison for $0 < y_+ < 80$ (D3) and different spanwise wavenumbers ($k=1$ to $8$ which corresponds to $\bar{k}_+$ between $0.0011$ to $0.0086$) - the legend for a), c) and e) is given in a);
Right column: Comparison between the domains $0 < y_+ < 80$ (D3) and $0 < y_+ < 590$ (D4)
for selected wavenumbers $k=1,3,5,7$.}
\label{velcomp_freq}
\end{figure}

% Comparison of eigenfunctions for D4 and D5
In order to gain more insight on the structure of the modes, Figures~\ref{velcomp_freq}a, c and e show the shape of 
the most energetic eigenfunction components $\pphi_{0kf}^{1,p}$ 
associated with the characteristic frequency 
$\bar{f}_{c+} \sim 0.0044$ for different spanwise wavenumbers. The location of the velocity maxima moves closer to the wall as the spanwise wavenumber increases, which is
in agreement with previous descriptions of wall-attached structures 
(\citet{kn:delalamo06}, \citet{kn:jfe10}). 
The values of the maxima are similar for both cross-stream components,
also in agreement with previous observations (\citet{kn:delalamo06}, \citet{kn:pof17}),
but the value tends to increase with the spanwise wavenumber for the streamwise and the wall-normal component, while 
it slightly decreases for the spanwise component. 
The monotonous evolution of the shape of the modes with the spanwise wavenumber suggests 
self-similarity, as was proposed in \citet{kn:pof17}.

Figures~\ref{velcomp_freq}b, d and f compare the eigenfunctions obtained on the domain $0 < y_+ < 80$ (for D3) 
and $0 < y_+ < 590$ (for D4) for several spanwise wavenumbers.
The eigenfunctions of D4 have been rescaled to have the same energy content as those of D3 on $0 < y_+ < 80$.
We observe that the eigenfunctions nearly coincide over their common definition 
domain, which is not a trivial result.
 The persistence of the eigenfunction shape with respect 
to the wall-normal extension 
 of the decomposition domain shows the coherence of the most energetic 
motions over the entire height of the boundary layer. 
It also means that the restrictions of the eigenfunctions on the larger
domain are orthogonal to each other
on the smaller domain (since they coincide with the
eigenfunctions there), which makes it possible to recover
the amplitude of the eigenfunction on the larger domain directly from
the projection of the velocity field in the smaller domain 
onto the corresponding eigenfunction.
This shows the relevance of the decomposition for estimation
purposes in a context of partial information (see for instance \cite{kn:jfe10}).

%Several most energetic eigenfunctions correspond to the frequency (..). Figure shows their eigenfunctions.

%Both frequency indices $f=13$ and $f=14$ were retained, since they both appear in the decomposition. 

%The plots are ordered by increasing spanwise wavenumber. 
%As expected, modes corresponding to close frequencies and the same spanwise wavenumber have similar distributions. 
%The curves are similar in the region $y_+ < 10$.
%Above that height, 

%However we note that the variations of the mode with the wavenumber are less marked above the values $k \sim 8-10$
%which corresponds to the range $\lambda_z+ \sim 150-180$.

\begin{figure}
\begin{minipage}{0.24\textwidth}
\includegraphics[width=0.95\textwidth]{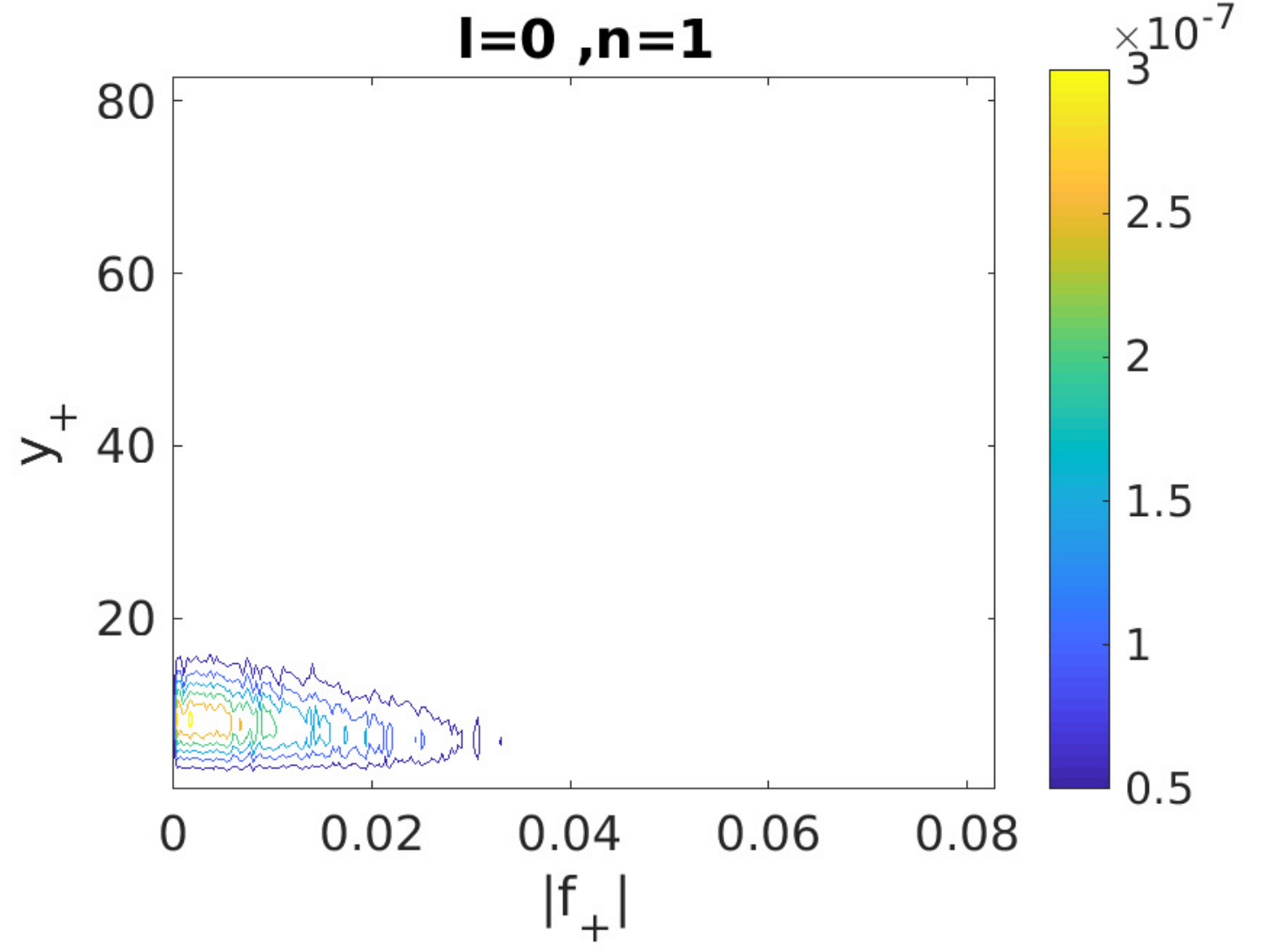}
\end{minipage}
\begin{minipage}{0.24\textwidth}
\includegraphics[width=0.95\textwidth]{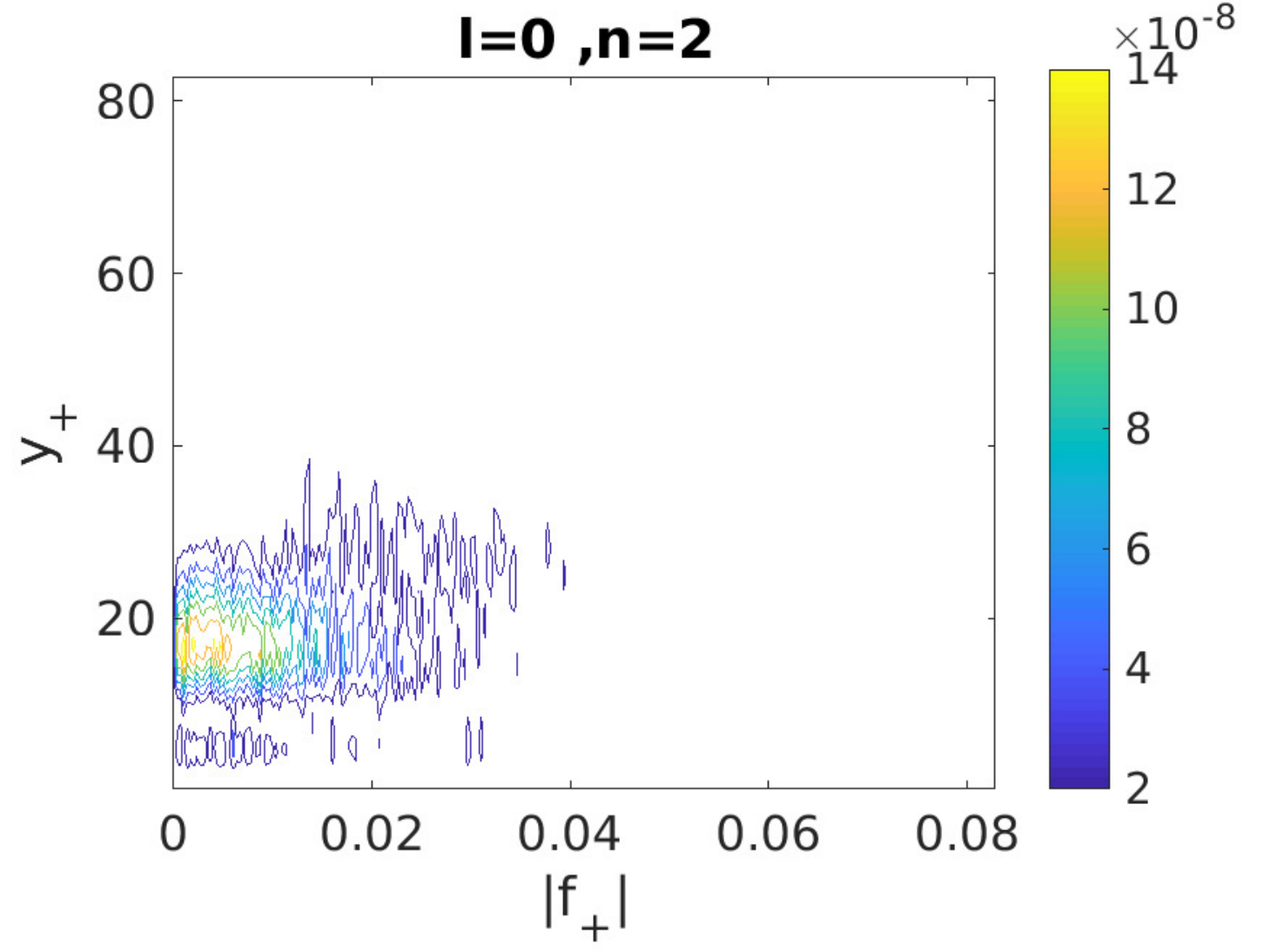}
\end{minipage}
\begin{minipage}{0.24\textwidth}
\includegraphics[width=0.95\textwidth]{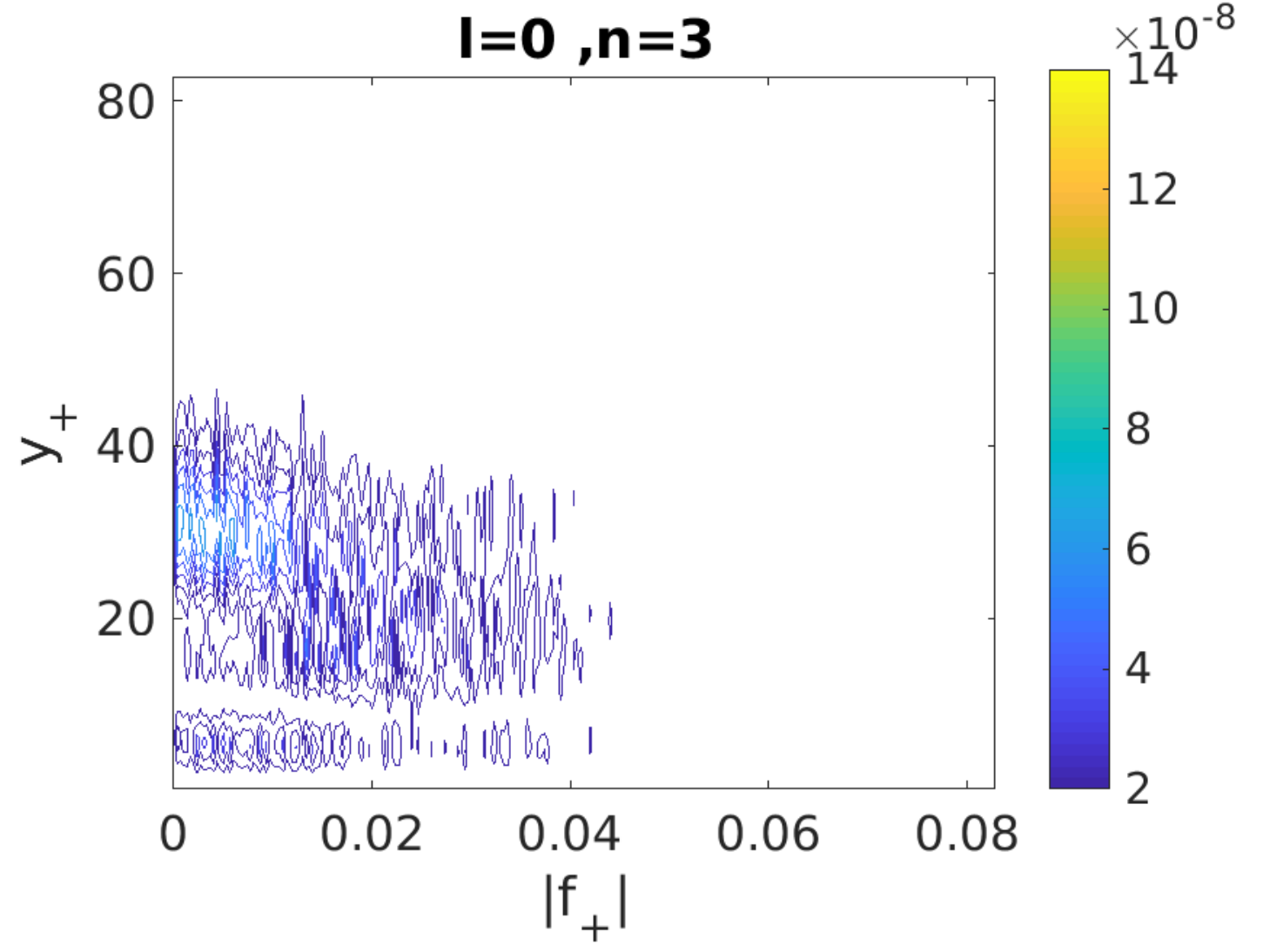}
\end{minipage}
\begin{minipage}{0.24\textwidth}
\includegraphics[width=0.95\textwidth]{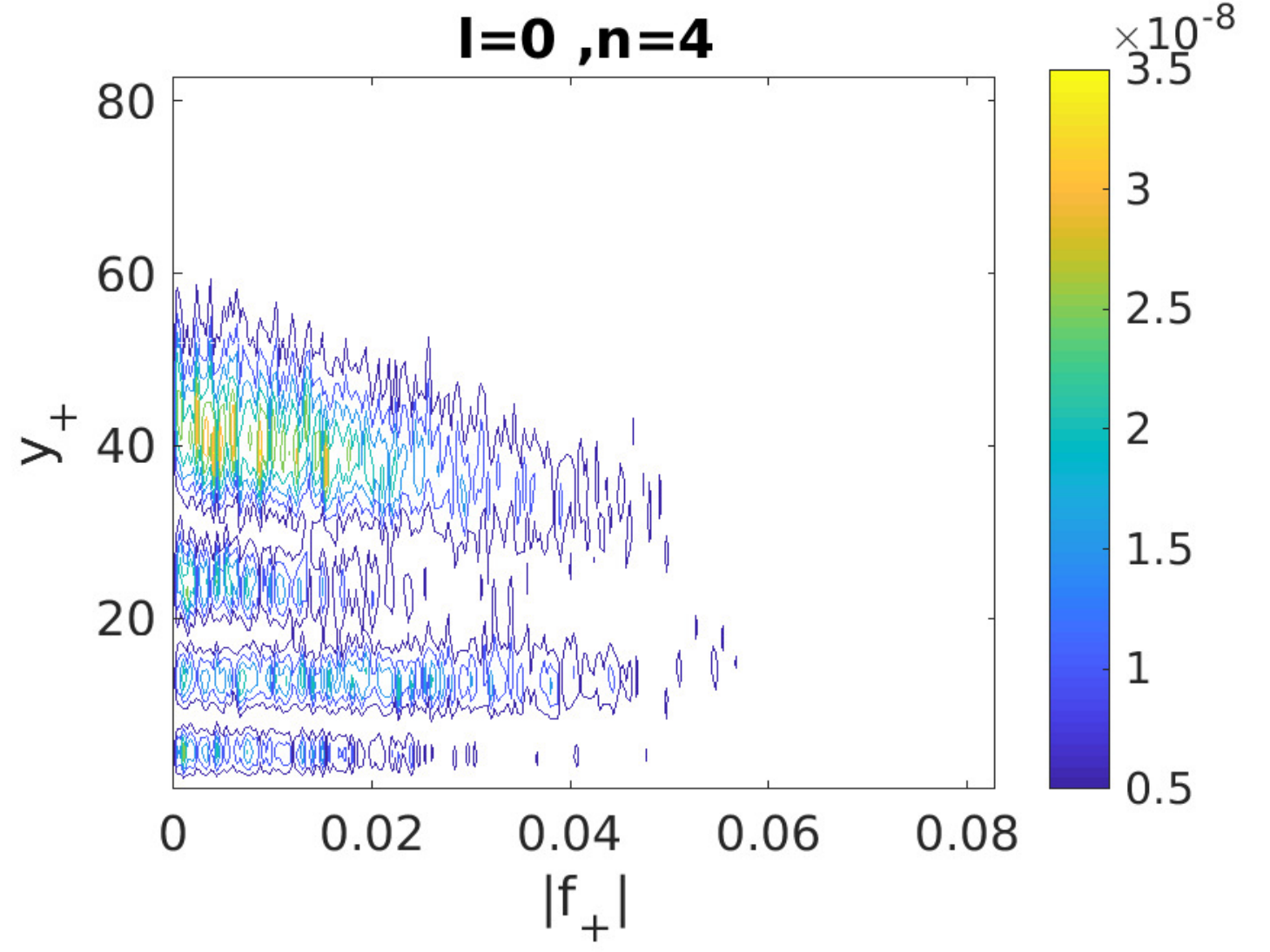}
\end{minipage}
\begin{minipage}{0.24\textwidth}
\includegraphics[width=0.95\textwidth]{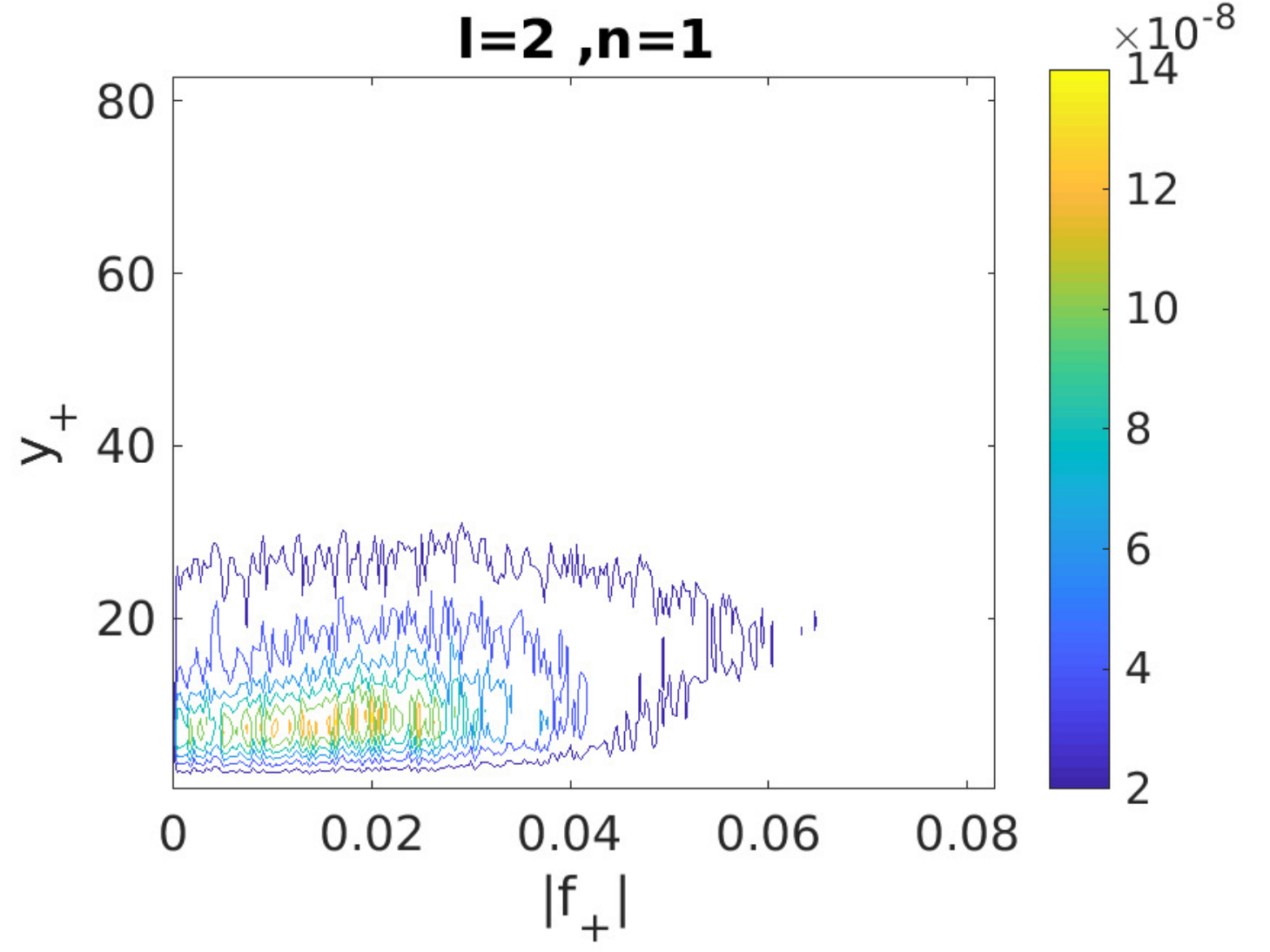}
\end{minipage}
\begin{minipage}{0.24\textwidth}
\includegraphics[width=0.95\textwidth]{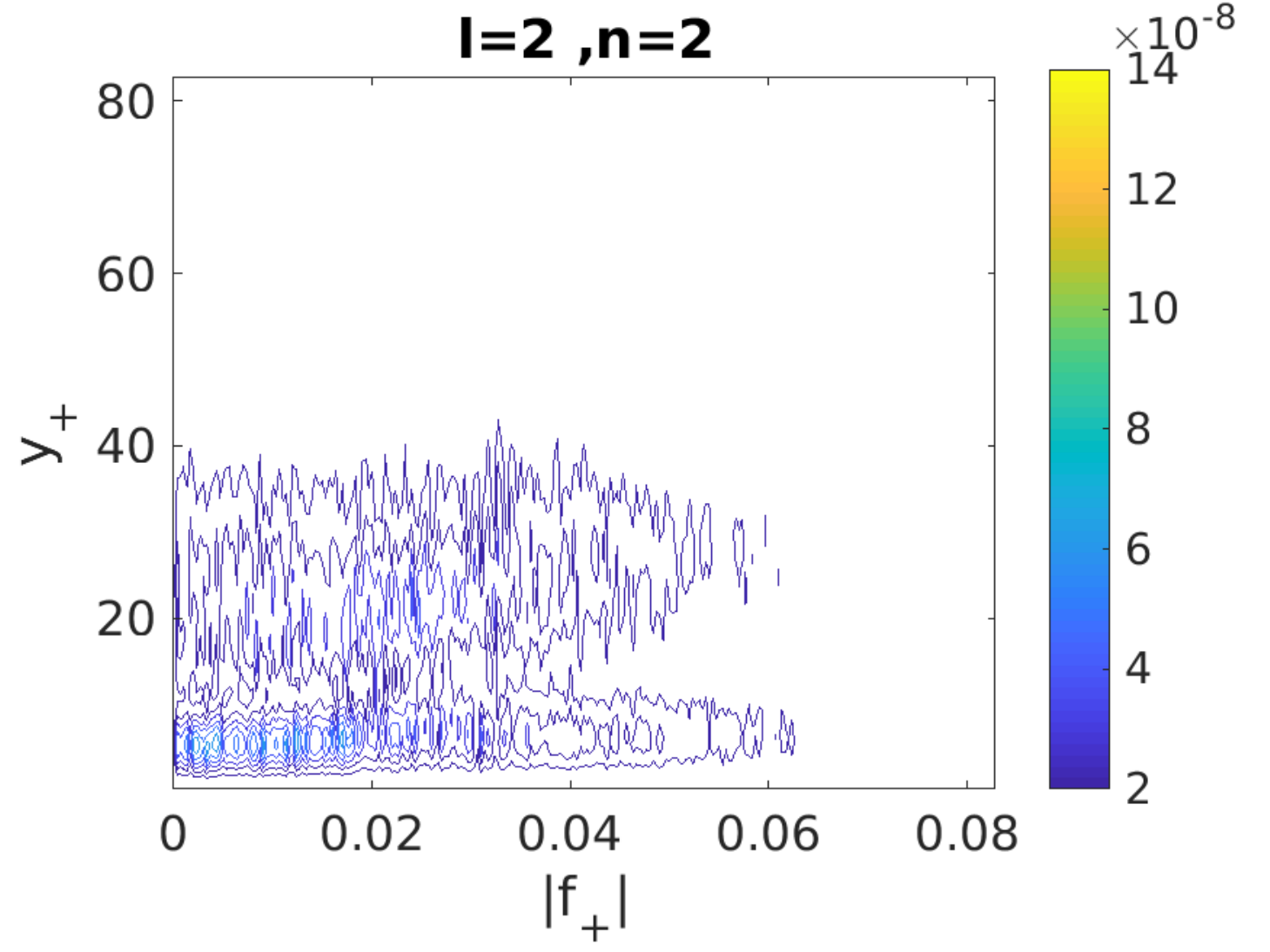}
\end{minipage}
\begin{minipage}{0.24\textwidth}
\includegraphics[width=0.95\textwidth]{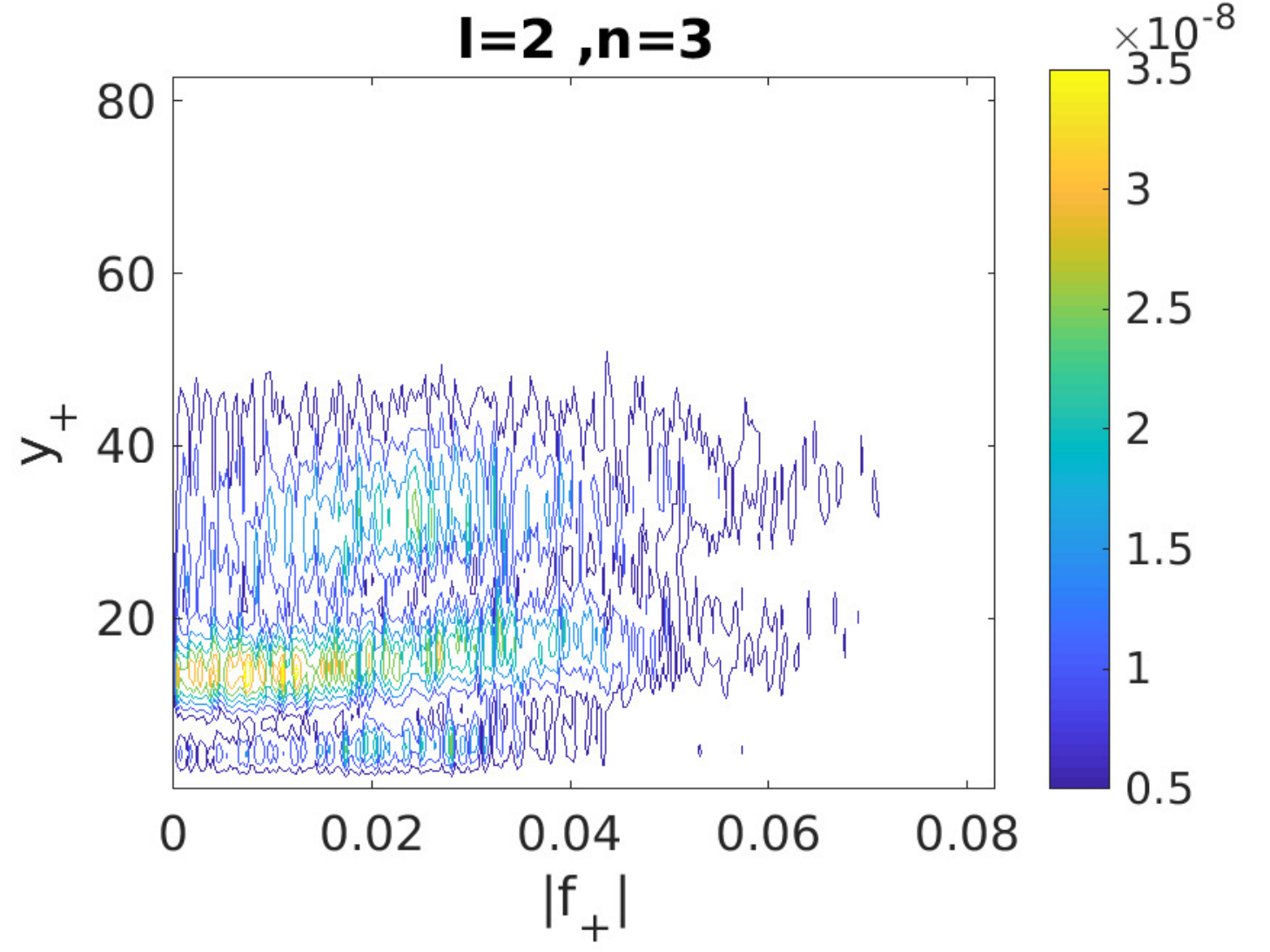}
\end{minipage}
\begin{minipage}{0.24\textwidth}
\includegraphics[width=0.95\textwidth]{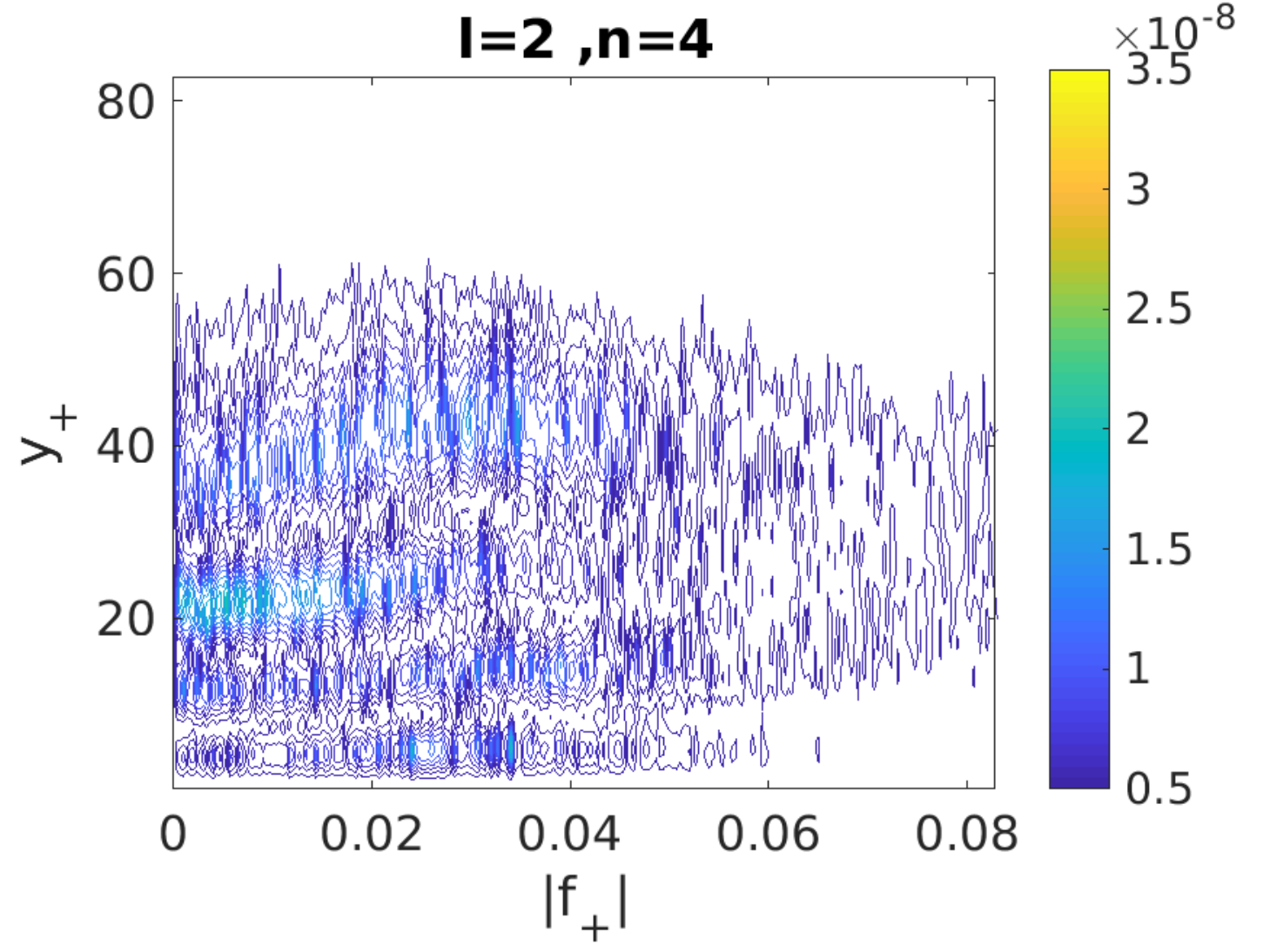}
\end{minipage}
\begin{minipage}{0.24\textwidth}
\includegraphics[width=0.95\textwidth]{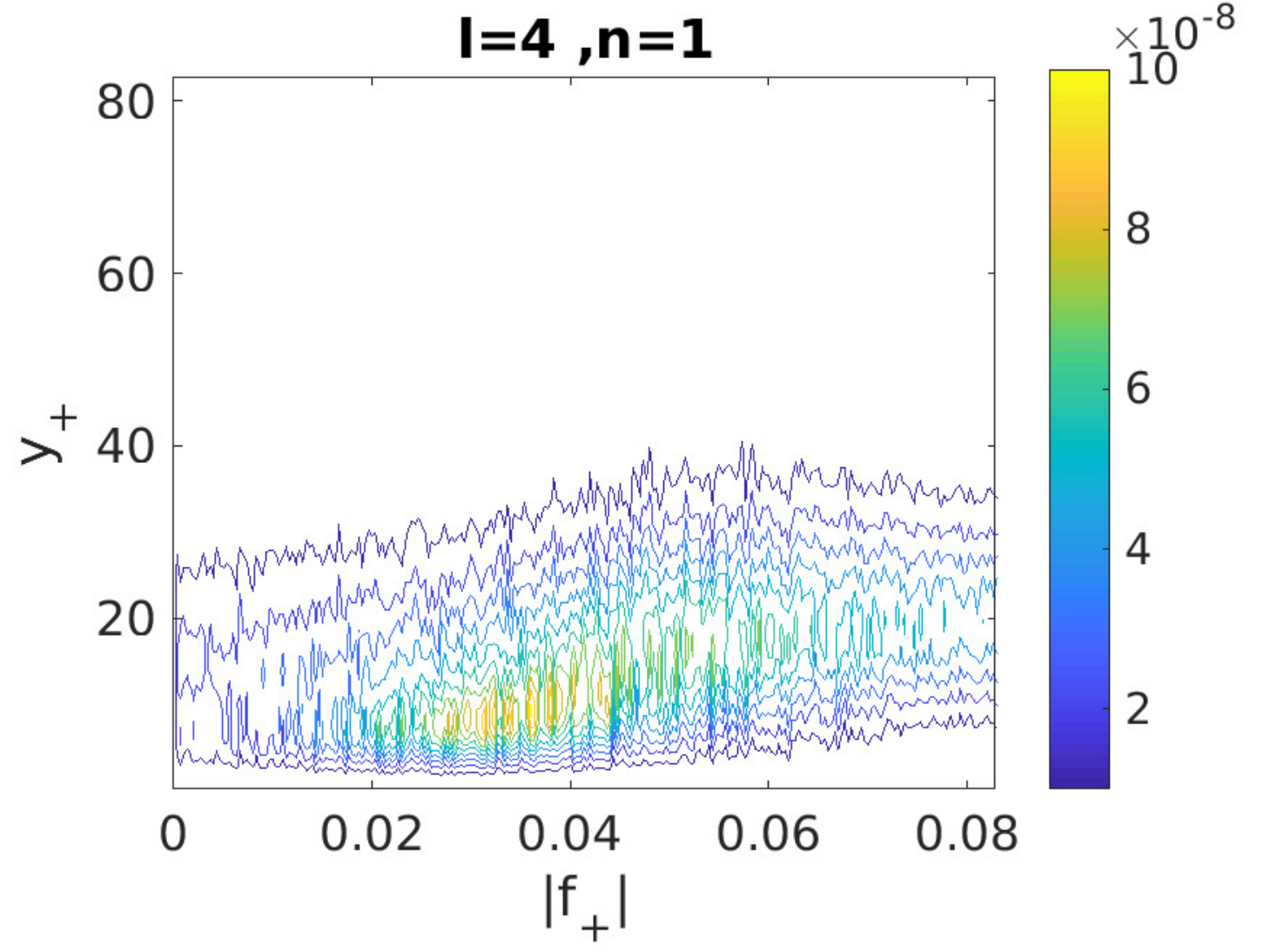}
\end{minipage}
\begin{minipage}{0.24\textwidth}
\includegraphics[width=0.95\textwidth]{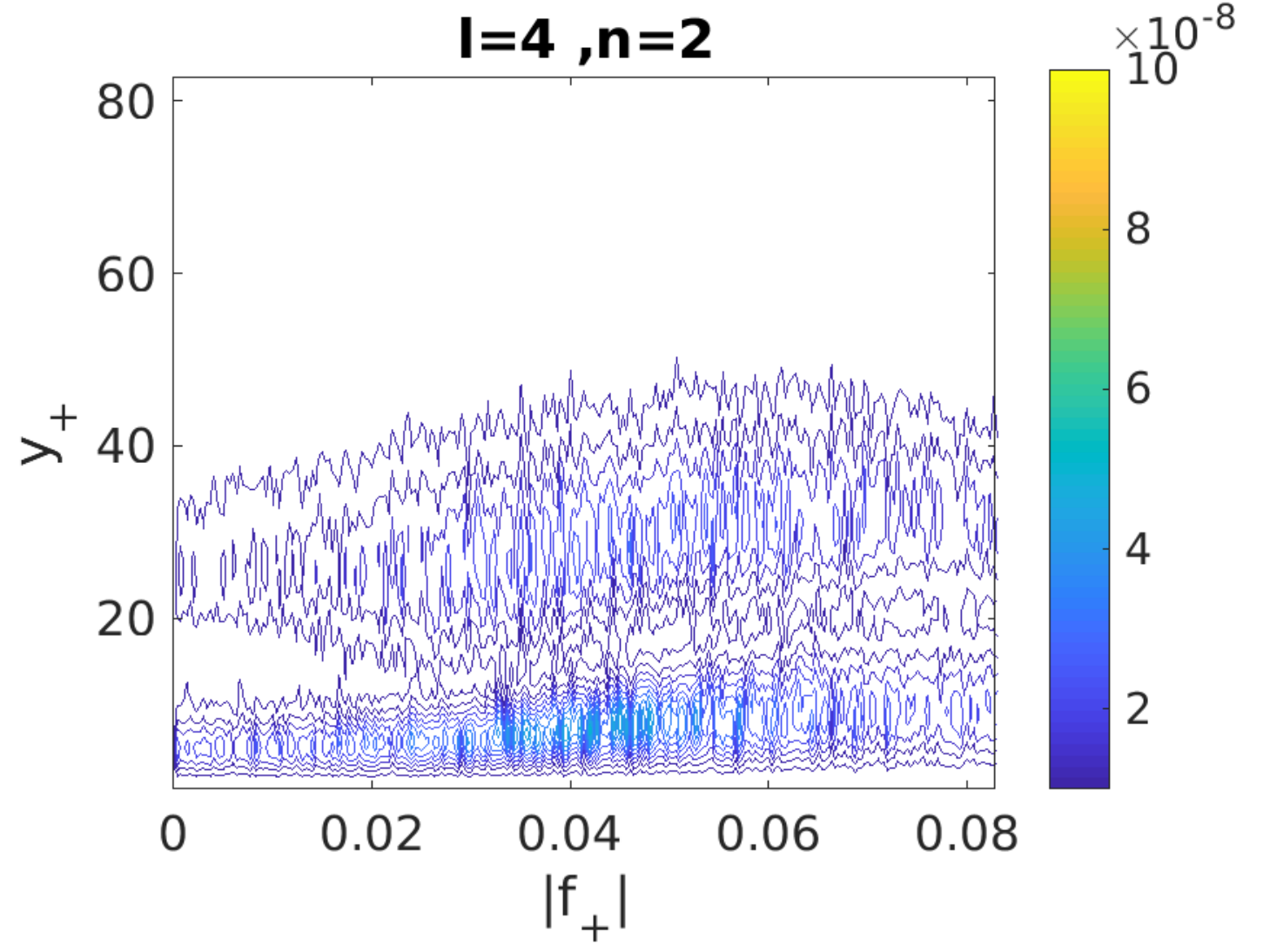}
\end{minipage}
\begin{minipage}{0.24\textwidth}
\includegraphics[width=0.95\textwidth]{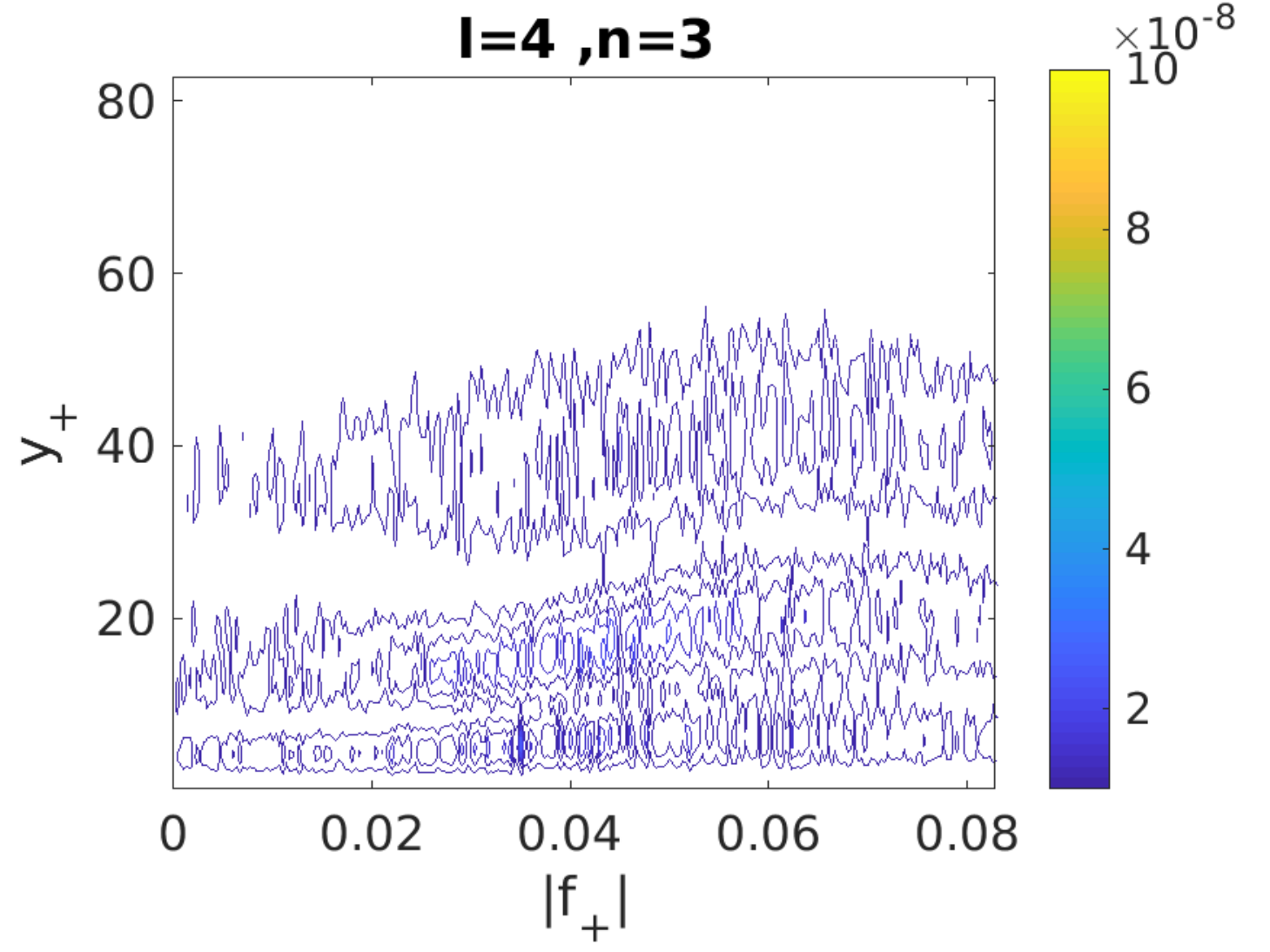}
\end{minipage}
\begin{minipage}{0.24\textwidth}
\includegraphics[width=0.95\textwidth]{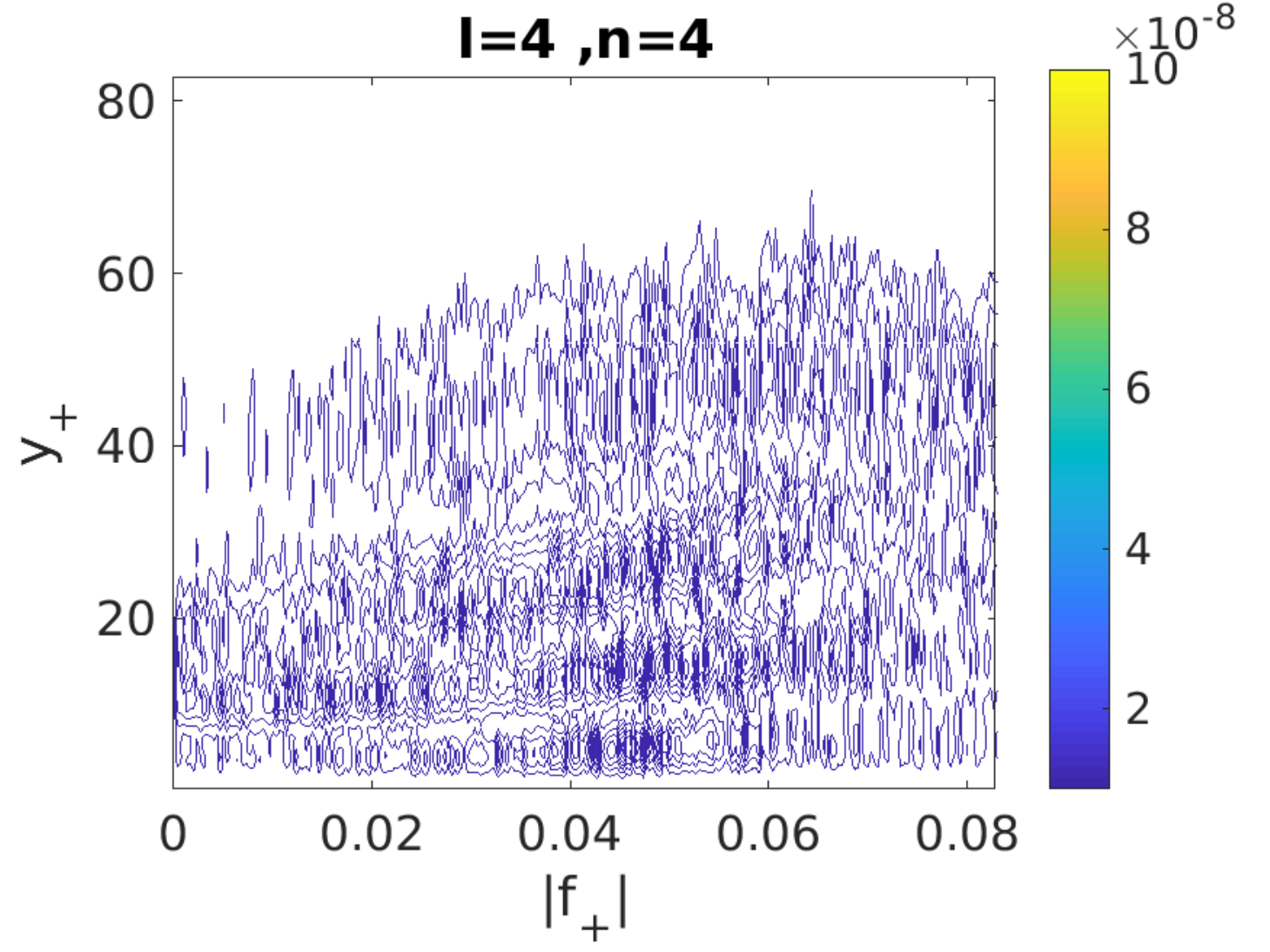}
\end{minipage}
\begin{minipage}{0.24\textwidth}
\includegraphics[width=0.95\textwidth]{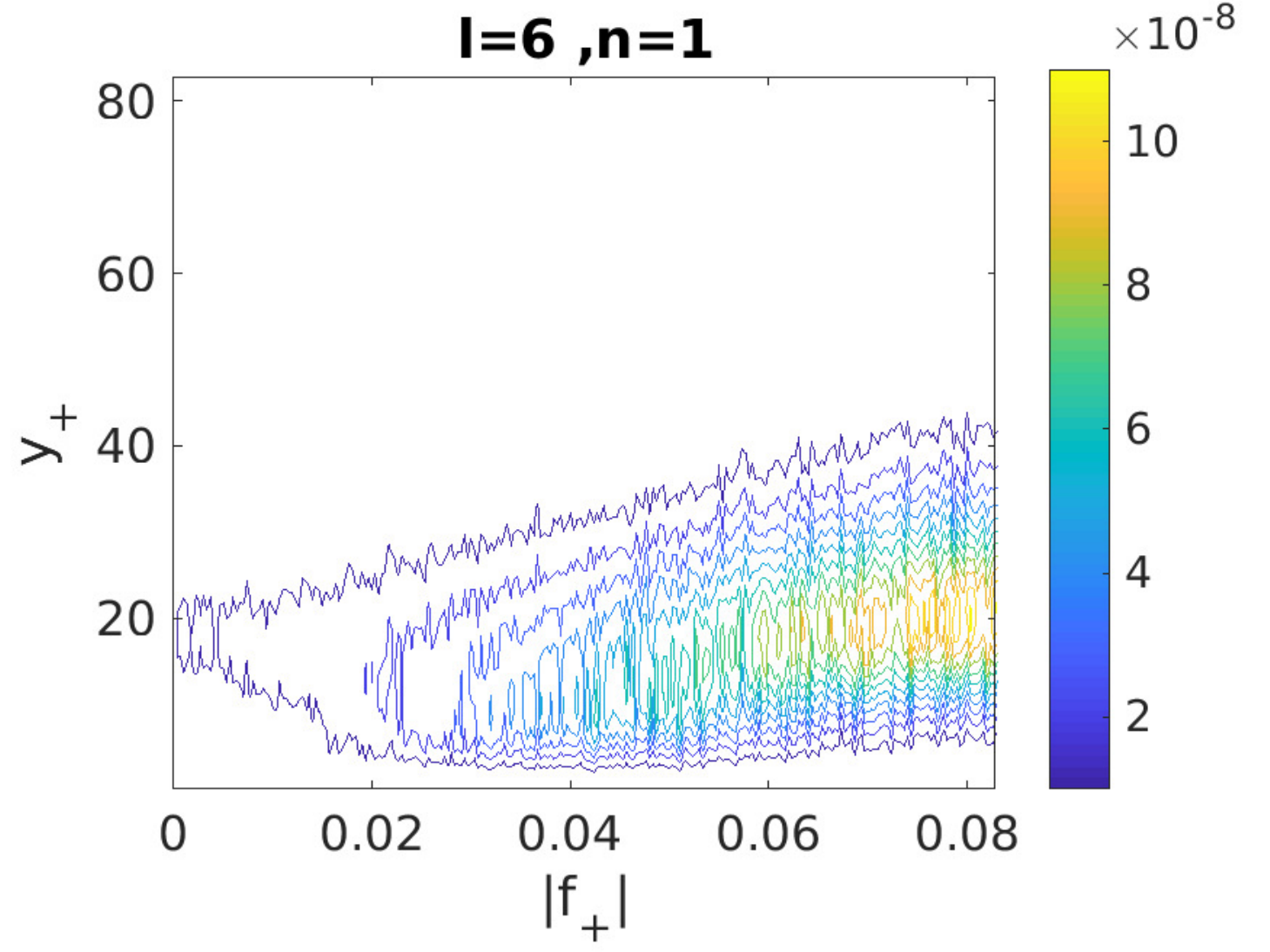}
\end{minipage}
\begin{minipage}{0.24\textwidth}
\includegraphics[width=0.95\textwidth]{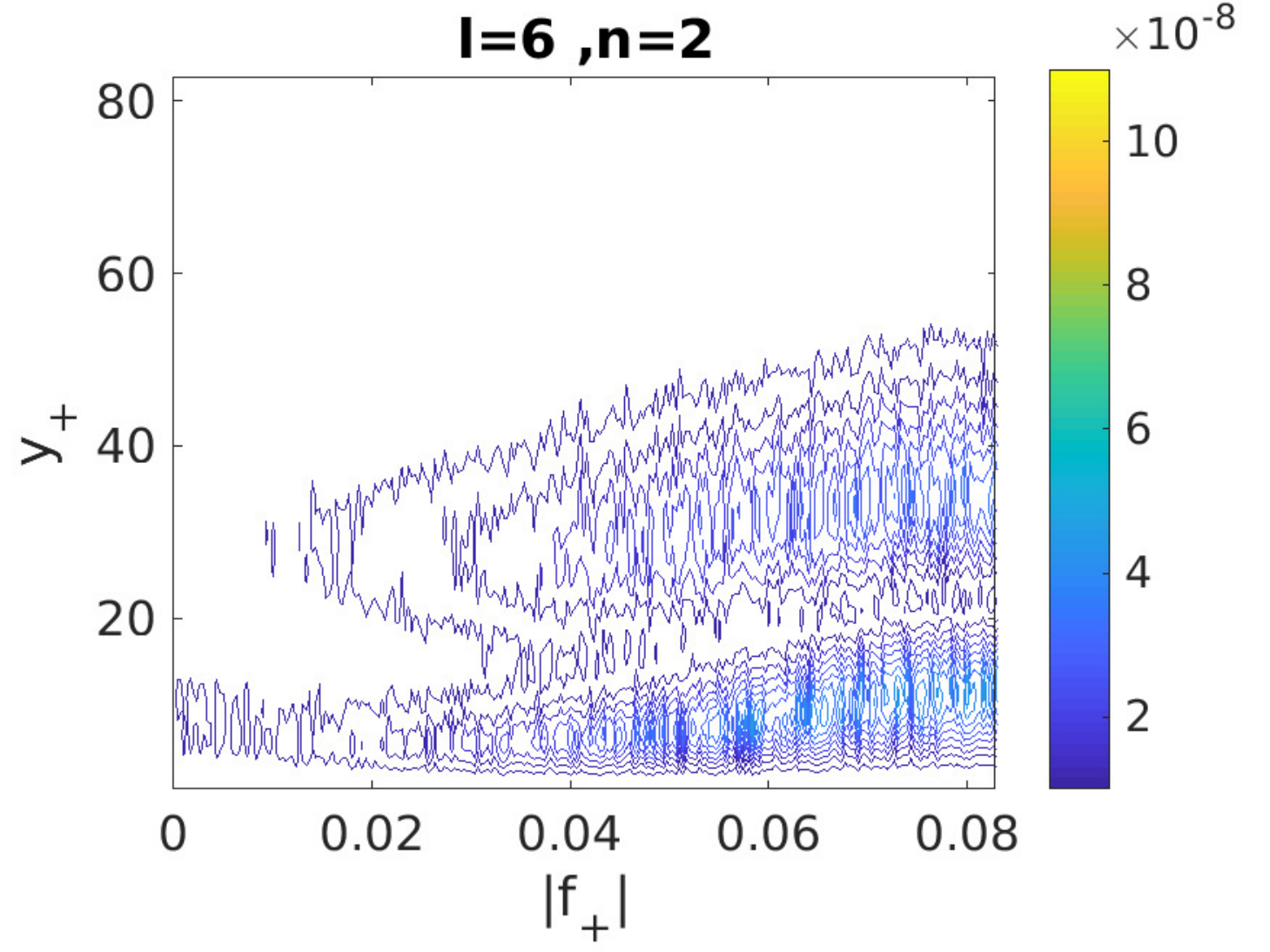}
\end{minipage}
\begin{minipage}{0.24\textwidth}
\includegraphics[width=0.95\textwidth]{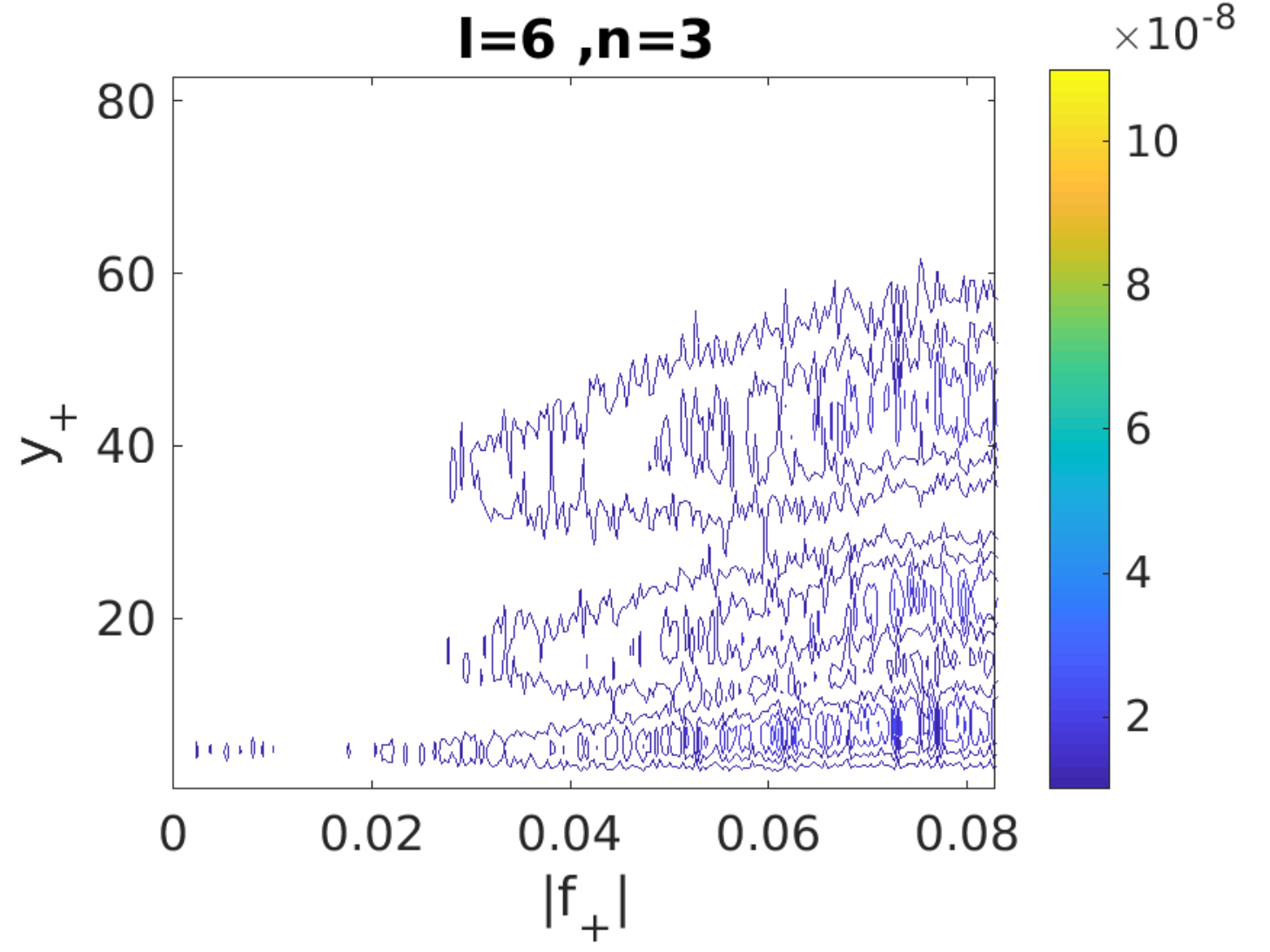}
\end{minipage}
\begin{minipage}{0.24\textwidth}
\includegraphics[width=0.95\textwidth]{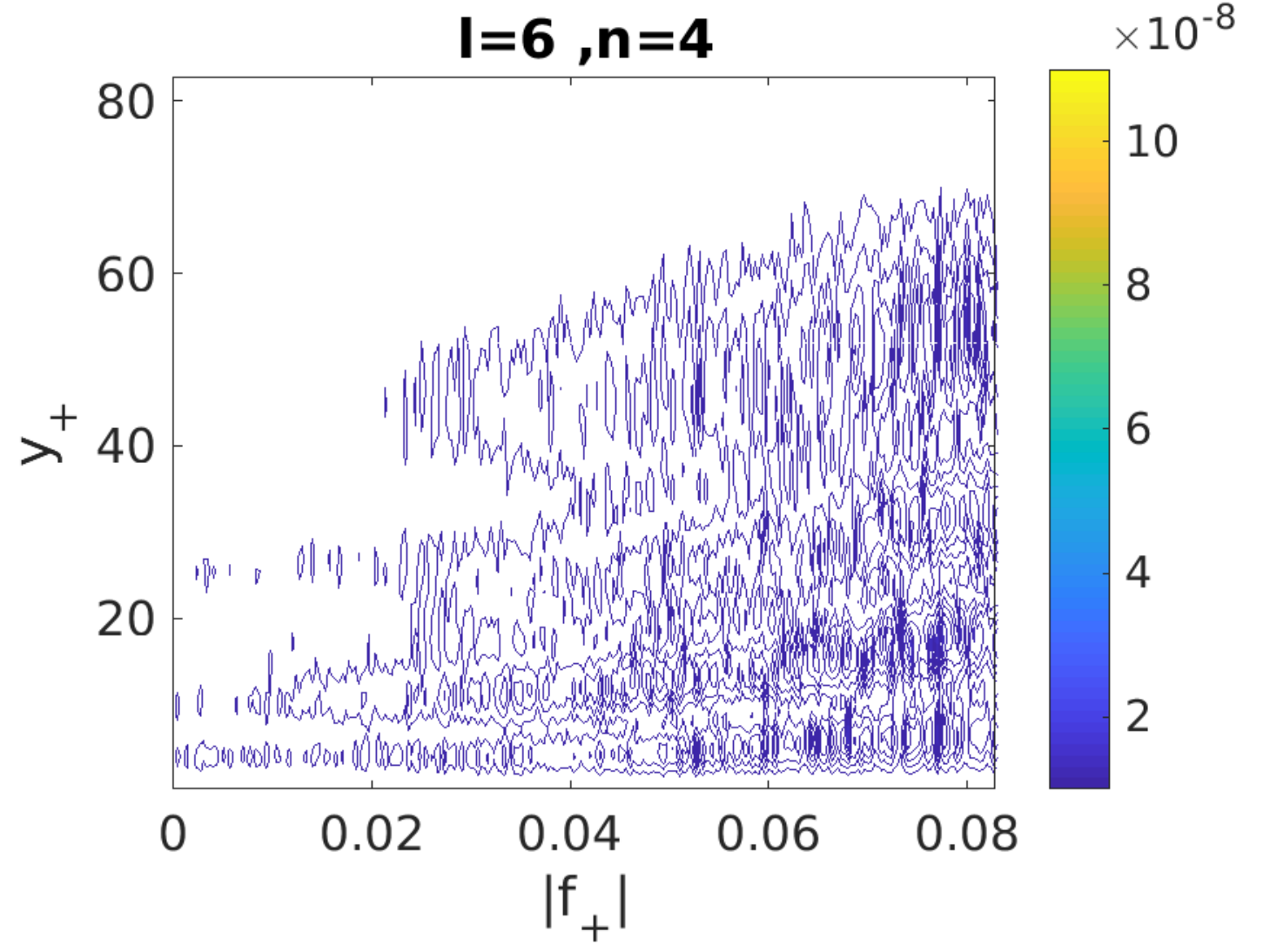}
\end{minipage}
\caption{ Structure intensity 
$I_{ln}(y,|\overline{f}|)=\sum_{k} \lambda_{lkf}^{n} |\pphi_{lkf}^{n}(y)|^{2} + \lambda_{lk-f}^{n} |\pphi_{lk-f}^{n}(y)|^{2}  $
as a function of 
the vertical direction $y$ and of the frequency $|\overline{f}|$ 
for the first quantum numbers $n \le 4$ and different streamwise wavenumbers $l$.
From left to  right: increasing quantum numbers  $n=1, 2, 3, 4$. From top to bottom: 
increasing streamwise wavenumbers $l=0, 2, 4, 6$.
}   
\label{intensity}
\end{figure}

% Add plot of mode evolution
\berengere{Figure~\ref{intensity} shows the hierarchical organization of the eigenfunctions
in various dimensions of the four-dimensional space. 
For each streamwise wavenumber and quantum mode, 
we define 
the intensity of the eigenfunctions as a function of 
height and absolute frequency (for the sake of clarity positive and negative frequencies are aggregated, although as seen in a previous
section there is no symmetry): 
$I_{ln}(y,|\overline{f}|)=\sum_{k}  (\lambda_{lkf}^{n} |\pphi_{lkf}^{n}(y)|^{2} + \lambda_{lk-f}^{n} |\pphi_{lk-f}^{n}(y)|^{2}). $
It is represented in Figure~\ref{intensity}  
for the first quantum numbers $n \le 4$ and selected streamwise wanumbers $l$.
Each quantum number $n$ is characterized by $n$ peaks at a given frequency.
The wall-normal location of the highest peak increases nearly linearly with the quantum number $n$,
and is about $10 n$ in wall units.
%an effect already noted in \cite{kn:jfe10}. 
At a given wavenumber in the horizontal space, 
energy is therefore transferred from the more energetic  to the less energetic modes 
towards the core region away from the wall. 
The range of frequencies associated with a high intensity increases only slightly with the quantum number $n$, but increases
significantly 
with the streamwise wavenumber $l$ with a shift towards higher frequencies, which is due to convection effects.
}

%{\berengere  }

%\subsection{Temporal evolution }

\begin{figure}
\begin{tabular}{c}
%\begin{subfigure}[b]{1.0\textwidth}
%\includegraphics[width=\linewidth]{Mode_shape_t=10_k=1331-eps-converted-to.pdf} \\
\includegraphics[width=\linewidth]{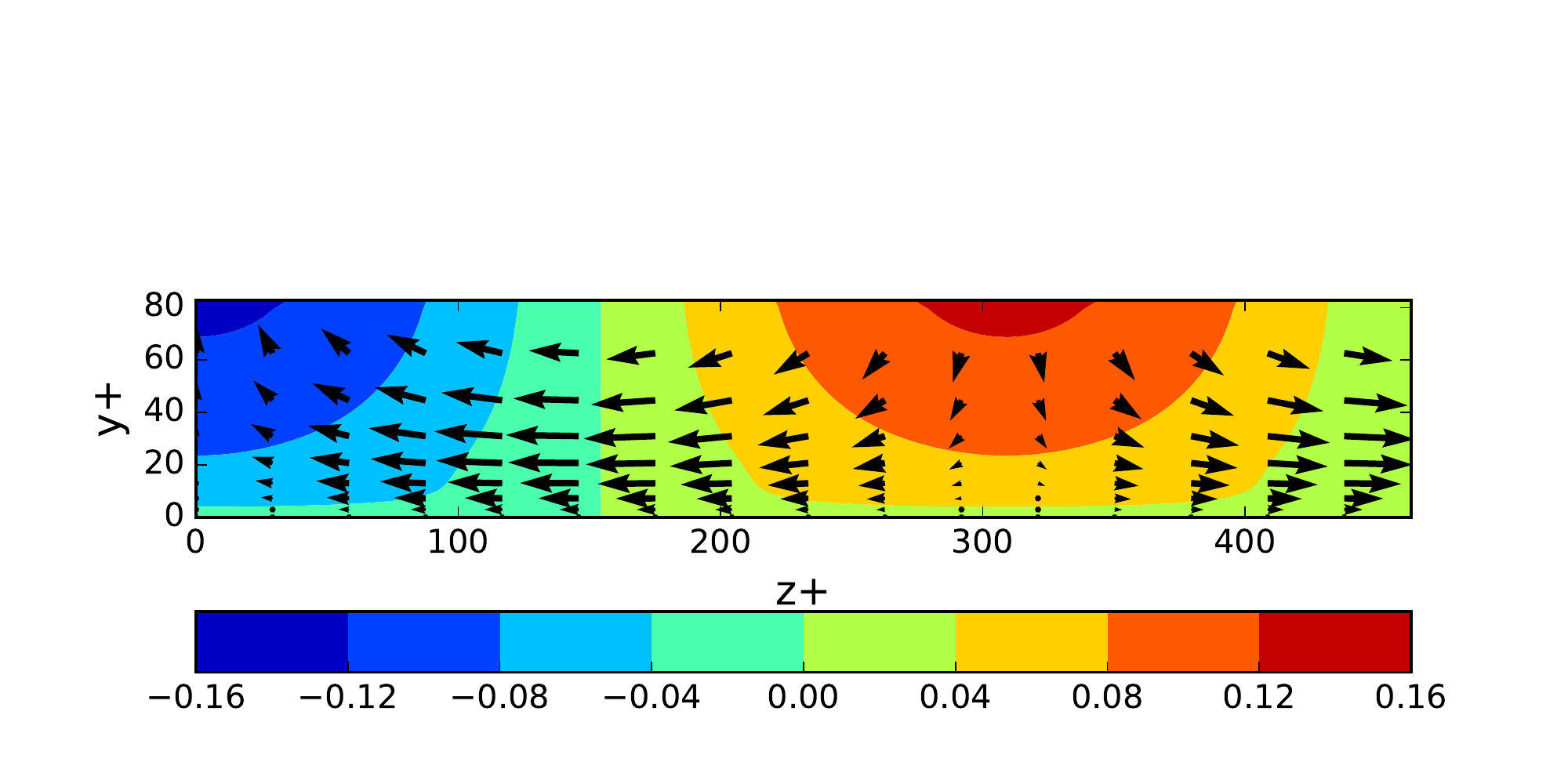}
\centerline{a) $l = 0, k = 3, f = 13$} \\
%\caption{$l = 0, k = 3, f = 13$} 
%\end{subfigure}
%\begin{subfigure}[b]{1.0\textwidth}
%\includegraphics[width=\linewidth]{Mode_shape_t=10_k=13161-eps-converted-to.pdf} \\
\includegraphics[width=\linewidth]{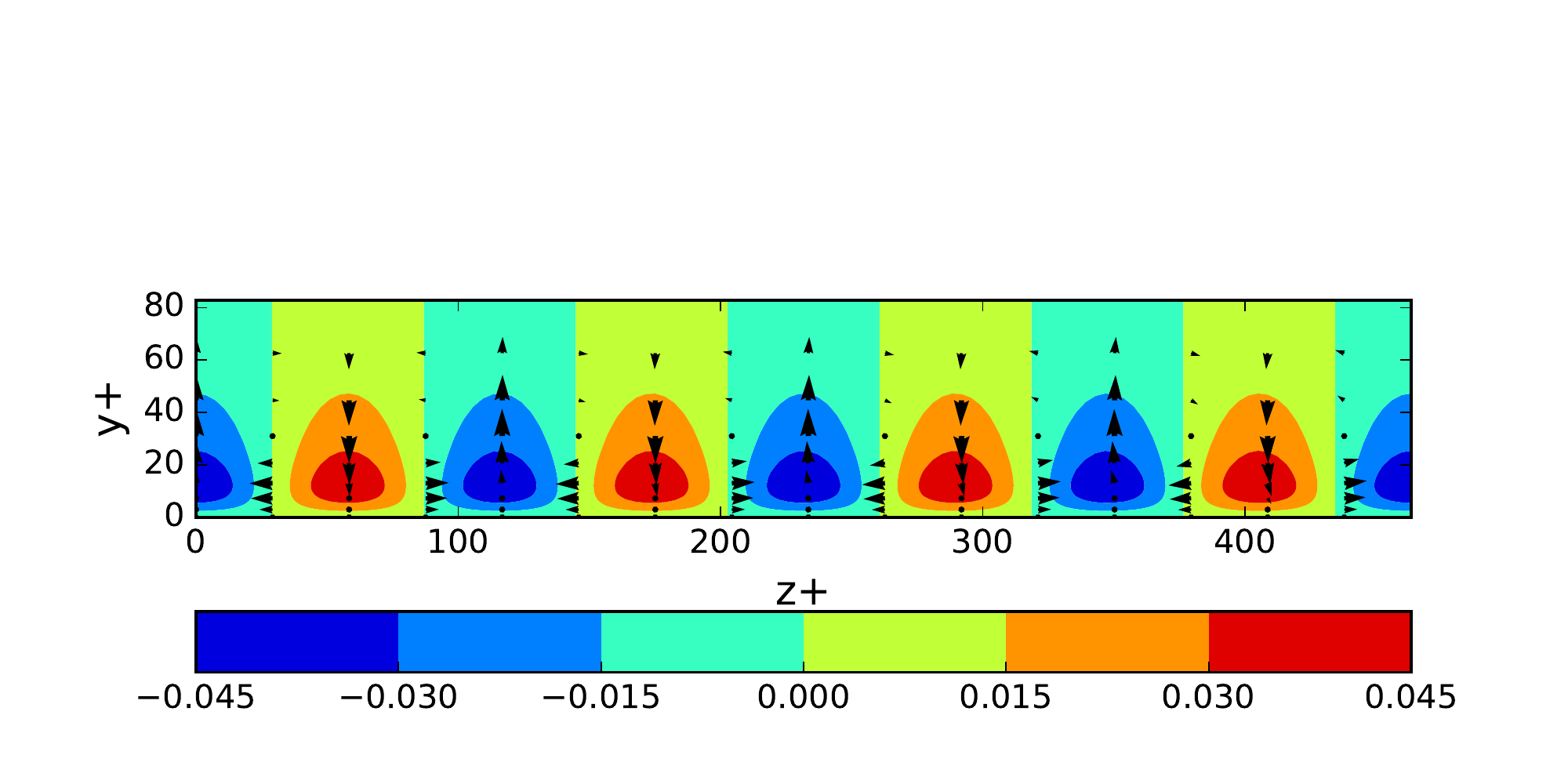}
\centerline {b) $l = 0, k = 16, f = 13$} \\
%\caption{$l = 0, k = 16, f = 13$} 
%\end{subfigure}
\end{tabular}
\caption{Contour plot of streamwise velocity component  
reconstructed from eigenfunction 
in $(z,y)$ plane along with a vector plot for $(w,v)$ velocity components for the non-trivial dominant eigenmode corresponding to a) $l = 0$, $k = 3$, $f = 13$, $n=1$ and b) $l = 0$, $k = 16$, $f = 13$, $n=1$.}
\label{mode_evol}
\end{figure}

\begin{figure}
\centering
\includegraphics[width=0.7\textwidth]{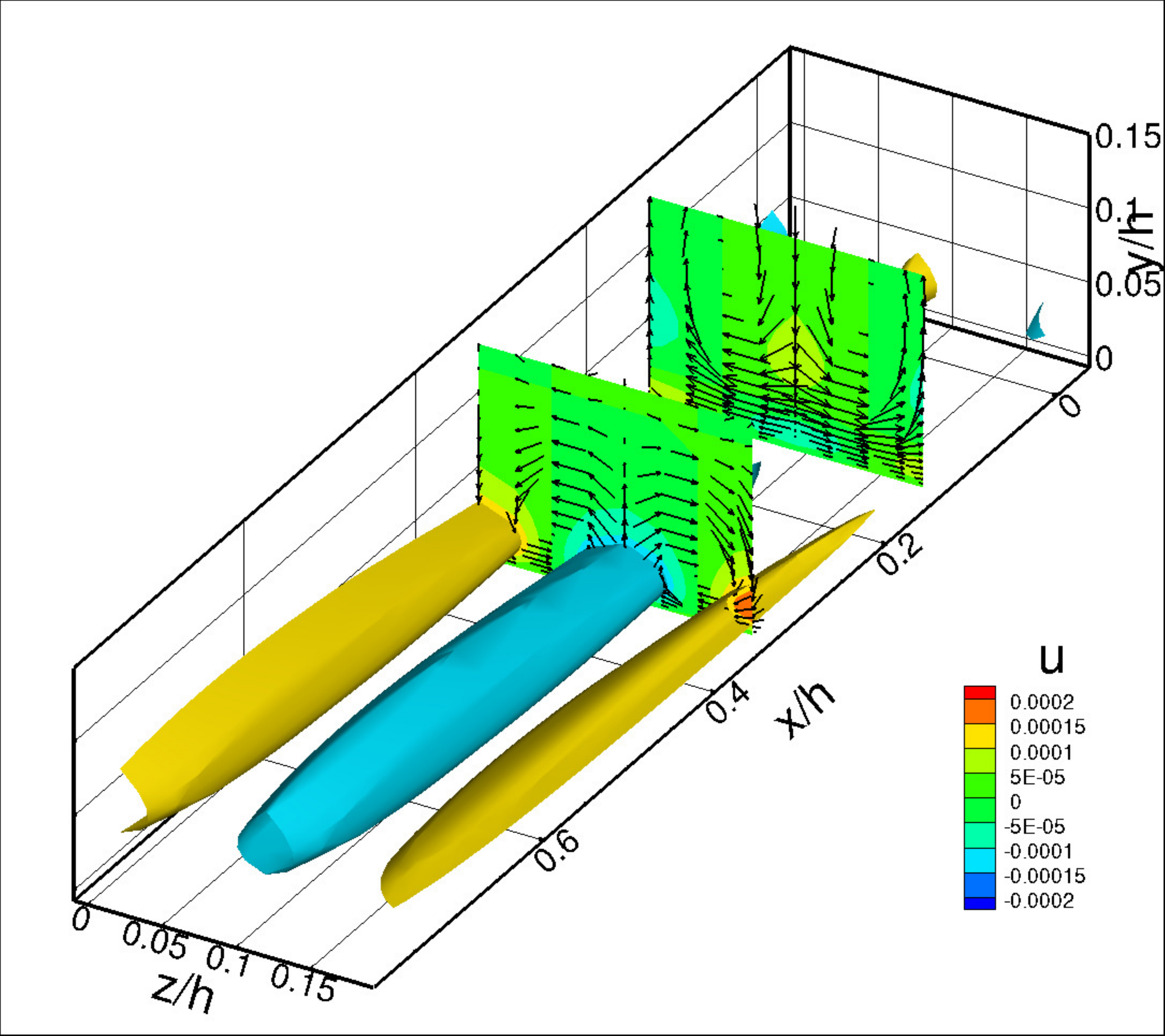}
\caption{Reconstruction of the most energetic mode corresponding to the
peak in the frequency spectrum in figure ~\ref{fig:cumul_f} for 
$l = 1$, $k = 16$, $f = -40$, $n=1$. Isosurfaces  of streamwise 
component in $(x,y,z)$ are shown along with two slice views at $x/h=0.15$
and $x/h=0.35$ showing a contour plot of the streamwise component and a vector
plot of the spanwise and wall-normal components. \berengeredeux{Units are arbitrary.}} 
\label{mode_evol3D}
\end{figure}

%In all that follows, results come from the D4 set.
Figure~\ref{mode_evol} shows the flow generated by selected modes 
in figure ~\ref{velcomp_freq} 
associated with $\bar{f}_{c+}$ for two different values of $\bar{k}_+$. As these dominant modes 
have no streamwise variation (i.e. $l = 0$), the three-dimensional shapes of the modes are represented by a contour plot of the streamwise velocity component ($u$) and a velocity plot of the in-plane quantities $(v,w)$. 
The modes are characterized by updrafts of low-speed fluid alternating with downdrafts 
of high-speed fluid, associated with vortical motions in the streamwise direction, 
in agreement with classic observations \citep{kn:corinobrod}.
 The wall-normal extension of the structure decreases as the spanwise wavenumber
increases, in agreement with Townsend's model \citep{kn:town} of wall-attached eddies 
(note the similarity of Figure~\ref{mode_evol} with Figure 9 in \cite{jimenez2013}).
Figure~\ref{mode_evol3D} shows that 
the most energetic streamwise mode convected at a velocity 
of $12 u_{\tau}$ also corresponds to 
 high and low-speed streaks alternating
in the streamwise as well as in the spanwise direction and
associated with vortical motions in the cross-stream plane. 
The wall-normal position of the vortex centers depends on
the vertical extension of the streaks.

% CHANGE THIS FIGURE

\begin{figure}
\begin{subfigure}[b]{0.5\textwidth}
\includegraphics[width=\linewidth]{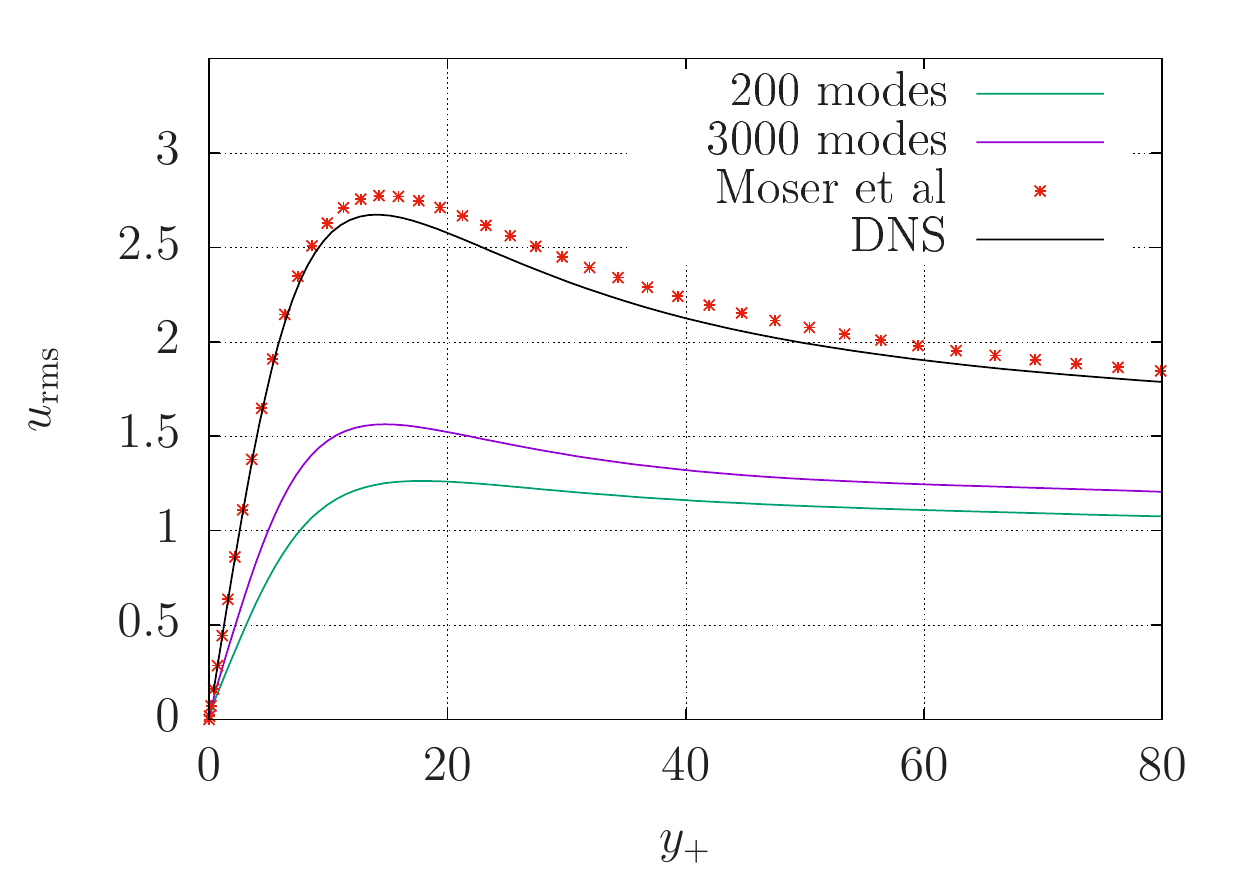}
\caption{$u_{\mathrm{rms}}$}
\end{subfigure}
\begin{subfigure}[b]{0.5\textwidth}
\includegraphics[width=\linewidth]{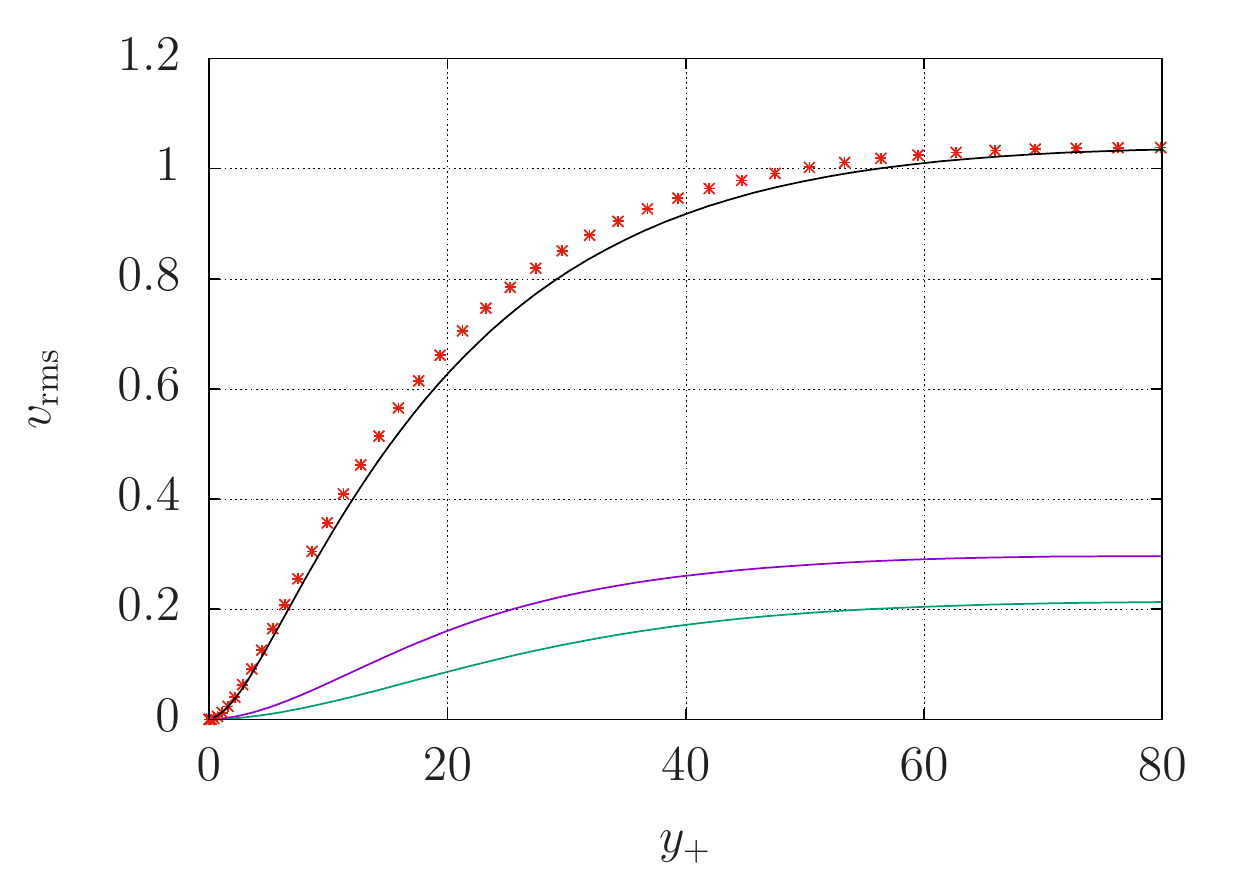}
\caption{$v_{\mathrm{rms}}$}
\end{subfigure}
\begin{subfigure}[b]{0.5\textwidth}
\includegraphics[width=\linewidth]{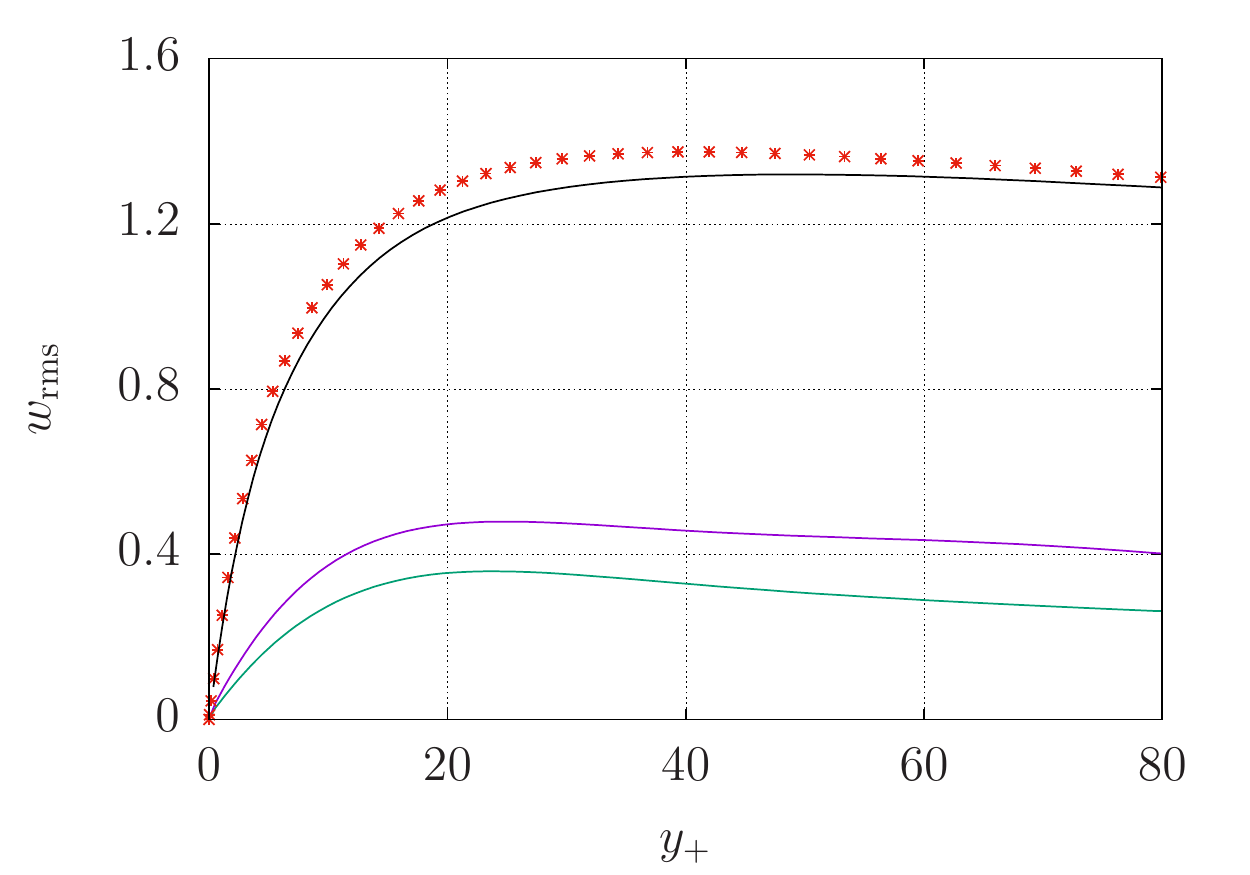}
\caption{$w_{\mathrm{rms}}$}
\end{subfigure}
\begin{subfigure}[b]{0.5\textwidth}
\includegraphics[width=\linewidth]{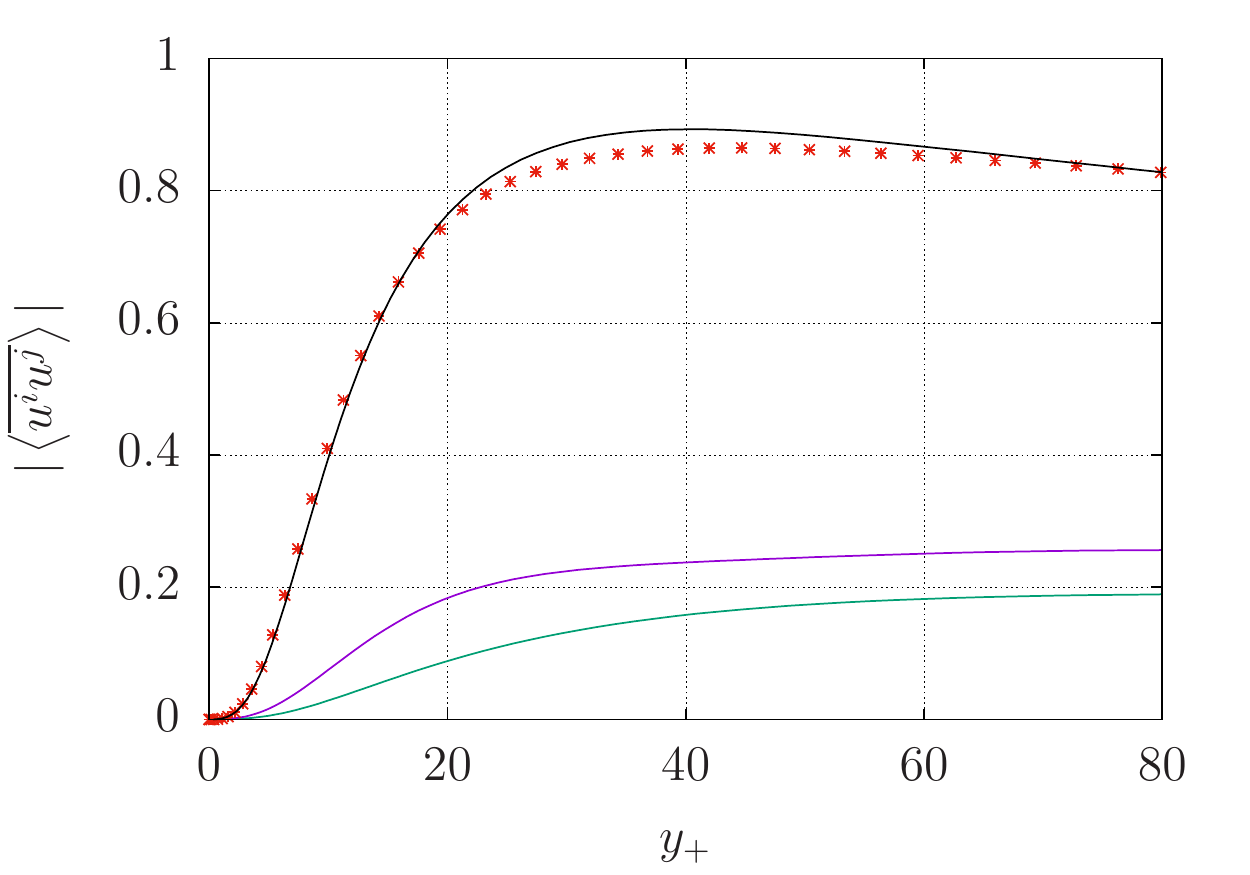}
\caption{$\left|\left<\overline{u^i u^j}\right> \right|$}
\end{subfigure}
\caption{ Reconstruction of the turbulent intensities and Reynolds stress using 200 and 3000 modes. Comparison
with \citet{moser1999}'s and our DNS data is also shown. The legend is indicated in figure 14 a). } 
\label{reynolds_stress}
\end{figure}

The contribution of the $M$ most energetic modes to the fluctuating stresses can be evaluated using
\begin{equation}
\left<\overline{u^i u^j}\right>_{M} = \sum_{N(l,k,f,n) \le M} \lambda_{lkf}^{n} \phi_{lkf}^{n,*i}\phi_{lkf}^{m,j},
\end{equation}
%where the summation is over 200 modes. 
where the subscript `$M$' is used to distinguish it as a reconstructed quantity using $M$ most energetic modes. The bar indicates average over space and time and $\left<.\right>$ denotes the usual ensemble operator. Note that the mean modes corresponding to $(0,0,0,n)$ have been excluded from the reconstruction as the stresses are associated with fluctuating components.
% the $(0,0,0,n)$ modes have been removed.
Figures~\ref{reynolds_stress}a, b and c show the three components of the root-mean square velocity (rms) and Figure~\ref{reynolds_stress}d shows the Reynolds stress as a function of $y_+$ reconstructed using 200 and then 3000 modes. The plots are compared with the present simulation results and the DNS results of \citet{moser1999}. 
 A non-negligible fraction
of the turbulent intensity is recovered with the most energetic 200 modes:
60\% for the streamwise velocity, 30\% for the cross-stream components, and about 25\% of the Reynolds stresses.
The gain obtained using 15 times as many modes (3000) is relatively small (40\% increase for the spanwise component and 30\% for
the wall-normal component), which highlights the complexity of the flow.

% Add a table of energy budget

\section{\berengeredeux{Energy budget} of modes}

\berengere{
It is of interest to determine the contributions made by the most energetic POD modes to the momentum equations.
In pioneering studies, \citet{kn:hongrubesin85} and \citet{kn:gatskiglauser92}
examined the contributions of the dominant POD mode. Hong and Rubesin showed that the dominant fluctuating mode
captured several essential characteristics of the fully turbulent flow. Gatski and Glauser considered
the various contributions of the first POD eigenmode to the different terms of the momentum equation and showed that a turbulent model for the transport term could be built from one-mode estimates.}

With the new variant of Proper Orthogonal Decomposition, projection of the Navier-Stokes equations onto the basis of eigenfunctions no longer yields a dynamical system for the temporal amplitudes of the spatial eigenfunctions, but makes it possible to evaluate the contributions made by the different terms of the momentum equations for each mode $N(l,k,f,n)$.

%into the Navier-Stokes equation
%\begin{equation}
%\dtuidt + \tilde{u}_{j} \ddx{\tui}{i} = - \ddx{p}{i} + \nu \deltau{u_{i}} 
%\end{equation}

%To bring about the influence of the mean flow, 
For consistency with standard analysis, we decompose the $p$th component of the velocity field 
into its mean part $U^{p}$ which, as we have seen above, can be
represented by the mode $(0,0,0,1)$ and a fluctuating part $u^{p}$.

We rewrite the equations as
\begin{equation}
\mathrm{E}_{\mathrm{p}} = -\dupdt - u^{j} \ddx{U^{p}}{j} - U^{j} \ddx{u^p}{j} - u^{j}\ddx{u^p}{j} - \ddx{p}{p} + \nu \deltau{u^{p}} = 0, \label{energy_budget}
\end{equation}
where $U^{1}=U$ (mean streamwise velocity) and $U^{j}=0$ if $j = 2,3$ and $\ddx{U^p}{j} \neq 0$ only if $p=1$ and $j=2$.

To obtain a budget for each mode $(l,k,f,n)$ we project Equation~\eqref{energy_budget} onto the corresponding mode, i.e. we take the Fourier transform in the $(l,k,f)$ space, take the inner product in the wall-normal direction, and apply an 
ensemble average. This gives
\begin{equation}
 \int_{0}^{L_{x}} \int_{0}^{L_{z}} \int_{0}^{T} 
\int_{0}^{Y} \left<\mathrm{E}_{\mathrm{p}} a_{lkf}^{*n}\right> \phi_{lkf}^{*n,p}(y) \expmbig \mathrm{d}y \mathrm{d}t \mathrm{d}z \mathrm{d}x,
\end{equation}
where $*$ denotes complex conjugation.

\begin{itemize}
\item The first term $-\dupdt$ yields a contribution ${-I}_{lkf}^{n}= - \mathrm{i} \overline{f} \lambda_{lkf}^{n}$. 
We note that this contribution is purely imaginary.
%The sum of all the other terms should therefore have a zero real part.

\item The second and third terms represent the interaction of the mode with the mean velocity profile. The second term 
 is associated with the classic production term and is expected to represent a source of energy for the fluctuations: 
\begin{equation}
P_{lkf}^{n} = -\int_0^Y \left< u^2 \dUdy a_{lkf}^{*n} \phi_{lkf}^{*n,1} \right> \mathrm{d}y = 
-\lambda_{lkf}^{n} \int_0^Y \dUdy \phi_{lkf}^{n,2} \phi_{lkf}^{*n,1} \mathrm{d}y. 
\end{equation}
The third term represents the convection effect of the mean field: 
\begin{equation*}
C_{lkf}^{n}= -\int_0^Y \left< U \ddx{u^p}{1} a_{lkf}^{*n} \phi_{lkf}^{*n,p} \right> \mathrm{d}y =
- \int_0^Y \sum_m U \ddx{\phi_{lkf}^{m,p}}{1} \phi_{lkf}^{*n,p} \left< a_{lkf}^{m} a_{lkf}^{*n} \right> \mathrm{d}y 
\end{equation*}
\begin{equation}
= - \mathrm{i} \hspace{0.01in} \bar{l} \lambda_{lkf}^{n}
\int_0^Y U \phi_{lkf}^{n,p} \phi_{lkf}^{*n,p} \mathrm{d}y.
\end{equation}

\item The last term on the left-hand-side of the equation corresponds to the 
viscous diffusion term $D_{lkf}^{n}$ defined as: 
\begin{equation}
 D_{lkf}^{n}= \lambda_{lkf}^{n} \int_{0}^{Y} \deltau{\phi_{lkf}^{n,p}} \phi_{lkf}^{*n,p} \mathrm{d}y 
= \lambda_{lkf}^{n}\left[ (-\bar{l}^{2} - \bar{k}^{2}) 
 \int_{0}^{Y} \phi_{lkf}^{n,p}  \phi_{lkf}^{*n,p} \mathrm{d}y
+ \int_{0}^{Y} \frac{d^{2} \phi_{lkf}^{n,p}}{dy^{2}} \phi_{lkf}^{*n,p} \mathrm{d}y \right].
\end{equation}
Integrating the last term by parts, one has
\begin{equation}
 D_{lkf}^{n} = (-\bar{l}^{2} - \bar{k}^{2}) 
- \int_0^{Y} \frac{d \phi_{lkf}^{n,p}}{\mathrm{d}y} \frac{d \phi_{lkf}^{*n,p}}{dy} \mathrm{d}y +\frac{d \phi_{lkf}^{n,i}(Y)}{\mathrm{d}y} \phi_{lkf}^{*n,i}(Y). 
\end{equation}
Except for the last term corresponding to a boundary effect, all 
contributions to $D_{lkf}^{n}$ are real and negative, which corresponds to an energy loss
for the mode, as expected.

Note that $I_{lkf}^{n}, P_{lkf}^{n}, C_{lkf}^{n}$ and $D_{lkf}^{n}$ only require 
 information about the mode $(l,k,f,n)$ (one can think of them as 
purely diagonal operators) and can be directly evaluated from the Proper Orthogonal Decomposition,
while the other two contributions require additional information about the coefficients $a_{lkf}^{n}$.

\item As in the classic derivation (\cite{kn:aubry88}) , the contribution from the pressure 
 term represents the influence of the pressure at the 
upper boundary of the wall layer and can be expressed as 
\[ Pr_{lkf}^{n} = \left<p_{lkf}(Y) a_{lkf}^{*n}\right> \phi_{lkf}^{*n,2}(Y), \]
where $p_{lkf}(Y)$ is the Fourier transform (along $x$, $z$ and $t$) of 
pressure at height $Y$, which needs to be evaluated from 
 the DNS. This term is proportional to the wall-normal intensity of the structure at the top of the layer 
and depends on the velocity-pressure correlation at the top of the layer. It represents an 
external forcing term which corresponds to the interaction of the wall layer with the
outer region \citep{kn:aubry88}.

\item Finally, the quadratic interactions can be evaluated as 
\begin{equation}
 Q_{lkf}^{n}= -\sum_{l'} \sum_{k'} \sum_{f'} \sum_{m} \sum_{p}
\left<a_{l'k'f'}^{m}a_{l-l'k-k'f-f'}^{p}a_{lkf}^{*n}\right> 
\int_0^Y \ddx{\phi_{l'k'f'}^{m,i}}{j} \phi_{l-l' k-k' f-f'}^{p,j} \phi_{lkf}^{*n,i} \mathrm{d}y.
\end{equation}
 $Q_{lkf}^{n}$ requires information about the triple correlations 
of the coefficients $a_{lkf}^{n}$,
and involves first derivatives of the eigenfunctions.
This term characterizes the energy transfer from the different modes to the mode
$(l,k,f,n)$. It can be seen as 
a non-isotropic extension of the energy transfer function
defined in isotropic turbulence \citep{kn:zhou93}, and involves triads of
modes in the $(l,k,f)$ space. 
 It also corresponds to the forcing term of the resolvent analysis 
\citep{kn:mckeon2017}.

 If all other terms are known, the quadratic interaction terms $Q_{lkf}^{n}$ 
can be determined from the budget equation:
\begin{equation}
 I_{lkf}^{n} = P_{lkf}^{n} + C_{lkf}^{n} + D_{lkf}^{n} + Pr_{lkf}^{n} + Q_{lkf}^{n}.
\label{budget}
\end{equation}
The real part of the equation represents a balance between 
 the production $P_{lkf}^{n}$ and 
dissipation term $D_{lkf}^{n}$, which depend only on the characteristics of 
the mode $(l,k,f,n)$, and the interaction terms due to pressure $Pr_{lkf}^{n}$ and convection by 
velocity fluctuations $Q_{lkf}^{n}$, which characterize
how the mode interacts with the full flow.
The imaginary part of the equation can be seen as 
a phase dispersion relation linking the frequency $f$ of 
the mode ($I_{lkf}^{n}$) with the different physical mechanisms.

\end{itemize}

\begin{figure}
\begin{subfigure}[b]{0.5\textwidth}
\includegraphics[width=\linewidth]{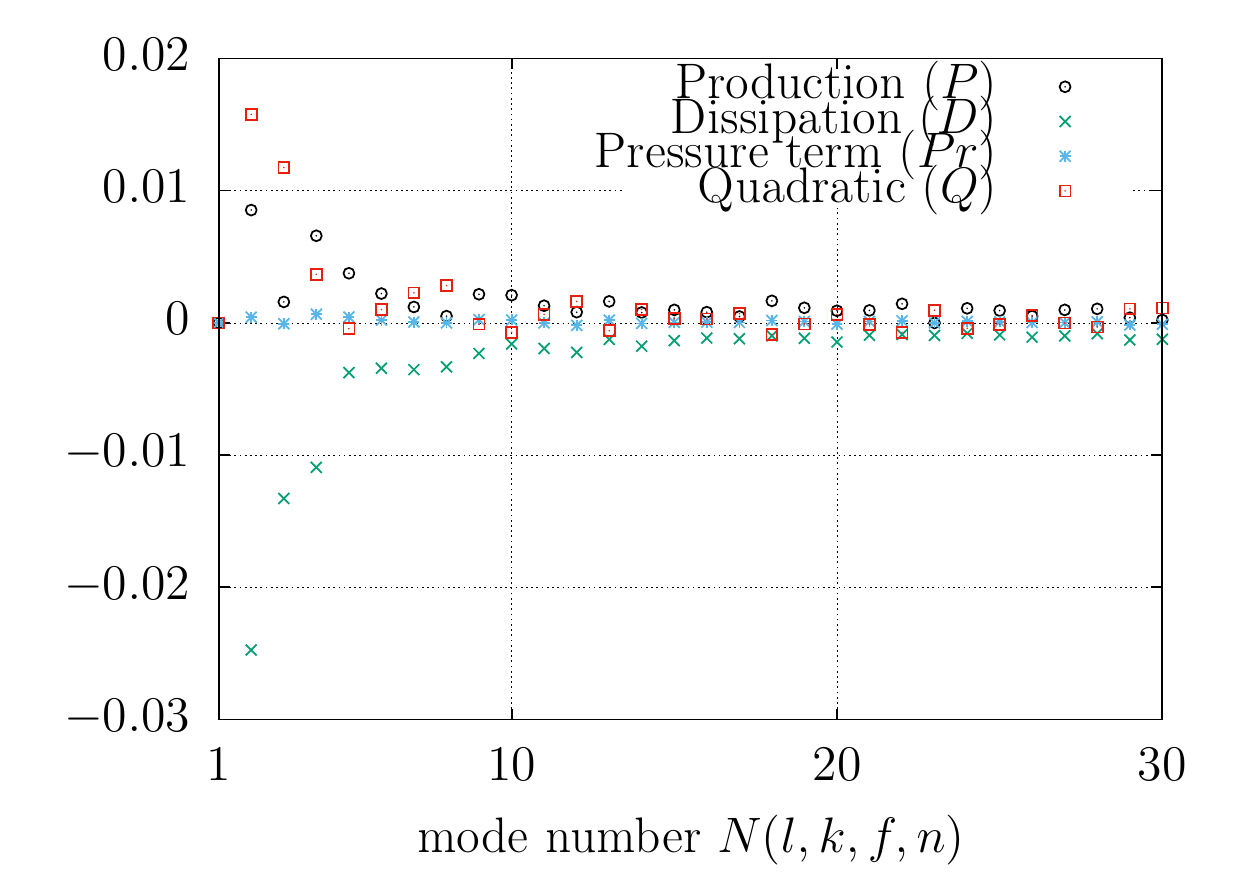}
\caption{Real part, modes 1-30 }
\end{subfigure}
\begin{subfigure}[b]{0.5\textwidth}
\includegraphics[width=\linewidth]{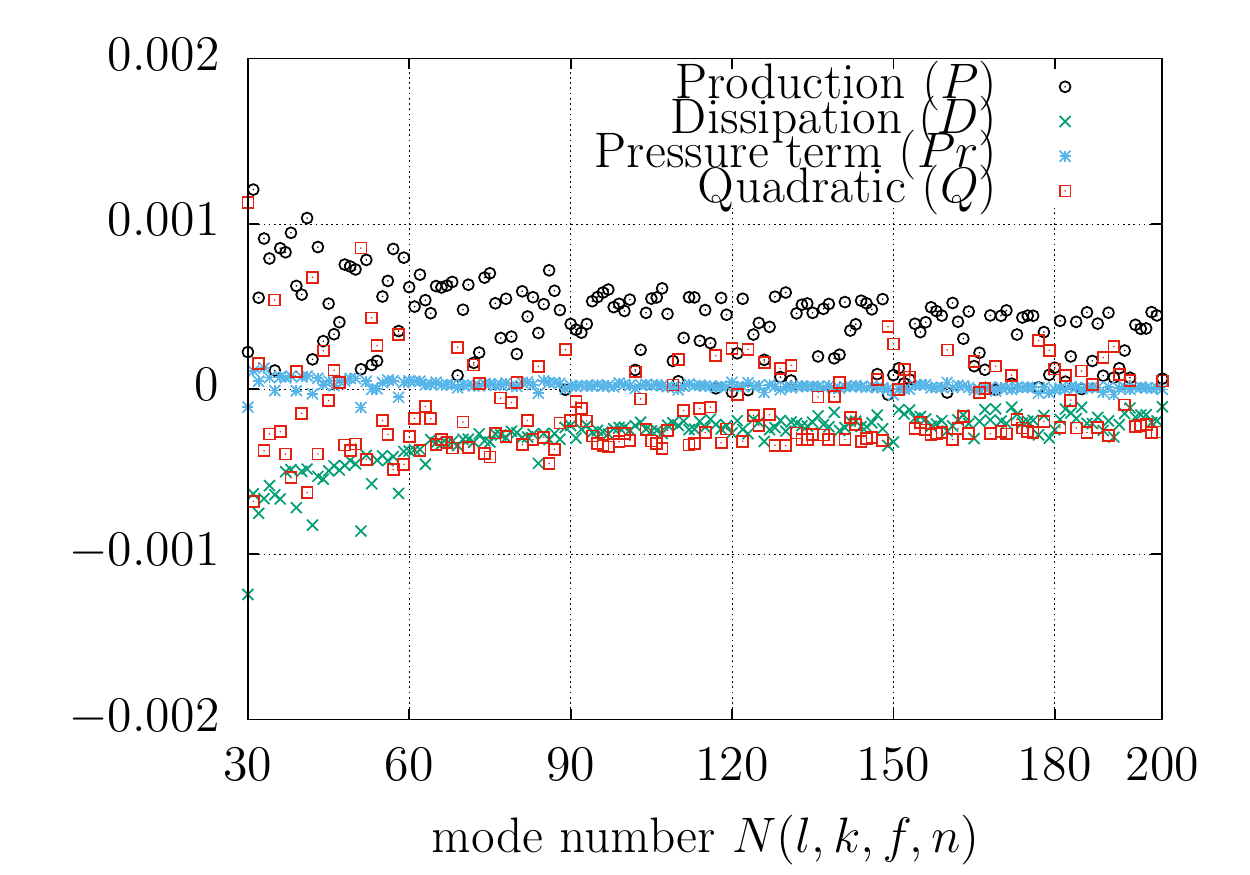}
\caption{Real part, modes 30-200}
\end{subfigure}
\begin{subfigure}[b]{0.5\textwidth}
\includegraphics[width=\linewidth]{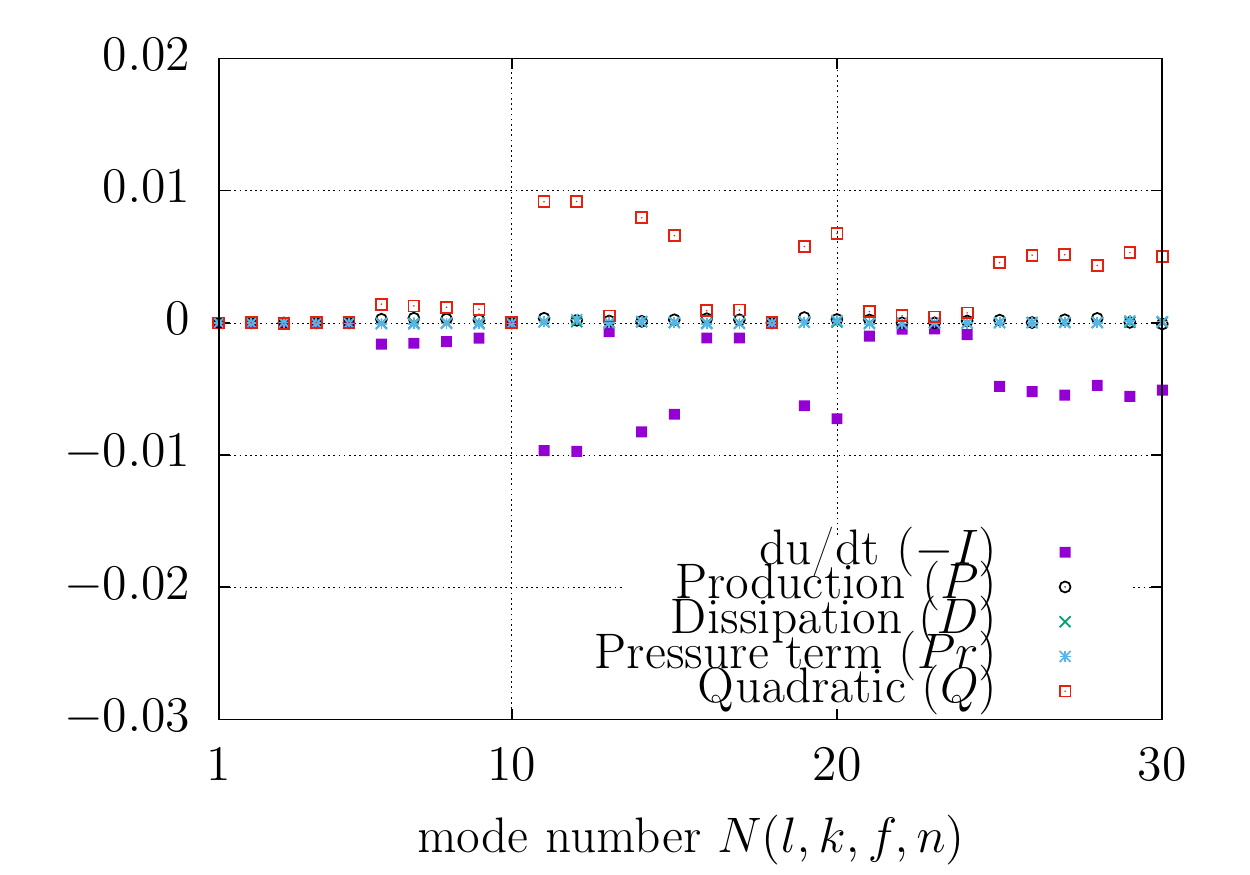}
\caption{Imaginary part, modes 1-30}
\end{subfigure}
\begin{subfigure}[b]{0.5\textwidth}
\includegraphics[width=\linewidth]{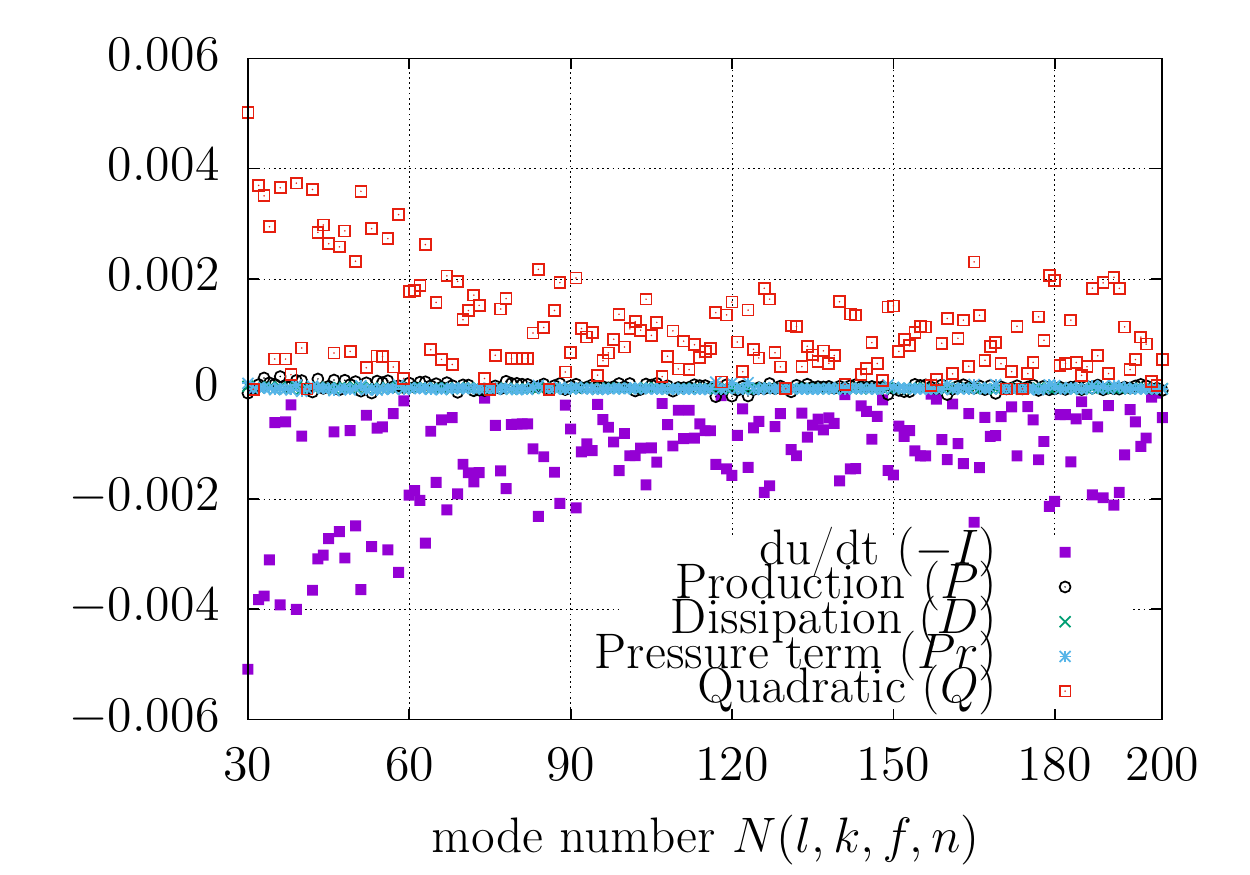}
\caption{Imaginary part, modes 30-200}
\end{subfigure}
\caption{Energy budget for the first 200 most energetic spatially 
fluctuating modes.}
\label{Lin_terms}
\end{figure}

Figure~\ref{Lin_terms} represents the contributions of the different terms
to the equations for the largest $200$ modes 
(the contributions of the modes corresponding to the mean flow were set to zero). 
These modes are all characterized by a zero streamwise 
wavenumber ($l=0$) and a quantum number $n=1$.
All terms are evaluated directly, except the
quadratic term $Q_{0kf}^{1}$ which is evaluated using Equation~\eqref{budget}.
We have checked (not shown here though) that the quadratic terms 
could not be evaluated correctly by direct computation limited to the first 
200 modes, as higher-order contributions were significant, which is typical of
the closure problem. 
\berengeredeux{ We note that the x-axis (N) 
represents modes ordered by energy, so that continuity in wavenumber or 
frequency space is not enforced, which may explain the "noisy" appearance
of the plots.} 

%~\eqref{NSeq}
 For the sake of clarity, the left plots show mode number $N$ from $1-30$ and the right plots show $N$ from $30-200$. 
The top row (Figures~\ref{Lin_terms}a and b) shows that the real part of the production term $P_{lkf}^{n}$ is essentially balanced 
by the sum of the real part of the dissipation and the quadratic terms.
For the less energetic modes $N > 30$ (Figure ~\ref{Lin_terms}b) we note that the quadratic terms 
are mostly negative, 
which is consistent with the idea of a positive energy transfer from
the large scales (most energetic modes) to the small scales.
Figure~\ref{Lin_terms} shows that the pressure interaction term 
%which represents the effect of the outer layer on the wall layer modes 
%at the boundary and can be considered as an external forcing term, 
is very small compared to the other terms, 
in agreement with \cite{kn:aubry88}'s derivation 
where it is modeled as a stochastic term of small amplitude.
If we neglect the influence of the pressure term we have 
\begin{equation}
- \mathrm{Re}[Q_{0kf}^{1}] \sim P_{0kf}^{1} + D_{0kf}^{1}.
\end{equation}

%expected as it is known that pressure acts as a weak stochastic forcing on the outer layer. 
%Note that the $\frac{\partial u^i}{\partial t}$ terms is purely imaginary, and thus is not plotted in figure~\ref{Lin_terms}a and b.

% CHANGE THIS FIGURE

Figures~\ref{Lin_terms}c and d show that for the most energetic modes, as 
 $l=0$, the frequency $f$ of the mode is directly related to 
the imaginary part (denoted by $\mathrm{Im}$) of the nonlinear contributions 
to the mode $(l,k,f,n)$:
\begin{equation}
 \mathrm{Im}[Q_{0kf}^{1}] \sim \mathrm{i} \hspace{0.01in} \bar{f} \lambda_{0kf}^{n}. 
\end{equation}
This corresponds to the following closure approximation: 
\begin{equation}
\begin{split}
-Q_{0kf}^{1} \sim  \sum_{l'} \sum_{k'} \sum_{f'} \sum_{m} \sum_{p} 
 \left[\left<a_{l'k'f'}^{m}a_{l-l'k-k'f-f'}^{p}a_{0kf}^{*n}\right> 
\int \ddx{\phi_{l'k'f'}^{m,i}}{j} \phi_{l-l' k-k' f'f'}^{p,j} \phi_{0kf}^{*1,i} \mathrm{d}y \right]   \\
\sim P_{0kf}^{1} + D_{0kf}^{1} - \mathrm{i} \bar{f} \lambda_{0kf}^{1}.  
\end{split}
\end{equation}
As observed above, 
we can see that triple correlations between the different POD modes are 
essential to characterize the energy transfer between the scales.
Computing these correlations is difficult, however  
if $Q_{lkf}^{n}$ can be determined from Equation~\eqref{budget}, 
the decomposition offers a new way  
to reconstruct the quadratic terms using
\begin{equation}
- u^{j}\ddx{u^p}{j} 
= \sum_{l} \sum_{k} \sum_{f} \sum_{n} \frac{a_{lkf}^{n}}{\lambda_{lkf}^{n}} Q_{lkf}^{n}
\phi_{lkf}^{n,p}(y) \expbig 
\end{equation}
Substitution of the velocity decomposition into the left-hand side of the equation 
requires performing a cumbersome 
convolution on all coefficients $a_{lkf}^{n}$, but 
the right-hand-side 
provides a straightforward expression of the quadratic terms in the 
stochastic POD basis of coefficients $a_{lkf}^{n}$, where coordinates are 
solely determined from second-order statistics.
At each scale level $(l,k,f,n)$, the effect of the quadratic terms is to stretch and to rotate
the corresponding velocity mode by a factor $Q_{lkf}^{n}$.
Direct modeling of the distribution of $a_{lkf}^{n}$ could 
therefore lead to new formulations of turbulence models
in the decomposition framework. 

%A first step would be to examine if and how such relationships can be extended to non-zero streamwise wavenumbers $l$ and higher quantum numbers $n$.

%Figure~\ref{Quad_comp} shows that the quadratic contribution 
%is far larger than its restriction to the 
%first 200 modes, which indicates that the effect of the higher-order modes is essential. 
%We can then determine how much of this contribution is captured by a selection of modes.
%\begin{figure}
%\begin{subfigure}[b]{0.5\textwidth}
%\includegraphics[width=\linewidth]{Quad_real1-eps-converted-to.pdf}
%\caption{Real part, modes 1-30 }
%\end{subfigure}
%\begin{subfigure}[b]{0.5\textwidth}
%\includegraphics[width=\linewidth]{Quad_real2-eps-converted-to.pdf}
%\caption{Real part, modes 30-200}
%\end{subfigure}
%\begin{subfigure}[b]{0.5\textwidth}
%\includegraphics[width=\linewidth]{Quad_imag1-eps-converted-to.pdf}
%\caption{Imaginary part, modes 1-30}
%\end{subfigure}
%\begin{subfigure}[b]{0.5\textwidth}
%\includegraphics[width=\linewidth]{Quad_imag2-eps-converted-to.pdf}
%\caption{Imaginary part, modes 30-200}
%\end{subfigure}
%\caption{Comparison of the full quadratic term in the energy budget with that reconstructed using 200 dominant modes (excluding mean) for each mode.}
%\label{Quad_comp}
%\end{figure}

\section{Conclusion}
 Spatio-temporal Proper Orthogonal Decomposition has been applied to the wall layer of a
 turbulent channel flow. 
The decomposition represents an efficient data reduction technique which is adapted to 
large simulation databases. It brings to light typical features of wall turbulence in
a straightforward manner, but also provides a fresh viewpoint on the flow 
organization. 
 Due to symmetry properties, the decomposition singles out 
 empirical eigenfunctions for each frequency and horizontal spatial wavenumber.
%The analysis was focused on the first eigenfunction, which captures about 80\% of the turbulent kinetic energy in the region $y_+ < 80$.
 Besides time scales superior to 3000 wall units, 
 which our limited implementation of POD did not allow us to characterize fully,
 we have shown that the most energetic modes were characterized by 
a time scale on the order of 250 wall units, \berengere{which could have significant implications for control}. 
  Convection velocities could be directly defined from the POD spectrum. 
 A global convection velocity on the order of 12$u_{\tau}$ was identified in the wall layer, in 
good agreement with previous approaches.
\berengere{Examination of the spectrum provided an 
assessment and validation of the Taylor's frozen turbulence hypothesis.}

About 30\% of the turbulent kinetic energy was captured by the 200 most energetic modes.
 The most energetic modes were found to have a self-similar shape
that appeared largely independent from the wall-normal 
extent of the decomposition 
domain, which shows the coherence of the motions over the height of the boundary layer. 
\berengere{The modes appeared to be hierarchically organized, with a number of peaks
and a maximal peak location  at $l=0$ directly proportional to the quantum number $n$, indicating
that energy cascading towards the higher-order modes is directed away from the wall into the core region.}

Finally, substitution of the decomposition into the Navier-Stokes equation and
 numerical computation of the different contributions of the modes to the turbulent kinetic energy budget
 highlighted the key role played by quadratic interactions and 
allowed us to propose a new closure formulation to model 
the contribution of these interactions. 
Such relationships, which need to be further explored in a careful manner, could be 
useful to derive new turbulence models.
We hope that this work will pave the way for comprehensive 
investigations of wall turbulent flows
using spatio-temporal Proper Orthogonal Decomposition. 

\section*{Acknowledgements}
\berengere{We are grateful to W.K. George for motivating discussions. 
We thank the anonymous referees for pointing out valuable references and making helpful suggestions.}
This work was supported by the Center of Data Science from the Paris-Saclay University.
Computations were carried out at IDRIS-GENCI (project 02262). 

\bibliographystyle{jfm}
\bibliography{pod_article_new,biblio}

\end{document}

%% file: Lionel1rev.tex
% \documentclass[12pt]{elsart_LM}
% 
% \usepackage{amsmath, amssymb, color, natbib, mathrsfs}
% 
% \newcommand{\bxi}    {\boldsymbol{\xi}}
% \newcommand{\bu}     {\boldsymbol{u}}
% \newcommand{\bx}     {\boldsymbol{x}}
% \newcommand{\by}     {\boldsymbol{y}}
% \newcommand{\nt}     {n_t}
% \newcommand{\nsLM}     {n_s}
% \newcommand{\bcoefs} {\boldsymbol{c}}
% \newcommand{\coefs}  {c}
% \newcommand{\icomp}  {\mathrm{i}}
% \newcommand{\transpose}		    {\mathsf{T}}
% \DeclareMathOperator*{\argmin}	{{\mathrm{ arg\,min}}}
% \newcommand{\normLM}[2]		    {\left\|{#1}\right\|_{#2}}
% 
% \begin{document}
% 
% \begin{frontmatter}
% 
% \title{Preliminary thoughts for a revision of the spatio-temporal POD paper with Srikanth and B\'ereng\`ere}
% 
% \author{Contribution of Lionel}
% 
% \end{frontmatter}
% 
% 
% \section{On the connection with DMD and snapshot POD}

We now briefly explore the connection of the spatio-temporal POD with the snapshot POD and the Dynamic Mode Decomposition, two established methods for representing and analyzing fluid flows. \Lionel{The numerical cost and requirements of the present method are also discussed.}

\subsection{Snapshot POD} \label{Sec_snapPOD}
Snapshot POD \citep{Sirovich_87} relies on an ergodicity assumption to identify the ensemble average operator $\left<\cdot\right>$ with the time-average. In practice, one has to rely on a finite set of samples (snapshots) and the average operator reduces to an empirical algebraic average. Snapshot POD then considers the empirical spatial autocorrelation tensor $K_U\left(\bx, \bx'\right) = \nt^{-1} \, \sum_{j=1}^{\nt}{u\left(\bx,t_j\right) \, u\left(\bx',t_j\right)}$, with $\bx$ the vector-valued continuous space variable. Autocorrelation tensor eigenfunctions $\phi_j\left(\bx\right)$, associated with the largest eigenvalues $\lambda_j$, are the dominant POD modes. The operator involved in the Fredholm equation of the snapshot POD method being Hilbert-Schmidt, its eigenfunctions are orthonormal with respect to the retained inner product.
% % , usually chosen as (unweighted) Euclidean, $\phi_i^\transpose \phi_j = \delta_{i j}$.

The spatial autocorrelation tensor $K_U$ is estimated from a finite set of samples and then only provides an approximation of the true autocorrelation tensor $K_U^\star \left(\bx, \bx'\right) = \mathrm{lim}_{\card{\Omega_t} \ra \infty} \, \card{\Omega_t}^{-1} \int_{\Omega_t}{u\left(\bx,t\right) \, u\left(\bx',t\right) \, \mathrm{d}t}$. In general, the spatial POD modes are then only an approximation of the true autocorrelation eigenvectors. In the spatio-temporal POD approach, we take advantage of the closed-form solution of the Fredholm equation in homogeneous dimensions. No approximation is then introduced and the spatio-temporal modes are the exact eigenvectors of the autocorrelation tensor along the homogeneous dimensions.

From a dynamical perspective, the temporal evolution of snapshot POD mode $j$ is given by the orthogonal projection of the flow field onto the corresponding eigenfunction: 
%$a_j\left(t\right) = \phi_j^\transpose \bu\left(t\right)$, 
$a_j\left(t\right) = (\phi_j , \bu\left(t\right))$, 
where $\bu\left(t\right) \equiv u\left(\bx,t\right)$ and 
%considering the %Euclidean inner product. 
 $\left( , \right)$ 
\berengeredeux{ 
represents the (typically weighted Euclidean) inner product.} 
The time evolution of mode $j$ then inherits many properties of the flow field such as its wide frequency content. It is important to note that, since it results from a projection, there is no guarantee that the dynamics of a snapshot POD mode is smooth in time.

In contrast, the present spatio-temporal POD relies on a spectral decomposition in the homogeneous dimensions, including time. Each mode then follows a smooth (harmonic) dynamics. The original flow field being approximated by a finite linear combination of smooth modes, it remains smooth in time. More generally, spatio-temporal POD enjoys structure in both time and homogeneous dimensions and allows to interpret modes as %\emph{atoms}, or 
coherent structures, localized in the frequency-spatial wavenumber space. In contrast, snapshot POD only enjoys structure in space and lacks structure in time, preventing identification of its modes with coherent structures which dynamics is essentially smooth.

\subsection{DMD} \label{Sec_DMD}
The Dynamic Mode Decomposition (DMD) is another popular method for modal decomposition of fluid flows. Different definitions have been considered in the literature but we here focus on the original formulation discussed in \cite{Schmid_JFM10}. Considering the flow field at a collection of $N_x$ points in space as a vector-valued observable, and assuming it is a state vector for the underlying physical system, the DMD can be closely related to the Koopman theory, \cite{Mezic_05, Rowley_09}. In a nutshell, the DMD estimates the eigenvectors and associated eigenvalues of the linear operator mapping the discrete flow field at a given time to a subsequent time $\TDMD$ in the future. 
Specifically, $\nt$ time-ordered snapshots of the flow field sampled every $\TDMD$ in time are collected in a matrix $U$. Each snapshot is of size $N_x$ and constitutes a vector-valued observable $\bu$ of the state vector of the underlying physical system. Letting $U_1$ be the first $\nt-1$ columns of $U$ and $U_2$ the last $\nt-1$ columns, the DMD is concerned with the characterization of the linear operator $A$ such that $U_2 \simeq A \, U_1$. Formulating the approximation as an optimization problem, the eigendecomposition of the matrix $A$ can be obtained: $A \, \Psi = \Psi \, \Lambda$, with $\Psi \in \mathbb{C}^{\ntmone \times \ntmone}$ the matrix of eigenvectors and $\Lambda \in \mathbb{C}^{\ntmone \times \ntmone}$ the diagonal matrix of eigenvalues $\lambda_j$. The flow field $u\left(\bx, t\right)$ is identified with $\bu\left(t\right)$ and can then be approximated at a time $t+l \, \TDMD$ as $$\bu\left(t+l \, \TDMD\right) \approx \Psi \, \mathrm{diag}\left(\Psi^{-1} \, \bu\left(t\right)\right) \, \boldsymbol{\lambda}^l = \Xi \, \boldsymbol{\lambda}^l, \qquad \forall \, l \in \mathbb{Z},$$ where $\boldsymbol{\lambda}^l = \left(\lambda_1^l \: \lambda_2^l\ldots \lambda_{\nt-1}^l \right)^\transpose$ and the matrix $\Xi = \left(\bxi_1 \: \bxi_2 \ldots \bxi_{\nt-1}\right)$ contains the so-called \emph{DMD modes} $\bxi_j$.

Assuming the collection of snapshots is linearly independent and subtracting the empirical algebraic time-average from the snapshot data $U$, the DMD is equivalent to the temporal Discrete Fourier Transform (DFT), \cite{ChenDMD12}. Specifically, denoting zero-mean quantities with a superscript $^\odot$, the vector-valued observable $\bu^{\odot}\left(t + l \, \TDMD\right)$ then obeys $$\bu^{\odot}\left(t + l \, \TDMD\right) = \sum_{j=1}^{\nt-1}{\exp\left(\frac{2 \pi \icomp l j}{\nt}\right) \, \bxi_j}.$$

In this context, DMD decomposes the flow field in monochromatic spatial modes $\bxi_j$ oscillating at a given frequency $j \slash \left(\TDMD \, \nt\right)$ in time.
Both the spatio-temporal POD and DMD then derive modes oscillating at a given frequency. The sampling frequency being a multiple integer of these frequencies, both methods represent the data with modes at the \emph{same frequencies}. 

%\srikanth{Dynamic mode decomposition (DMD) is another recent technique that aims at obtaining
%coherent structures from data that capture the temporal information thereby overcoming the
%drawbacks of classic spatial POD. 
%\citep{kn:colonius18}  have shown that Spectral POD is closely
%related to DMD, such that they can be considered as ensemble averaged DMD modes in the case of
%stationary flows. 
%Classic DMD is applied to one flow realization, and thus a mode computed at
%given frequency will have statistical variability over a set of realizations. 
%Spectral POD
%can be viewed as an optimal basis that account for the variability of modes over an ensemble of
%realizations. 
%Towne et al. have also shown connections between Spectral POD and resolvent analysis.
%We would like to note that spatio-temporal POD is just a fully spectral POD as the flow is homogeneous along streamwise and spanwise directions and thus the Fourier transform is applied in both space and time, and correlation function is
%computed in the only inhomogeneous wall-normal direction.
%}

\berengere{However, an essential difference between DMD and POD is that there is no notion of ensemble average
in DMD (only one sample, corresponding to the set of snapshots, is considered), and DMD modes are extracted from mapping one snapshot in the time series to the next one.
\srikanth{Standard DMD is applied to one flow realization, and thus a mode computed at
given frequency would have statistical variability over a set of realizations.} 
In contrast, in the present decomposition, ensemble average is a key feature of the procedure and requires
several samples. POD modes are extracted from the autocorrelation tensor  built from the different samples.  
They constitute an \srikanth{optimal basis that account for the variability of modes over an ensemble of realizations.} 
We point out that in our numerical implementation, the different samples are obtained by breaking down a single 
series of snapshots into different non-overlapping blocks.   
%We note that the connection of DMD and snapshot POD with a POD involving Fourier transform in time is also discussed in \cite{kn:colonius18} who have introduced a method called spectral POD, which also has many similarities with our present implementation.
%However they use the method of snapshots for all the spatial variables, whereas we apply Fourier transform in the
%two homogeneous spatial directions and use the direct  POD method in the remaining inhomogeneous direction.
} 

%However, the spatio-temporal POD here relies on blocks of size $\nt$ in time while DMD is typically computed on the whole dataset at once, of length $\TDMD \, \nsLM \, \nt$ in time (non-overlapping blocks). The lower part of the spectrum is hence wider with DMD, spanning from a frequency $1 \slash \left(\TDMD \, \nsLM \, \nt\right) = 1 \slash \left(T \, \nt\right)$ upwards, while spatio-temporal POD describes a spectrum starting at a frequency of $1 \slash \left(\TDMD \, \nt\right) = 1 \slash T$, with $T = \TDMD \, \nt$.

%Let $\dumu_{l k f}\left(y; x, z, t\right) = u_{l k f}\left(y\right) \, \expbig$ be a spatio-temporal mode evaluated at time $t$. It oscillates at a frequency $f \slash T$ in time. Considering the value of this mode $\TDMD$ in the future, it can be written as $\dumu_{l k f}\left(y; x, z, t+\TDMD\right) = \dumu_{l k f}\left(y; x, z, t\right) \, \exp\left(2 \pi \icomp f \TDMD \slash T\right) = A \, \dumu_{l k f}\left(y; x, z, t\right)$ where reference is made to the operator $A$ mapping $\TDMD$ into the future. The mode $\dumu_{l k f}$ is then an eigenfunction of the mapping operator $A$, with associated eigenvalue $\lambda = \exp\left(2 \pi \icomp f \TDMD \slash T\right)$. It is hence also a DMD mode. 

%\textbf{MORE !!}

%% file: Lionel2rev.tex
% \documentclass[12pt]{elsart_LM}
% 
% \usepackage{amsmath, amssymb, color, natbib, mathrsfs}
% 
% \newcommand{\bxi}    {\boldsymbol{\xi}}
% \newcommand{\bu}     {\boldsymbol{u}}
% \newcommand{\bx}     {\boldsymbol{x}}
% \newcommand{\by}     {\boldsymbol{y}}
% \newcommand{\nt}     {n_t}
% \newcommand{\nsLM}     {n_s}
% \newcommand{\bcoefs} {\boldsymbol{c}}
% \newcommand{\coefs}  {c}
% \newcommand{\icomp}  {\mathrm{i}}
% \newcommand{\transpose}		    {\mathsf{T}}
% \DeclareMathOperator*{\argmin}	{{\mathrm{ arg\,min}}}
% \newcommand{\normLM}[2]		    {\left\|{#1}\right\|_{#2}}
% 
% \begin{document}
% 
% \begin{frontmatter}
% 
% \title{Preliminary thoughts for a revision of the spatio-temporal POD paper with Srikanth and B\'ereng\`ere}
% 
% \author{Contribution of Lionel}
% 
% \end{frontmatter}
% 
% 
% \section{On the connection with DMD and snapshot POD}

\subsection{Requirements and cost of the decomposition} \label{Sec_Costs}

The spatio-temporal POD discussed in this paper relies on the same information as the snapshot POD or the DMD. Owing to the DFT in time and homogeneous spatial directions, it however involves distinctive differences such as the requirement of collecting time-resolved snapshots. This is in contrast with the other methods which only require time-sampling in pairs (DMD) or a set of snapshots sampled in time according to an approximately ergodic measure (snapshot POD). 

\berengere{Obtaining time-resolved data in the full spatial domain  from a numerical simulation 
is  not  a problem,  but requires 
 integration over a sufficiently long period of time, especially 
if it needs to be broken down into different samples.
In contrast, in an experiment, obtaining different samples over relatively long periods of time 
is relatively easy, but 
access to full spatial resolution may be more difficult, especially for time-resolved data, 
despite recent advances in PIV techniques (see for instance \citet{westerwheel13}). 
However, it should be borne in mind that ensemble average and Fourier transform commute,
so that full information in the spatial domain does not have 
to be acquired simultaneously, since  the autocorrelation tensor for given 
spatial separations in  homogeneous directions
can be computed independently.  
If there is one homogeneous direction (say $x$), the autocorrelation tensor
\Lionel{$K_U(x_1,y_1,x_2,y_2)= f(x_2-x_1,y_1,y_2)$}
can be evaluated independently for each separation $x_2-x_1$
from two simultaneous planes $x=x_{1}$ and $x=x_2$: the information does
not have to be acquired for all positions $x$ at the same time, only pairs
of simultaneous positions are necessary. For the same reason,
in the case of two homogeneous directions ($x$ and $z$), one can show that 
 the full spatial autocorrelation tensor at all separations \Lionel{$(x_1-x_2,z_1-z_2)$}
can be entirely recovered from two orthogonal planes 
$x=\mathrm{cst}$ and $z=\mathrm{cst}$ (as was done in the WALLTURB experiment,
\citet{wallturb}).}

We now briefly discuss the numerical cost of the solution method presented in this paper. We consider the common situation where the sampling is uniform in time and sufficient finely resolved with respect to the finest scales one is interested in. A Discrete Fourier Transform is then applied in homogeneous dimensions. We use a multidimensional Fast Fourier Transform (FFT) which essentially amounts to the composition of a sequence of one-dimensional FFTs along each homogeneous dimension. The numerical cost of this step then retains the $O\left(N \, \log\left(N\right)\right)$ scaling of FFT techniques, where $N$ is here $N_l, N_k$ or $\nt$. \Lionel{Efficient algorithms also exist in case of non-uniform sampling, \textit{e.g.}, \cite{RuizAntolin_Townsend_18}, and retain the $O\left(N \, \log\left(N\right)\right)$ scaling.}
The remaining step involves a POD in the non-homogeneous dimension for each atom of triad $\left(l, k, f\right)$. In the present case, it then reduces to a set of \emph{independent} one-dimensional PODs. %, embarrassingly parallelizable.

Finally, we would like to stress that the eigenvalue problems involved in every method discussed here (spatio-temporal POD, spatial POD, DMD) can be advantageously solved by 
\berengere{recent} %modern 
numerical techniques.  In particular,
 sketching and randomized methods \citep{Halko_etal_SIAMREV_11,kannan_vempala_2017} can very significantly alleviate the numerical cost and the memory requirement in cases only part of the spectrum is required, as is often the case in our applications.

% \section{Disclaimer}
% This whole contribution is FAR from polished and requires extensive work\ldots
% 
% \bibliographystyle{lionel_style}
% \bibliography{biblio,pod_article_BP}
% 
% 
% \end{document}

%% file: final_arXiv.bbl
\begin{thebibliography}{53}
\expandafter\ifx\csname natexlab\endcsname\relax\def\natexlab#1{#1}\fi
\def\au#1{#1} \def\ed#1{#1} \def\yr#1{#1}\def\at#1{#1}\def\jt#1{\textit{#1}}
  \def\bt#1{#1}\def\bvol#1{\textbf{#1}} \def\vol#1{#1} \def\pg#1{#1}
  \def\publ#1{#1}\def\arxiv#1{#1}\def\org#1{#1}\def\st#1{\textit{#1}}

\bibitem[Alamo \& Jimenez(2009)]{kn:alamojimenez09}
{\sc \au{Alamo, J. C.~Del} \& \au{Jimenez, J.}} \yr{2009}  \at{Estimation of
  turbulent convection velocities and corrections to taylor's approximation}.
  \jt{Journal of Fluid Mechanics}  \bvol{640},  \pg{5--26}.

\bibitem[Alamo {\em et~al.\/}(2006)Alamo, Jimenez, Zandonade \&
  Moser]{kn:delalamo06}
{\sc \au{Alamo, J. C.~Del}, \au{Jimenez, J.}, \au{Zandonade, P.} \& \au{Moser,
  R.~D.}} \yr{2006}  \at{Self-similar vortex clusters in the turbulent
  logarithmic region}.  \jt{Journal of Fluid Mechanics}  \bvol{561},
  \pg{329--358}.

\bibitem[Arndt {\em et~al.\/}(1997)Arndt, Long \& Glauser]{kn:arndt97}
{\sc \au{Arndt, R.E.A}, \au{Long, D.F.} \& \au{Glauser, M.N.}} \yr{1997}
  \at{The proper orthogonal decomposition of pressure surrounding a turbulent
  jet}.  \jt{J. Fluid Mech.}  \bvol{340},  \pg{1--33}.

\bibitem[Aubry {\em et~al.\/}(1988)Aubry, Holmes, Lumley \& Stone]{kn:aubry88}
{\sc \au{Aubry, N.}, \au{Holmes, P.}, \au{Lumley, J.~L.} \& \au{Stone, E.}}
  \yr{1988}  \at{The dynamics of coherent structures in the wall region of the
  wall boundary layer}.  \jt{Journal of Fluid Mechanics}  \bvol{192},
  \pg{115--173}.

\bibitem[Berkooz {\em et~al.\/}(1993)Berkooz, Holmes \& Lumley]{berkooz1993}
{\sc \au{Berkooz, G.}, \au{Holmes, P.} \& \au{Lumley, J.~L.}} \yr{1993}
  \at{The proper orthogonal decomposition in the analysis of turbulent flows}.
  \jt{Annual review of fluid mechanics}  \bvol{25}~(1),  \pg{539--575}.

\bibitem[Blackwelder \& Haritodinis(1983)]{kn:blackwelder83}
{\sc \au{Blackwelder, R.~F.} \& \au{Haritodinis, J.~H.}} \yr{1983}  \at{Scaling
  of the bursting frequency in turbulent boundary layers}.  \jt{Journal of
  Fluid Mechanics}  \bvol{132},  \pg{87--103}.

\bibitem[Chen {\em et~al.\/}(2012)Chen, Tu \& Rowley]{ChenDMD12}
{\sc \au{Chen, K.K.}, \au{Tu, J.H.} \& \au{Rowley, C.W.}} \yr{2012}
  \at{Variants of dynamic mode decomposition: boundary condition, koopman, and
  fourier analyses}.  \jt{J. Nonlinear Sci.}  \bvol{22}~(6),  \pg{887--915}.

\bibitem[Choi {\em et~al.\/}(1998)Choi, Debisschop \& Clayton]{kn:kschoi98}
{\sc \au{Choi, K.S.}, \au{Debisschop, J.R.} \& \au{Clayton, B.R.}} \yr{1998}
  \at{Turbulent boundary layer control by means of spanwise wall oscillations}.
   \jt{AIAA Journal}  \bvol{36}~(7),  \pg{1157--1163}.

\bibitem[Citriniti \& George(2000)]{kn:citriniti00}
{\sc \au{Citriniti, J.} \& \au{George, W.}} \yr{2000}  \at{Reconstruction of
  the global velocity field in the axisymmetric mixing layer utilizing the
  proper orthogonal decomposition}.  \jt{J. Fluid Mech.}  \bvol{418},
  \pg{137--166}.

\bibitem[Corino \& Brodkey(1969)]{kn:corinobrod}
{\sc \au{Corino, E.~R.} \& \au{Brodkey, R.~S.}} \yr{1969}  \at{A visual
  investigation of the wall region in turbulent flow}.  \jt{Journal of Fluid
  Mechanics}  \bvol{37},  \pg{1--30}.

\bibitem[Delville {\em et~al.\/}(1999)Delville, Ukeiley, Cordier, Bonnet \&
  Glauser]{kn:delville99}
{\sc \au{Delville, J.}, \au{Ukeiley, L.}, \au{Cordier, L.}, \au{Bonnet, J.P.}
  \& \au{Glauser, M.}} \yr{1999}  \at{Examination of large-scale structures in
  a turbulent plane mixing layer. part 1. proper orthogonal decomposition}.
  \jt{J. Fluid Mech.}  \bvol{391},  \pg{91--122}.

\bibitem[Dennis(2015)]{DENNIS2015}
{\sc \au{Dennis, David~J.C.}} \yr{2015}  \at{{Coherent structures in
  wall-bounded turbulence}}.  \jt{{Anais da Academia Brasileira de Ciencias}}
  \bvol{87},  \pg{1161 -- 1193}.

\bibitem[Frisch(1995)]{frisch}
{\sc \au{Frisch, U.}} \yr{1995} {\em Turbulence\/}.  \publ{Cambridge University
  Press}.

\bibitem[Gatski \& Glauser(1992)]{kn:gatskiglauser92}
{\sc \au{Gatski, M.} \& \au{Glauser, M.}} \yr{1992} Proper orthogonal
  decomposition based turbulence modeling.  \bt{In {\em Instability, Transition
  and Turbulence\/} (ed. \ed{M.Y. Hussaini, A.~Kumar \& C.L. Streett})}.
  \publ{Springer}.

\bibitem[George(2017)]{george2017}
{\sc \au{George, W.~K.}} \yr{2017}  \at{A 50-year retrospective and the
  future}.  \bt{In {\em Whither Turbulence and Big Data in the 21st
  Century?\/}},  \pg{pp. 13--43}.  \publ{Springer}.

\bibitem[Glauser \& George(1987)]{kn:glausergeorge}
{\sc \au{Glauser, M.N.} \& \au{George, W.K.}} \yr{1987} An orthogonal
  decomposition of the axisymmetric jet mixing layer utilizing cross-wire
  measurements.  \bt{In {\em Proceedings of the Sixth Symposium on Turbulent
  Shear Flow\/}}.  \publ{Toulouse}.

\bibitem[Glauser {\em et~al.\/}(1983)Glauser, Leib \&
  George]{kn:glauserleibgeorge}
{\sc \au{Glauser, M.N.}, \au{Leib, S.J.} \& \au{George, W.K.}} \yr{1983} An
  application of lumley's orthogonal decomposition to the axisymmetric jet
  mixing layer.  \bt{In {\em Bulletin of American Physical Society\/}}.
  \publ{DFD Meeting, Houston, Texas}.

\bibitem[Halko {\em et~al.\/}(2011)Halko, Martinsson \&
  Tropp]{Halko_etal_SIAMREV_11}
{\sc \au{Halko, N.}, \au{Martinsson, P.G.} \& \au{Tropp, J.A.}} \yr{2011}
  \at{Finding structure with randomness: probabilistic algorithms for
  constructing approximate matrix decompositions}.  \jt{SIAM Rev.}
  \bvol{53}~(2),  \pg{217--288}.

\bibitem[Hamilton {\em et~al.\/}(1995)Hamilton, Kim \&
  Waleffe]{kn:kimhamwaleffe}
{\sc \au{Hamilton, J.~M.}, \au{Kim, J.} \& \au{Waleffe, F.}} \yr{1995}
  \at{Regeneration mechanisms of near-wall turbulence structures}.  \jt{Journal
  of Fluid Mechanics}  \bvol{287},  \pg{317--348}.

\bibitem[Hong \& Rubesin(1985)]{kn:hongrubesin85}
{\sc \au{Hong, S.K.} \& \au{Rubesin, M.W.}} \yr{1985}  \bt{Application of
  large-eddy interaction model to channel flow}. {\em Tech. Rep.\/} 86691.
  \org{NASA Technical Memorandum}.

\bibitem[Jim{\'e}nez(2013)]{jimenez2013}
{\sc \au{Jim{\'e}nez, J.}} \yr{2013}  \at{Near-wall turbulence}.  \jt{Physics
  of Fluids}  \bvol{25}~(10),  \pg{101302}.

\bibitem[Jimenez \& Moin(1991)]{kn:jimemoin}
{\sc \au{Jimenez, J.} \& \au{Moin, P.}} \yr{1991}  \at{The minimal flow unit in
  near-wall turbulence}.  \jt{Journal of Fluid Mechanics}  \bvol{225},
  \pg{213--240}.

\bibitem[Kannan \& Vempala(2017)]{kannan_vempala_2017}
{\sc \au{Kannan, R.} \& \au{Vempala, S.}} \yr{2017}  \at{Randomized algorithms
  in numerical linear algebra}.  \jt{Acta Numer.}  \bvol{26},  \pg{95--135}.

\bibitem[Kim {\em et~al.\/}(1971)Kim, Kline \& Reynolds]{kn:kim71}
{\sc \au{Kim, H.~T.}, \au{Kline, S.~J.} \& \au{Reynolds, W.~C.}} \yr{1971}
  \at{The production of turbulence near a smooth wall in a turbulent boundary
  layer}.  \jt{Journal of Fluid Mechanics}  \bvol{50}~(1),  \pg{133--160}.

\bibitem[Kreplin \& Eckelmann(1979)]{kn:kreplin79}
{\sc \au{Kreplin, H.~P.} \& \au{Eckelmann, H.}} \yr{1979}  \at{Propagation of
  perturbations in the viscous sublayer and adjacent wall region}.  \jt{Journal
  of Fluid Mechanics}  \bvol{95},  \pg{305--322}.

\bibitem[Krogstad {\em et~al.\/}(1998)Krogstad, Kaspersen \&
  Rinestead]{kn:krogstad98}
{\sc \au{Krogstad, P.~A.}, \au{Kaspersen, J.~H.} \& \au{Rinestead, S.}}
  \yr{1998}  \at{Convection velocities in a turbulent boundary layer}.
  \jt{Physics of Fluids}  \bvol{10},  \pg{949}.

\bibitem[Lo\`eve(1977)]{kn:loeve}
{\sc \au{Lo\`eve, M.}} \yr{1977} {\em Probability Theory\/}.  \publ{Berlin,
  Germany: Springer}.

\bibitem[Lumley(1965)]{kn:lumley65}
{\sc \au{Lumley, J.L.}} \yr{1965}  \at{On the interpretation of temporal
  spectra in high intensity shear flows}.  \jt{Phys. Fluids}  \bvol{8},
  \pg{1056}.

\bibitem[Lumley(1967)]{kn:lumleyPOD}
{\sc \au{Lumley, J.~L.}} \yr{1967}  \at{The structure of inhomogeneous
  turbulent flows}.  \bt{In {\em Atmospheric Turbulence and Radio Wave
  Propagation\/}},  \pg{pp. 221--227}.  \publ{Nauka, Moscow}.

\bibitem[McKeon(2017)]{kn:mckeon2017}
{\sc \au{McKeon, B.~J.}} \yr{2017}  \at{The engine behind (wall) turbulence:
  perspectives on scale interactions}.  \jt{Journal of Fluid Mechanics}
  \bvol{817},  \pg{1}.

\bibitem[Mezi\'c(2005)]{Mezic_05}
{\sc \au{Mezi\'c, I.}} \yr{2005}  \at{Spectral properties of dynamical systems,
  model reduction and decompositions}.  \jt{Nonlinear Dyn.}  \bvol{41}~(1),
  \pg{309--325}.

\bibitem[Moin \& Moser(1989)]{moin1989}
{\sc \au{Moin, P.} \& \au{Moser, R.~D.}} \yr{1989}  \at{Characteristic-eddy
  decomposition of turbulence in a channel}.  \jt{Journal of Fluid Mechanics}
  \bvol{200},  \pg{471--509}.

\bibitem[Moser {\em et~al.\/}(1999)Moser, Kim \& Mansour]{moser1999}
{\sc \au{Moser, R.~D.}, \au{Kim, J.} \& \au{Mansour, N.~N.}} \yr{1999}
  \at{Direct numerical simulation of turbulent channel flow up to
  {$\mathrm{Re}_\tau$= 590}}.  \jt{Physics of Fluids}  \bvol{11}~(4),
  \pg{943--945}.

\bibitem[Podvin(2001)]{podvin2001}
{\sc \au{Podvin, B.}} \yr{2001}  \at{On the adequacy of the ten-dimensional
  model for the wall layer}.  \jt{Physics of Fluids}  \bvol{13}~(1),
  \pg{210--224}.

\bibitem[Podvin \& Fraigneau(2014)]{podvin2014}
{\sc \au{Podvin, B.} \& \au{Fraigneau, Y.}} \yr{2014}  \at{{POD}-based wall
  boundary conditions for the numerical simulation of turbulent channel flows}.
   \jt{Journal of Turbulence}  \bvol{15}~(3),  \pg{145--171}.

\bibitem[Podvin \& Fraigneau(2017)]{kn:pof17}
{\sc \au{Podvin, B.} \& \au{Fraigneau, Y.}} \yr{2017}  \at{A few thoughts on
  proper orthogonal decomposition in turbulence}.  \jt{Physics of Fluids}
  \bvol{29},  \pg{531}.

\bibitem[Podvin {\em et~al.\/}(2010)Podvin, Fraigneau, Jouanguy \&
  Laval]{kn:jfe10}
{\sc \au{Podvin, B.}, \au{Fraigneau, Y.}, \au{Jouanguy, J.} \& \au{Laval,
  J.~P.}} \yr{2010}  \at{On self-similarity in the inner wall layer of a
  turbulent channel flow}.  \jt{Journal of Fluids Engineering}  \bvol{132}~(4),
   \pg{41202}.

\bibitem[Podvin \& Lumley(1998)]{kn:jfm98}
{\sc \au{Podvin, B.} \& \au{Lumley, J.~L.}} \yr{1998}  \at{A low-dimensional
  approach for the minimal flow unit}.  \jt{Journal of Fluid Mechanics}
  \bvol{362},  \pg{121--155}.

\bibitem[Quadrizio \& Ricco(2004)]{kn:quadrizioricco04}
{\sc \au{Quadrizio, M.} \& \au{Ricco, P.}} \yr{2004}  \at{Critical assessment
  of drag reduction through spanwise wall oscillations}.  \jt{Journal of Fluid
  Mechanics}  \bvol{521},  \pg{251--271}.

\bibitem[Robinson(1991)]{kn:robinson}
{\sc \au{Robinson, S.~K.}} \yr{1991}  \at{Coherent motions in the turbulent
  boundary layer}.  \jt{Annual Review of Fluid Mechanics}  \bvol{23},
  \pg{601--639}.

\bibitem[Rowley {\em et~al.\/}(2009)Rowley, Mezi\'c, Bagheri, Schlatter \&
  Henningson]{Rowley_09}
{\sc \au{Rowley, C.W.}, \au{Mezi\'c, I.}, \au{Bagheri, S.}, \au{Schlatter, P.}
  \& \au{Henningson, D.S.}} \yr{2009}  \at{Spectral analysis of nonlinear
  flows}.  \jt{J. Fluid Mech.}  \bvol{641},  \pg{115--127}.

\bibitem[Ruiz-Antol\'in \& Townsend(2018)]{RuizAntolin_Townsend_18}
{\sc \au{Ruiz-Antol\'in, D.} \& \au{Townsend, A.}} \yr{2018}  \at{A nonuniform
  fast {F}ourier transform based on low rank approximation}.  \jt{SIAM J. Sci.
  Comput.}  \bvol{40}~(1),  \pg{A529–A547}.

\bibitem[Schmid(2010)]{Schmid_JFM10}
{\sc \au{Schmid, P.J.}} \yr{2010}  \at{Dynamic mode decomposition of numerical
  and experimental data}.  \jt{J. Fluid Mech.}  \bvol{656},  \pg{5--28}.

\bibitem[Sirovich(1987)]{Sirovich_87}
{\sc \au{Sirovich, L.}} \yr{1987}  \at{Turbulence and the dynamics of coherent
  structures}.  \jt{Quart. J. Appl. Math.}  \bvol{45},  \pg{561--590}.

\bibitem[Smits {\em et~al.\/}(2011)Smits, McKeon \& Marusic]{kn:smits11}
{\sc \au{Smits, A.~J.}, \au{McKeon, B.~J.} \& \au{Marusic, I.}} \yr{2011}
  \at{High–reynolds number wall turbulence}.  \jt{Annual Review of Fluid
  Mechanics}  \bvol{43}~(1),  \pg{353--375}.

\bibitem[Stanislas {\em et~al.\/}(2011)Stanislas, Jimenez \& Marusic]{wallturb}
{\sc \au{Stanislas, M.}, \au{Jimenez, J.} \& \au{Marusic, I.}}, ed. \yr{2011}
  {\em Progress in Wall Turbulence: Understanding and Modeling\/}.
  \publ{Springer}.

\bibitem[Stanislas {\em et~al.\/}(2008)Stanislas, Perret \&
  Foucaut]{kn:stanislas08}
{\sc \au{Stanislas, M.}, \au{Perret, L.} \& \au{Foucaut, J.~M.}} \yr{2008}
  \at{Vortical structures in the turbulent boundary layer: a possible route to
  a universal representation}.  \jt{Journal of Fluid Mechanics}  \bvol{602},
  \pg{327--382}.

\bibitem[Towne {\em et~al.\/}(2018)Towne, Schmidt \& Colonius]{kn:colonius18}
{\sc \au{Towne, A.}, \au{Schmidt, O.} \& \au{Colonius, T.}} \yr{2018}
  \at{Spectral proper orthogonal decomposition and its relationship to dynamic
  mode decomposition and resolvent analysis}.  \jt{J. Fluid Mech.}  \bvol{847},
   \pg{821--867}.

\bibitem[Townsend(1976)]{kn:town}
{\sc \au{Townsend, A.~A.}} \yr{1976} {\em The Structure of turbulent shear
  flow\/}.  \publ{Cambridge University Press}.

\bibitem[Ukeiley {\em et~al.\/}(2001)Ukeiley, Cordier, Manceau, Delville,
  Glauser \& Bonnet]{kn:ukeiley01}
{\sc \au{Ukeiley, L.}, \au{Cordier, L.}, \au{Manceau, R.}, \au{Delville, J.},
  \au{Glauser, M.} \& \au{Bonnet, J.}} \yr{2001}  \at{Examination of large-
  scale structures in a turbulent plane mixing layer. part 2. dynamical systems
  model}.  \jt{J. Fluid Mech.}  \bvol{441},  \pg{67--108}.

\bibitem[Wallace(2014)]{kn:wallace14}
{\sc \au{Wallace, J.~M.}} \yr{2014}  \at{Space-time correlations in turbulent
  flow: A review}.  \jt{Theoretical and Applied Mechanics Letters}  \bvol{4},
  \pg{0022003}.

\bibitem[Westerwheel {\em et~al.\/}(2013)Westerwheel, Elsinga \&
  Adrian]{westerwheel13}
{\sc \au{Westerwheel, J.}, \au{Elsinga, G.~E.} \& \au{Adrian, R.~J.}} \yr{2013}
   \at{Particle image velocimetry for complex and turbulent flows}.  \jt{Ann.
  Review Fluid Mech.}  \bvol{45},  \pg{409--436}.

\bibitem[Zhou(1993)]{kn:zhou93}
{\sc \au{Zhou, J.}} \yr{1993}  \bt{Interacting scales and energy transfer in
  isotropic turbulence}. {\em Tech. Rep.\/} CR-191477.  \org{NASA}.

\end{thebibliography}
